\documentclass[aps,twocolumn,showpacs,superscriptaddress,amsmath,amssymb]{revtex4}
\usepackage{graphicx}
\usepackage{graphics}
\usepackage{epsfig}
\usepackage{verbatim}
\usepackage{amssymb}

\newcommand{\bea}{\begin{eqnarray}}
\newcommand{\beq}{\begin{equation}}
\newcommand{\eea}{\end{eqnarray}}
\newcommand{\eeq}{\end{equation}}
\begin{document}
\title{Shape transitions in two-body systems: two-electron quantum dots 
in a magnetic field}
\author{N. S. Simonovi\'c}
\affiliation{Institute of Physics,
University of Belgrade, P.O. Box 57, 11001 Belgrade, Serbia}
\author{R.G. Nazmitdinov}
\affiliation{Bogoliubov Laboratory of Theoretical Physics,
Joint Institute for Nuclear Research, 141980 Dubna, Russia}
\affiliation{Departament de F{\'\i}sica,
Universitat de les Illes Balears, E-07122 Palma de Mallorca, Spain}
\affiliation{Dubna State University, 141982 Dubna, Moscow region, Russia}

\begin{abstract}
We present a thorough analysis of the electron density distribution (shape) of
two electrons, confined in the three-dimensional harmonic oscillator potential,
as a function of the perpendicular magnetic field.
Explicit algebraic expressions are derived in terms of the system's parameters 
and the magnetic field strength to trace the shape transformations in the 
ground and low-lying excited states.
We found that the interplay of the classical and quantum properties lead to 
a quantum shape transition from a lateral to a vertical localization of electrons 
in low-lying excited states at relatively strong Coulomb interaction with 
alteration of the magnetic field. In contrast, in that regime in the 
ground states the electrons form always a ring type distribution in the lateral 
plane. The analytical results demonstrate a good agreement with quantum numerical 
results  near the transition point and at high magnetic field.
\end{abstract}
\maketitle

\section{Introduction}
Shape transitions  in quantum systems belong to symmetry breaking phenomenon 
which is specific to
finite systems \cite{bir}, when, under varying external or internal parameters, 
a finite system exhibits a change of shape. This change can be spontaneous: the 
system acquires the chosen form because it is energetically profitable. In this connection,
due to finite number of particles, quantum fluctuations play essential role.
They transform phase transitions observed in bulk systems to crossover from one 
type of symmetry to another one, and/or from one shape to another. In the latter case, 
although a partial symmetry may be preserved, the change of the density distribution 
can affect the physical properties of a
finite quantum system.

Nowadays, technological achievements in nanotechnology have opened wide opportunities 
for studying in detail such crossovers  in mesoscopic systems, which properties are 
defined by the interplay between microscopic (quantum) and macroscopic (classical) 
constituents. A semiconductor quantum dot (QD), where a few electrons are confined 
electrostatically to a nanometer sized region, provide a convenient correlation range 
ground to study this interplay.

Long ago Wigner predicted that electrons interacting by means of Coulomb forces could 
create a crystallized structure in a three-dimensional (3D) space at low enough 
densities and temperatures \cite{Wigner}. At these conditions the potential energy 
dominates over the kinetic energy and defines equilibrium configurations of electronic systems.
 Such crystallization in QDs is expected to result in the formation of the so-called Wigner
molecule in a two-dimensional (2D) case  \cite{Maksym,stef,yan}.
Generally, electrons in the Wigner molecule should
localize in space, i.e., they occupy fixed sites in a rotating
frame \cite{Mueller}. Within a 2D approach, a criterion for the
onset of the Wigner crystallization in QDs could be the appearance of a
local electron density minimum at the center of a dot
\cite{Creffield}. For a circular QD, created in a thin
semiconductor layer, this leads to a radial modulation in the
electron density, resulting in the formation of rings (see, e.g., \cite{peet,mar}).

\begin{figure*}[bth]
\vspace{-5cm}
\begin{center}
\hspace{3cm}
\includegraphics[scale=.35]{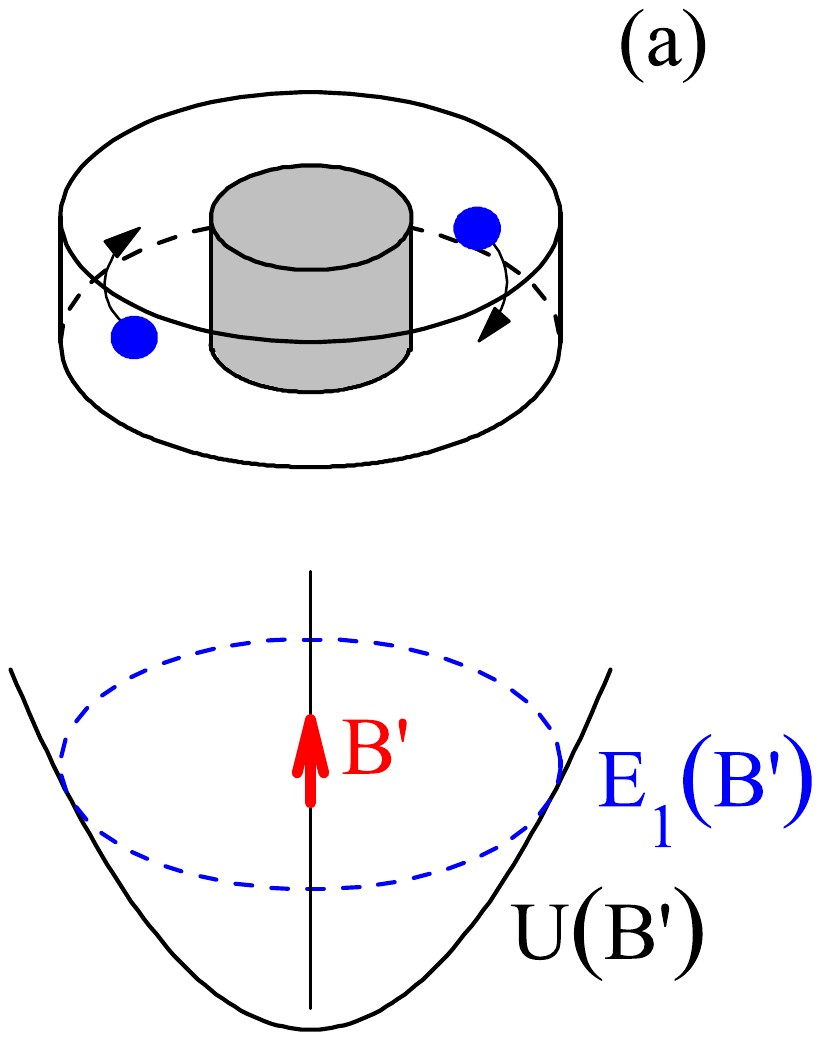}
\hspace{-2cm}
\includegraphics[scale=.35]{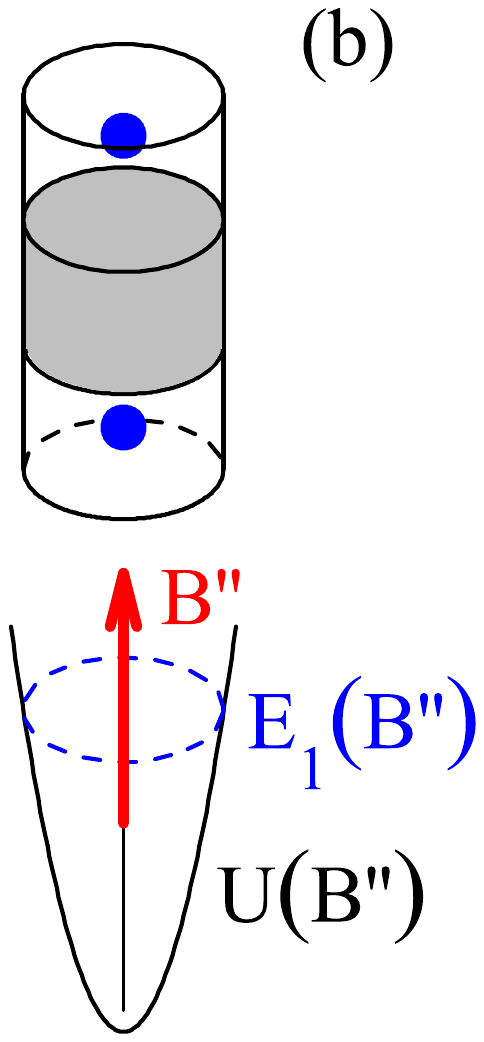}
\end{center}
\vspace{-.2in}
\caption{(Colour online) Schematic sketch of an axially symmetric two-electron QD
with a strong Coulomb repulsion between electrons in the presence
of: (a) a weak perpendicular magnetic field ($B^\prime$) and (b) a
sufficiently strong field ($B^{\prime\prime}$) leading to the
shape transition. Blue bullets represent electrons, whereas the
shaded area is the strong Coulomb interaction domain. Paraboloids
represent the lateral confining potential $U$ for electrons at
given field strengths, and the dashed circles mark the lowest
energy level.} 
\label{fig:shape-tr}
\end{figure*}

Another driving force for the Wigner crystallization
in a QD could be induced by a strong magnetic field
\cite{Matulis}.
The presence of a strong perpendicular magnetic field in QDs with a
non-negligible thickness, however, may change nontrivially the
electron density distribution. In our preliminary studies
\cite{NSPC,NS2013} we have shown that, when the field strength
exceeds a specific value, the preferable positions of electrons in
a two-electron QD are at the opposite sides of its center along
the vertical direction (i.e., perpendicularly to the layer plane),
see Fig.~\ref{fig:shape-tr}. Namely, by increasing the magnetic
field, the strength of the effective lateral confinement  of
electrons increases, too; that leads to a lateral squeezing of the
QD. At a sufficiently strong field the lateral size of the QD (at the
lowest states) becomes smaller than its vertical size (thickness),
and electrons  choose the vertical arrangement as a more stable, 
due to the Coulomb repulsion,. Thus, at a specific field
strength the QD experiences a shape transition such that the
electron density distribution changes from a ring or disk type
(depending on whether the Wigner molecule is formed or not) to a
vertical type. Taking into account the existence of these
transitions, a general criterion for the formation of a 3D Wigner
molecule in a two-electron QD may be a dislocation (radial or
vertical) of the electron density maximum from the dot center.

To study the combined role of a two-body interaction and the external magnetic field, 
related to shape transitions, we shall concentrate on the simplest nontrivial case,
on two identical particles (electrons) in a 3D axially symmetric harmonic oscillator potential.
The purpose of the present paper is to provide a thorough
analysis of the interplay between classical and quantum-mechanical properties in order 
to get deep insight into
a spontaneous symmetry breaking phenomenon under strong Coulomb interaction and/or
strong magnetic field. Although there are numerous results on two-electron QDs
in a magnetic field (for a review see \cite{bir,yan,n1}), up to now there is no a consistent analysis of
shape transitions in this system, which includes a detailed investigation of
classical and quantum-mechanical features.
Accurate numerical results will be complemented by analytical results in order 
to provide a physical insight in details of numerical calculations. Note, that 
the analytical results could establish
a theoretical framework for accurate analysis of confined many-electron systems, 
where the exact treatment of a 3D case becomes computationally intractable.

\section{Preliminaries}
\label{sec:model}
Two-electron quantum dots  provide  reliable experimental
 data (see, e.g., \cite{ihn,nish1,nish2})  as well as a prospective theoretical platform to
study various aspects of quantum correlations (see, e.g., \cite{frid,nl,gar} and references therein)
with a high accuracy. For small QDs with a few electrons an
effective trapping potential is quite well approximated by a parabolic confinement
(see discussion in \cite{bir}).
Note, if a 2D harmonic oscillator and the Coulomb potential are combined like in a QD, most of 
 symmetries are expected to be broken (see, e.g., \cite{dro,rad}).
In case of the axially symmetric 3D quantum dot
we have a nonintegrable motion, in general, too. However,  it was shown that the transition from chaotic
to regular dynamics and vice versa can be controlled with aid of the magnetic field \cite{NSR,SN}.
This result has been investigated  thoroughly one decade later \cite{ap},  and also rederived by 
using Killing tensors (see discussion in \cite{car}).

The system Hamiltonian for the case of the magnetic field B directed along the symmetry axis z
reads
%
%
\begin{equation}
\label{hamtot}
H = \sum_{i=1}^2 \bigg[ \frac{1}{2m^*\!}\,
\Big({\bf p}_i - e {\mathbf A}_i \Big)^{\! 2}
+ U({\mathbf r}_i) \bigg]
+ V_C+H_{\it spin}\,.
\end{equation}
Here the term $V_C=\alpha/{\vert{\mathbf r}_1 \!- {\mathbf
r}_2\vert}$ with $\alpha = e^2/4\pi\epsilon_0\epsilon_r$ describes
the Coulomb repulsion between electrons. The constants $m^*$, $e$,
$\varepsilon_0$ and $\varepsilon_r$ are the effective electron
mass, unit charge, vacuum and relative dielectric constants of a
semiconductor, respectively. For the magnetic field we choose the
vector potential with a gauge ${\mathbf A}_i = \frac{1}{2} {\mathbf
B} \times {\mathbf r}_i = \frac{1}{2}B(-y_i,x_i,0)$, and $H_{\it
spin}=g^*\mu_B({\bf s}_1+{\bf s}_2)ø{\bf B}$ is the Zeeman term,
where $\mu_B=|e|\hbar/2m_ec$ is the Bohr magneton. The confining
potential is approximated by a 2D circular harmonic
oscillator in $xy$-plane and the vertical confinement $V_z$:
$U(\mathbf{r}_i) = \frac{1}{2}\, m^* [\omega_0^2(x_i^2 + y_i^2) +
\omega_z^2 z_i^2]$. Here $\hbar\omega_0$ and $\hbar\omega_z$ are
the energy scales of the confinement in the $xy$-plane (lateral
confinement) and in the $z$-direction (vertical confinement),
respectively.  In the present analysis we neglect the spin
interaction, since the corresponding energy is small compared to
the confinement and the Coulomb energies. Herewith, we consider
only a vertical parabolic confinement $V_z$, while different forms
for the vertical confinement (e.g, \cite{n3}) may be analysed as
well.

By introducing the center of mass (CM) and relative coordinates,
$\mathbf{R} = \frac{1}{2}(\mathbf{r}_1 + \mathbf{r}_2)$ and
$\mathbf{r}_{12} = \mathbf{r}_1 - \mathbf{r}_2$, Hamiltonian
(\ref{hamtot}) separates into the CM and the relative motion terms, $H =
H_\mathrm{CM} + H_\mathrm{rel}$. Applying the above given gauge
condition, these terms take the forms
\begin{equation}
H_\mathrm{CM} = \frac{{\mathbf P}^2}{2{\cal M}} + \frac{1}{2}\,
{\cal M}\, [\Omega^2(X^2 + Y^2) + \omega_z^2 Z^2] - \omega_L L_z,
\label{cmham}
\end{equation}
\begin{equation}
H_{\rm rel} = \frac{{\mathbf p}_{12}^2}{2\mu} + \frac{1}{2}\,
\mu\, [\Omega^2 (x_{12}^2 + y_{12}^2) + \omega_z^2 z_{12}^2] +
\frac{\alpha}{r_{12}} - \omega_L l_z, \label{relham}
\end{equation}
where ${\cal M} = 2m^*$ and $\mu = m^*/2$ are the total and
reduced masses, $\omega_L = eB/2m^*$ is the Larmor frequency,
$\Omega = (\omega_0^2 + \omega_L^2)^{1/2}$ is the effective
lateral confinement frequency, and $L_z$ and $l_z$ are the
$z$-projections of the angular momenta for the CM and the relative
motions. Due to the axial symmetry the operators $L_z$ and $l_z$
are integrals of motion, and the corresponding (magnetic) quantum
numbers $M$ and $m$ are good quantum numbers. 

The separability of the Hamiltonian (\ref{hamtot}) allows us to write
its eigenenergies (QD's energy levels) as sums $E =
E_\mathrm{CM} + E_\mathrm{rel}$, and the corresponding
eigenfunctions in the form of products $\Psi(\mathbf{r}_1,
\mathbf{r}_2) = \psi_{\rm CM}(\mathbf{R})\,
\psi(\mathbf{r}_{12})$. Correspondingly, $E_\mathrm{CM}$ and $\psi_{\rm
CM}(\mathbf{R})$ are the eigenenergies and the eigenfunctions of
$H_\mathrm{CM}$, while $E_\mathrm{rel}$ and
$\psi(\mathbf{r}_{12})$ are the eigenenergies and the eigenfunctions
of $H_\mathrm{rel}$. Since the Coulomb interaction
enters only into the relative motion term, the
components of CM motion are decoupled. The CM  problem
is analytically solvable, which gives  the eigenenergies of $H_\mathrm{CM}$
in the form
\begin{equation}
E_\mathrm{CM} = \hbar\Omega(N + 2|M| + 1) + \hbar\omega_z(N_z +
1/2) - \omega_L M, \label{eq:Ecm}
\end{equation}
where $N, N_z = 0,1,2,\ldots$ are, in addition to $M$, good
quantum numbers. The corresponding eigenfunctions
$\psi_\mathrm{CM}(\mathbf{R})$ are products of the Fock-Darwin states
\cite{Fock} and oscillator functions in $z$-direction. On the
other hand, the relative motion  is not fully separable 
due to the Coulomb coupling (except the specific frequency ratios
$\omega_z/\Omega$ \cite{SN}). The corresponding eigenenergies and
eigenstates can be determined numerically or using approximate
methods (see Sec.~\ref{sec:spectra}). They are characterized by a
certain parity (even/odd) and by the good quantum number
$m$-value. We will see, however, that for low-lying states it is
possible to introduce two vibrational quantum numbers (related to
the radial and the vertical modes), which provide a full
classification.

For further analysis it is convenient to use the scaled
coordinates and momenta for the relative motion
$\tilde{\mathbf{r}}_{12} = \mathbf{r}_{12}/\ell_0^\mathrm{rel}$,
$\tilde{\mathbf{p}}_{12} =
\ell_0^\mathrm{rel}\mathbf{p}_{12}/\hbar$, where
$\ell_0^\mathrm{rel} = (\hbar/\mu\omega_0)^{1/2}$ is the
characteristic length for the relative motion in the confining
potential. The strength parameter $\alpha$ of the Coulomb
repulsion goes over to $\kappa = \alpha/(\hbar \omega_0
\ell_0^\mathrm{rel})$ (the so-called Wigner parameter \cite{bir} is then $R_W
= \sqrt{2}\kappa$). For example, in the case of GaAs QD
($m^*=0.067 m_e$, $\varepsilon=12$) with the confining frequency
$\hbar\omega_0 = 3\,$meV one has $\kappa = 1.45$ ($R_W = 2.05$).
For the sake of simplicity, below we drop the tilde in the scaled
variables.

In the scaled cylindrical coordinates the relative motion Hamiltonian
takes the form (in units of $\hbar
\omega_0$)
\begin{equation}
{\cal H} \equiv \frac{H_{\rm rel}}{\hbar\omega_0} = \frac{1}{2}\,
(p_{\rho_{12}}^2 + p_{z_{12}}^2) +
V_\mathrm{eff}(\rho_{12},z_{12}) - {\tilde\omega_L} m,
\label{relhamsc}
\end{equation}
where
\begin{equation}
V_{\rm eff} = \frac{m^2}{2\rho_{12}^2} +
\frac{1}{2}\,{\tilde\Omega}^2 \rho_{12}^2 +
\frac{1}{2}\,{\tilde\omega_z}^2 z_{12}^2 + \frac{\kappa}{r_{12}}
\label{effpot}
\end{equation}
and $r_{12} = (\rho_{12}^2+z_{12}^2)^{1/2}$, $\rho_{12}^2 =
x_{12}^2 + y_{12}^2$, $\tilde\Omega = \Omega/\omega_0$,
$\tilde\omega_z = \omega_z/\omega_0$, $\tilde\omega_L =
\omega_L/\omega_0$ and $m = l_z/\hbar$.

Below we provide a detailed analysis of the effective
potential (\ref{effpot}) and study its dependence on the
magnetic field strength. Assuming that at low-lying states of a
two-electron QD the maxima of probability density
$|\psi(\mathbf{r}_{12})|^2$ are located approximately at the
positions of minima of $V_\mathrm{eff}$, the analysis of this
potential can be useful in the study of localization of electrons
in the QD and formation of the Wigner molecule. 
In Sec.\ref{sec:localization} we will show that this assumption is 
well justified, indeed.

\section{Analysis of the effective potential}
\label{sec:effpot}

\subsection{Stationary points}

\begin{figure}[th]
\vspace{-5cm}
\begin{center}
\begin{minipage}{2in}
\includegraphics[scale=.32]{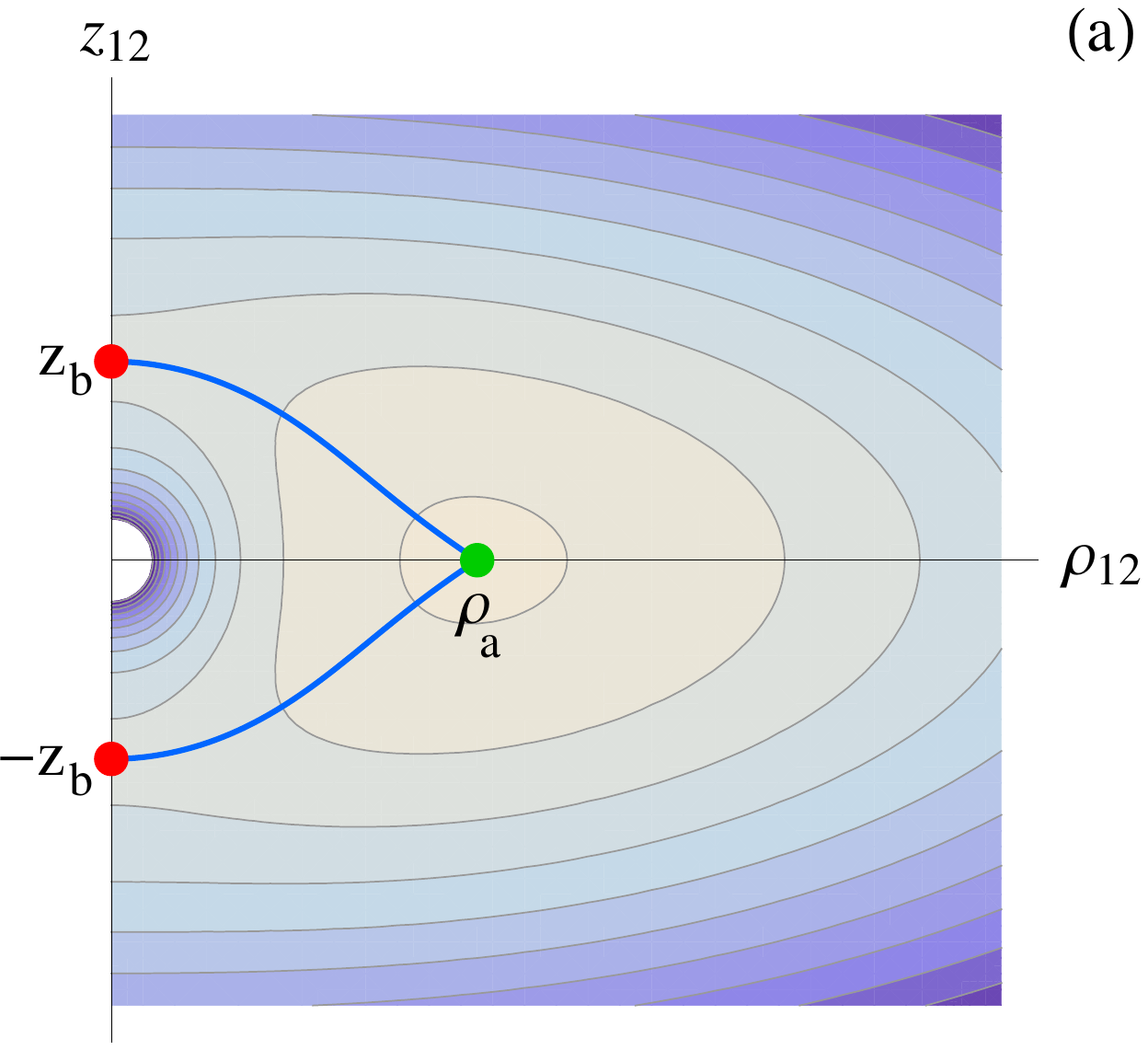}
\end{minipage}
\hspace{-.1in}
\begin{minipage}{1.3in}
\includegraphics[scale=.32]{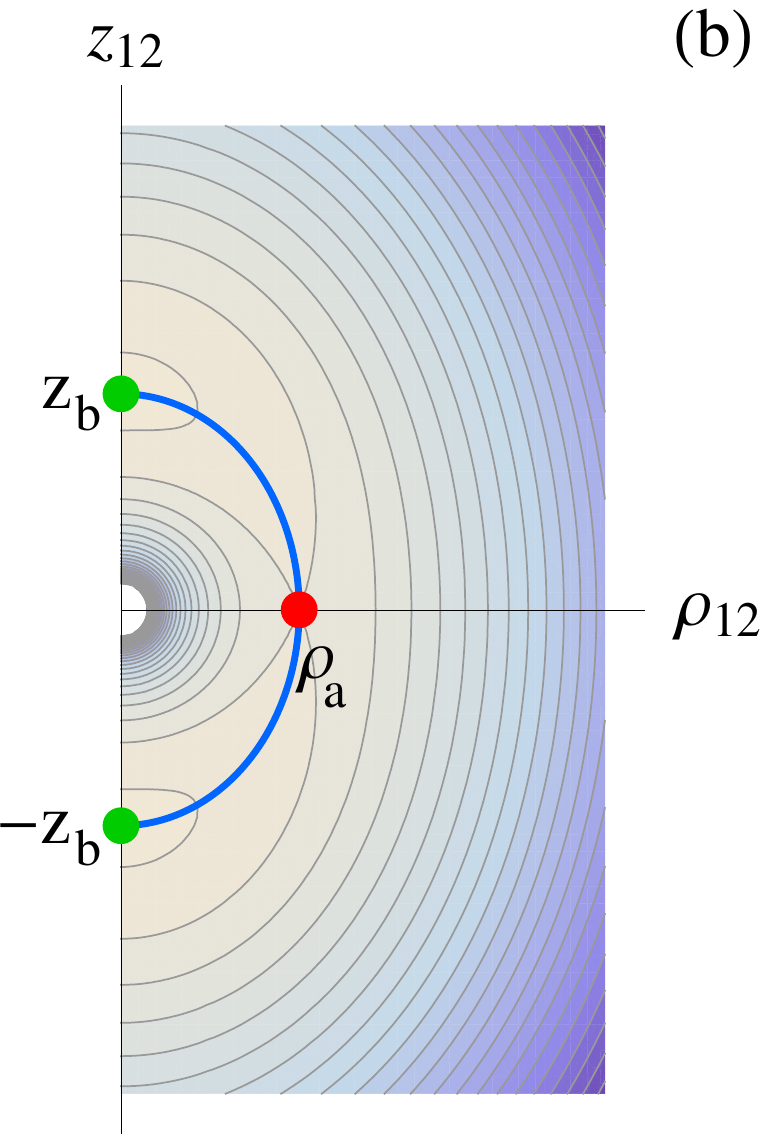}
\end{minipage}
\vspace{-4cm}
\\
\vspace{-1cm}
\begin{minipage}{2in}
\includegraphics[scale=.32]{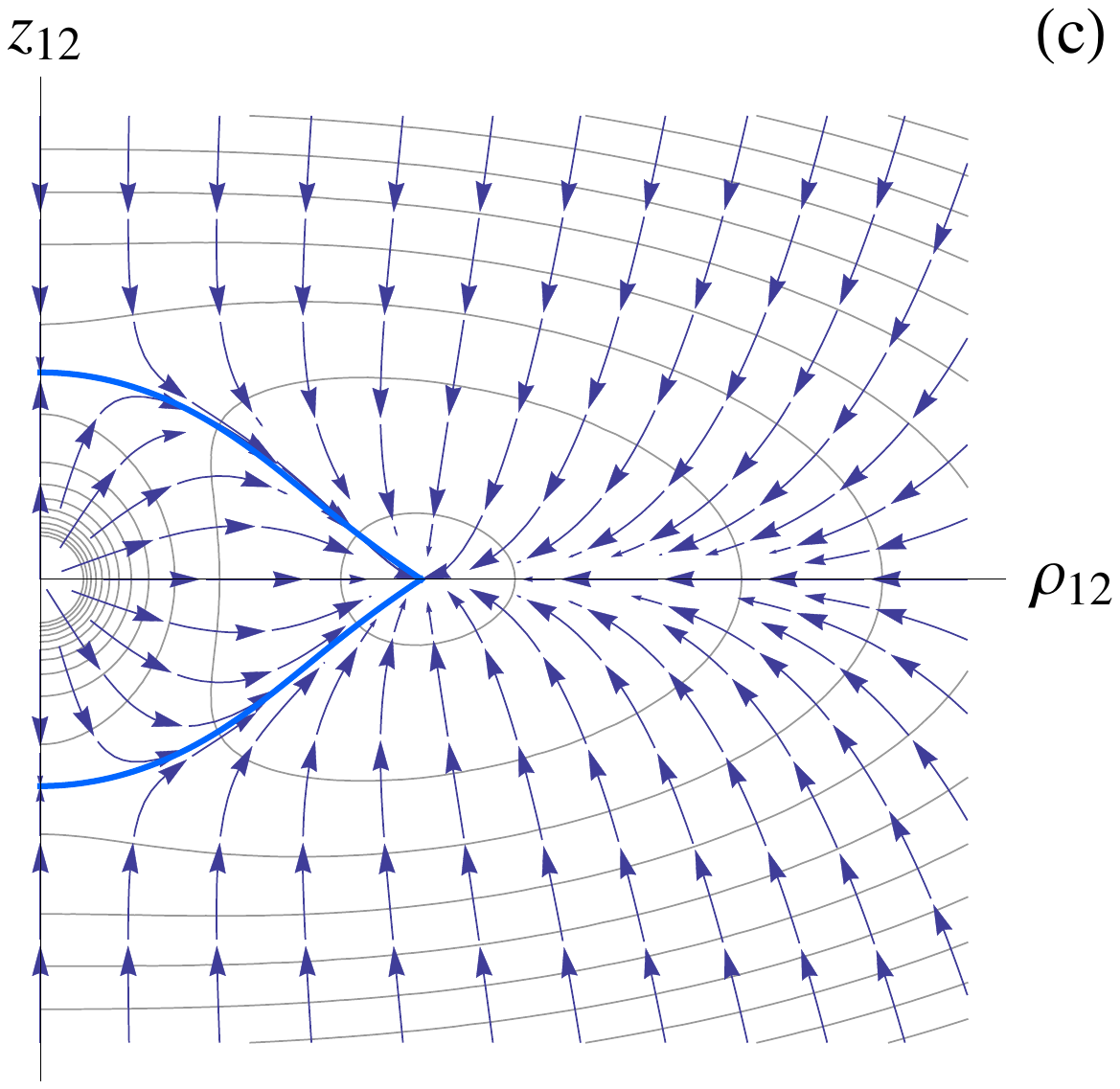}
\end{minipage}
%
\begin{minipage}{1.3in}
\includegraphics[scale=.32]{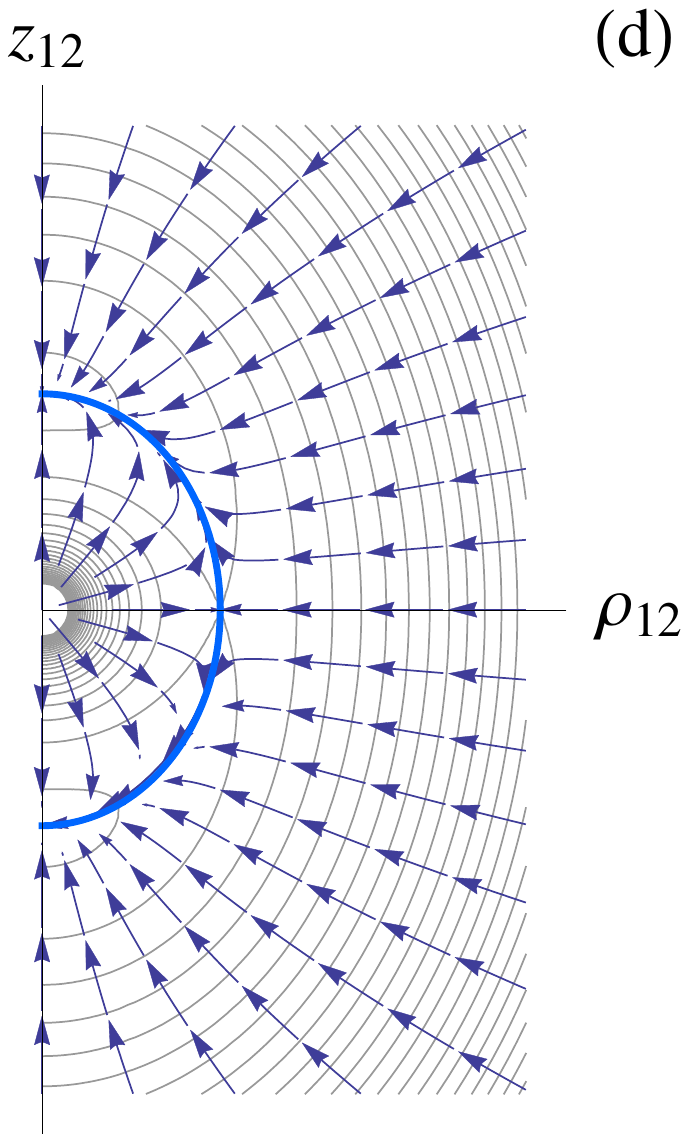}
\end{minipage}
\end{center}
\vspace{-0.5cm}
\caption{(Color online) Contour plots of the effective potential (\ref{effpot})
for $m = 0$ (top) and the vector diagrams of the corresponding force
field (bottom) at:  ${\tilde\Omega} < {\tilde\Omega}_\mathrm{bif}$, (a,\,c); 
and  ${\tilde\Omega} > {\tilde\Omega}_\mathrm{bif}$, (b,\,d). 
Positions of the stable and unstable
stationary points of the effective potential (minima and saddle
points) are marked by green and red dots, respectively. The blue
lines represent the minimum energy paths, that connect the
unstable and stable stationary points.} \label{fig:pot+ff-m=0}
\end{figure}

\begin{figure}[th]
\vspace{-5cm}
\begin{center}
\begin{minipage}{2in}
\includegraphics[scale=.32]{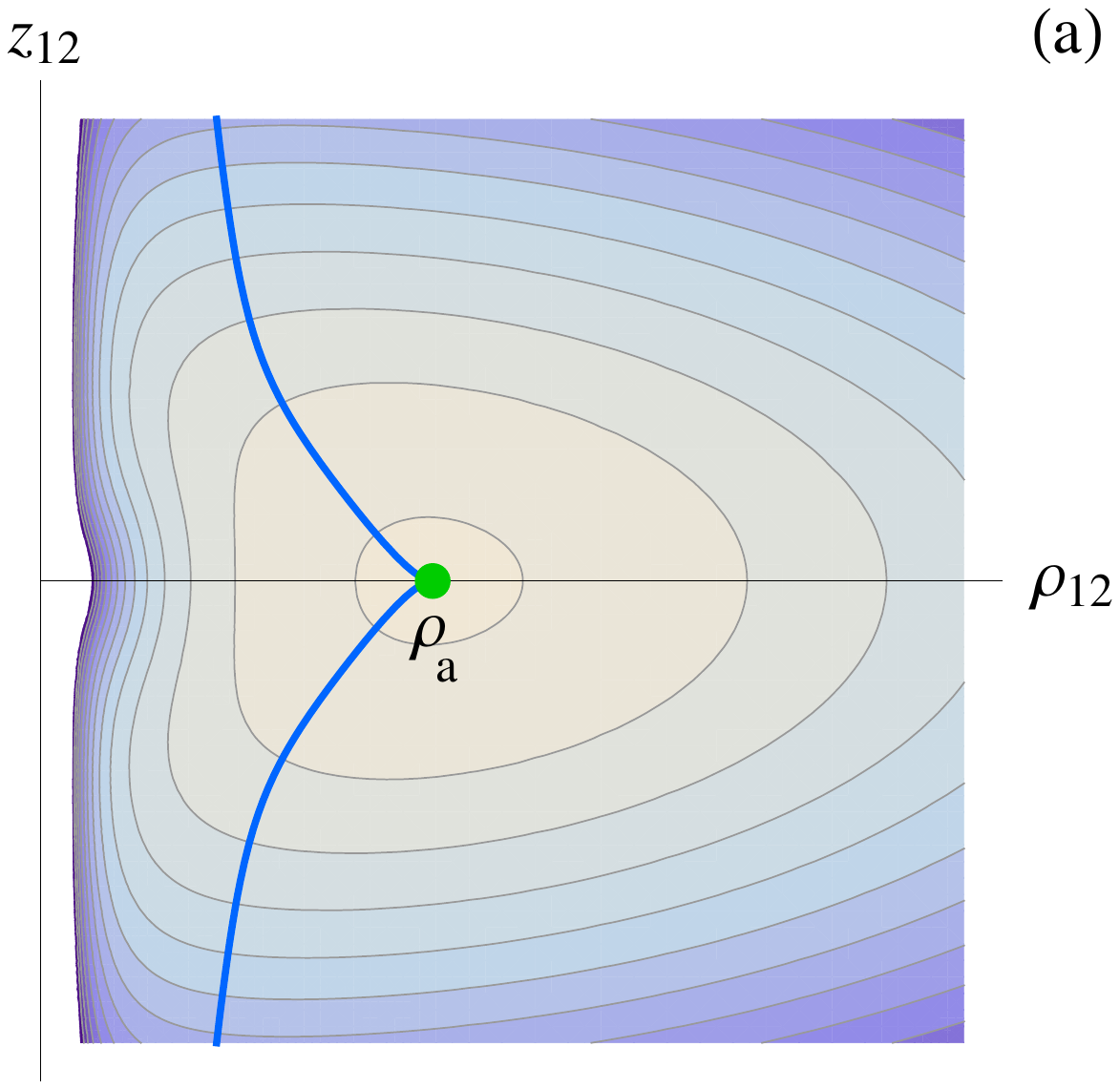}
\end{minipage}
\hspace{-.1in}
\begin{minipage}{1.3in}
\includegraphics[scale=.32]{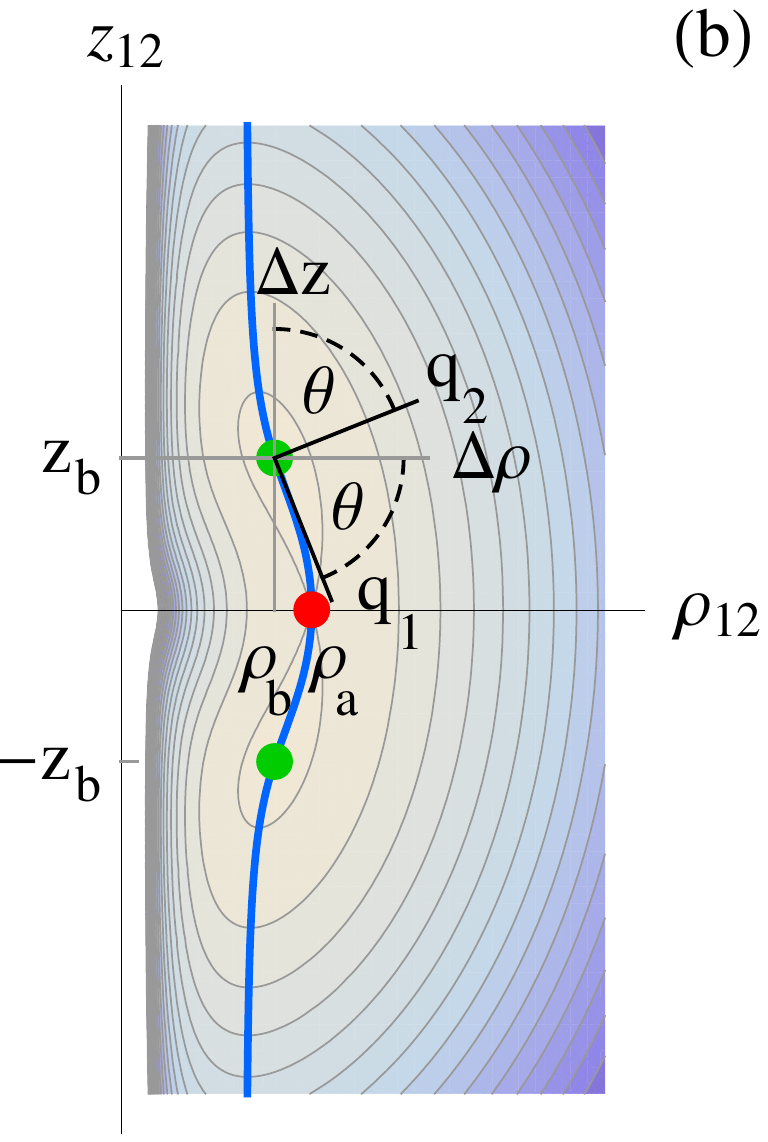}
\end{minipage}
\vspace{-5cm}
\\
\begin{minipage}{2in}
\includegraphics[scale=.32]{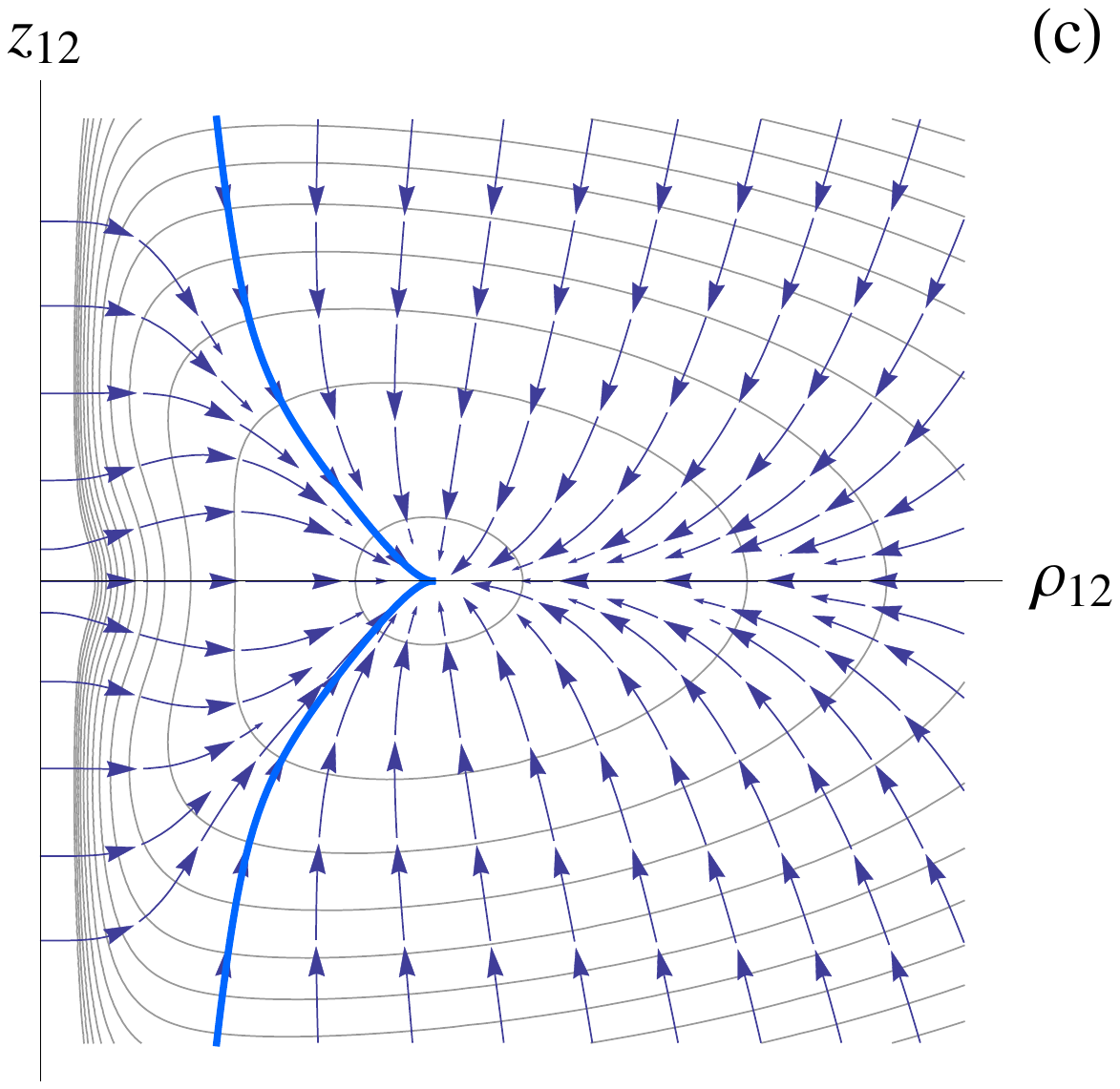}
\end{minipage}
%
\begin{minipage}{1.3in}
\includegraphics[scale=.32]{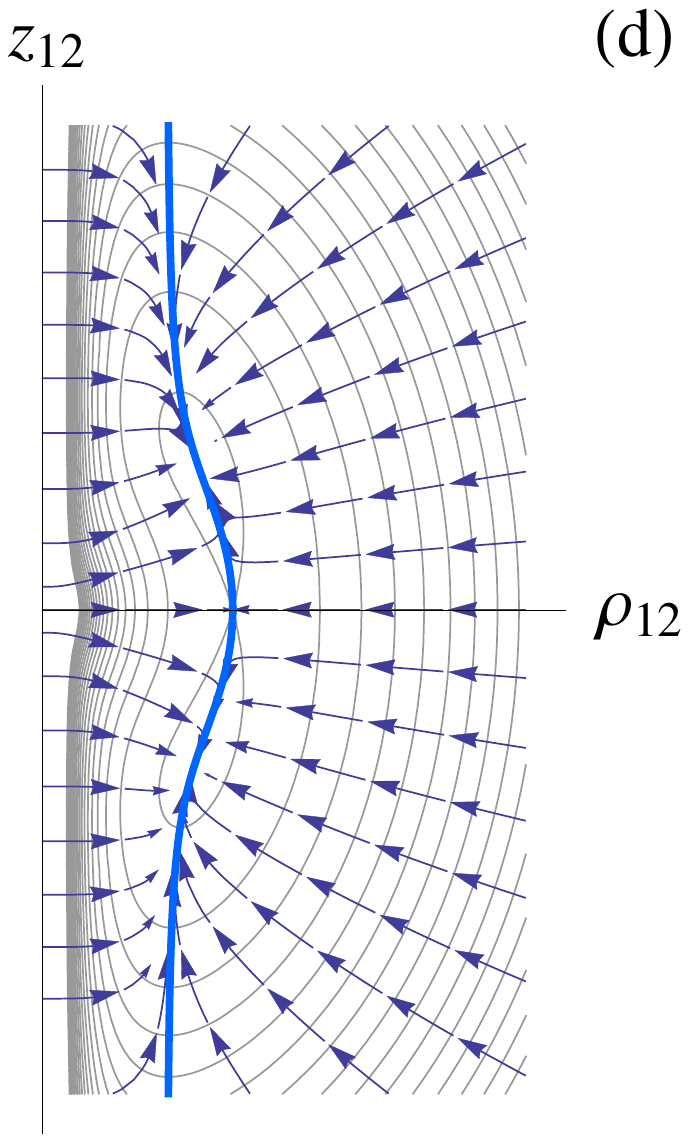}
\end{minipage}
\end{center}
\vspace{-.1in}
\caption{(Color online)
Similar to Fig.~\ref{fig:pot+ff-m=0}, for $m>0$.
The normal coordinates $(q_1,q_2)$,
describing small oscillations around the minimum ($\rho_b,z_b$) in
the case ${\tilde\Omega} > {\tilde\Omega}_\mathrm{bif}$, are
obtained by rotating the local cylindrical coordinates
$(\Delta\rho,\Delta z)$ on the angle $\theta$, determined by
Eq.~(\ref{c-diff}).} \label{fig:pot+ff-m>0}
\end{figure}

The positions of minima and all stationary points of the
effective potential (\ref{effpot}) are determined from the
conditions $\partial V_{\rm eff}/\partial\rho_{12} = 0$, $\partial
V_{\rm eff}/\partial z_{12} = 0$. The real solutions of this
system of equations are: (i) $(\rho_{12},z_{12}) = (\rho_a,0)$,
where $\rho_a$ is the positive root of equation
\beq
{\tilde\Omega}^2 \rho_a^4 - \kappa\;\! \rho_a - m^2 = 0,
\label{sp1}
\eeq
which in the case $m = 0$ is $\rho_a =
(\kappa/{\tilde\Omega}^2)^{1/3}$ (otherwise see Appendix \ref{sec:rhoa});
and (ii) $(\rho_{12},z_{12}) = (\rho_b,\pm z_b)$, where
\bea &&\rho_b = \biggl(\frac{m^2}{{\tilde\Omega}^2 -
{\tilde\omega}_z^2}\biggr)^{1/4}, \label{rhob}
\\
&&z_b = \sqrt{r_b^2 - \rho_b^2}, \quad r_b =
\biggl(\frac{\kappa}{{\tilde\omega}_z^2}\biggr)^{1/3}.\label{rb}
\eea
Obviously, for $m = 0$ we have $\rho_b = 0$ and $z_b = r_b$, whereas
for $m \neq 0$ the real solutions exist if ${\tilde\Omega} >
{\tilde\omega}_z$ $\wedge$ $r_b \ge \rho_b$. Therefore, the effective
potential (\ref{effpot}) has always a stationary point
located at $\rho_{12}$-axis (solution (i), see
Figs.~\ref{fig:pot+ff-m=0}(a,b) and \ref{fig:pot+ff-m>0}(a,b)).
Two additional points, located out of this axis (solution (ii),
see the same figures), exist: if $m = 0$; and for $m \neq 0$ if
${\tilde\Omega}$ is larger than the value
\beq {\tilde\Omega}_\mathrm{bif}(m) = \sqrt{{\tilde\omega}_z^2 +
m^2\bigg(\frac{{\tilde\omega}_z^2}{\kappa} \bigg)^{4/3}} \equiv
\sqrt{{\tilde\omega}_z^2 + \frac{m^2}{r_b^4}}, \label{bifur} \eeq
following from the condition $r_b \ge \rho_b$.

Depending on the parameters $\tilde\Omega$, $\tilde\omega_z$,
$\kappa$, as well as on the value of quantum number $m$, these
stationary points are minima or saddle points. Their type
is determined from the values of the second derivatives of the
effective potential. From the condition $\partial^2 V_{\rm
eff}/\partial z_{12}^2 \vert_{({\rho_{12} = \rho_a},\, {z_{12} =
0})} = 0$ and Eq.~(\ref{sp1}) it follows that the stationary point
(i) changes the character at ${\tilde\Omega} =
{\tilde\Omega}_\mathrm{bif}$. At the other values of ${\tilde\Omega}$
the sign of the second derivative indicates   that
the stationary point (i) is the minimum if ${\tilde\Omega} <
{\tilde\Omega}_\mathrm{bif}$; while it is the saddle point if
${\tilde\Omega} > {\tilde\Omega}_\mathrm{bif}$. Contrary, the
stationary points (ii) are minima if ${\tilde\Omega} >
{\tilde\Omega}_\mathrm{bif}$, and they are saddle points if
${\tilde\Omega} < {\tilde\Omega}_\mathrm{bif}$ and $m = 0$. If $m
\neq 0$ these points do not exist at ${\tilde\Omega} <
{\tilde\Omega}_\mathrm{bif}$, because this inequality is equivalent
to the condition $r_b < \rho_b$.

Different types of stationary points on the surface are connected
by a valley whose bottom defines the so-called minimum energy path
(MEP) \cite{Quapp} (blue line in Figs.~\ref{fig:pot+ff-m=0} and
\ref{fig:pot+ff-m>0}). In physical chemistry, such a path on the
potential energy surface of a molecular system is usually called
the reaction coordinate. The MEP has the property that any point
on the path is at an energy minimum in all directions
perpendicular to the path \cite{Sheppard}. As a consequence, the
resulting force acting on the system at a point of the MEP is
required to act along the path. This is evident in
Figs.~\ref{fig:pot+ff-m=0}(c,d) and \ref{fig:pot+ff-m>0}(c,d),
showing the vector diagrams of the force field $-{\vec \nabla}_{\!12}
V_\mathrm{eff}$.

\subsection{Bifurcation of the potential minimum}
\label{sec:bifur}

As it is mentioned in Sec. \ref{sec:model}, the magnetic field can
be used to vary the effective lateral confinement frequency of a
given QD. For typical QDs with $\omega_0 < \omega_z$ at zero
magnetic field (i.e., $\Omega = \omega_0$) it is $\Omega <
\omega_z$, and thus $\tilde\Omega < \tilde\Omega_\mathrm{bif}$.
Therefore, for $B = 0$ the stationary point (i) is the minimum of the
effective potential (\ref{effpot}) (see
Figs.~\ref{fig:pot+ff-m=0}(a), \ref{fig:pot+ff-m>0}(a)). With the
increase of the magnetic field over the value $B_{\rm bif}$ (which
corresponds to $\tilde\Omega = \tilde\Omega_\mathrm{bif}$) this
stationary point transforms to the saddle point, and two new minima
(stationary points (ii)) appear, divided by a potential barrier
(see Figs.~\ref{fig:pot+ff-m=0}(b), \ref{fig:pot+ff-m>0}(b)).
Since the change of a type of the stationary point (i) takes place
simultaneously with the appearance of new minima, one observes the
{\em bifurcation} of the stationary point (i) at $B =
B_\mathrm{bif} \equiv
(2m^*/e)\,\omega_0\,{\tilde\omega}_L^\mathrm{bif}$, where
${\tilde\omega}_L^{\rm bif} = ({\tilde\Omega}_\mathrm{bif}^2 -
1)^{1/2}$.

\begin{figure*}
\vspace{-7cm}
\begin{center}
\hspace{0.25cm}
\includegraphics[scale=.42]{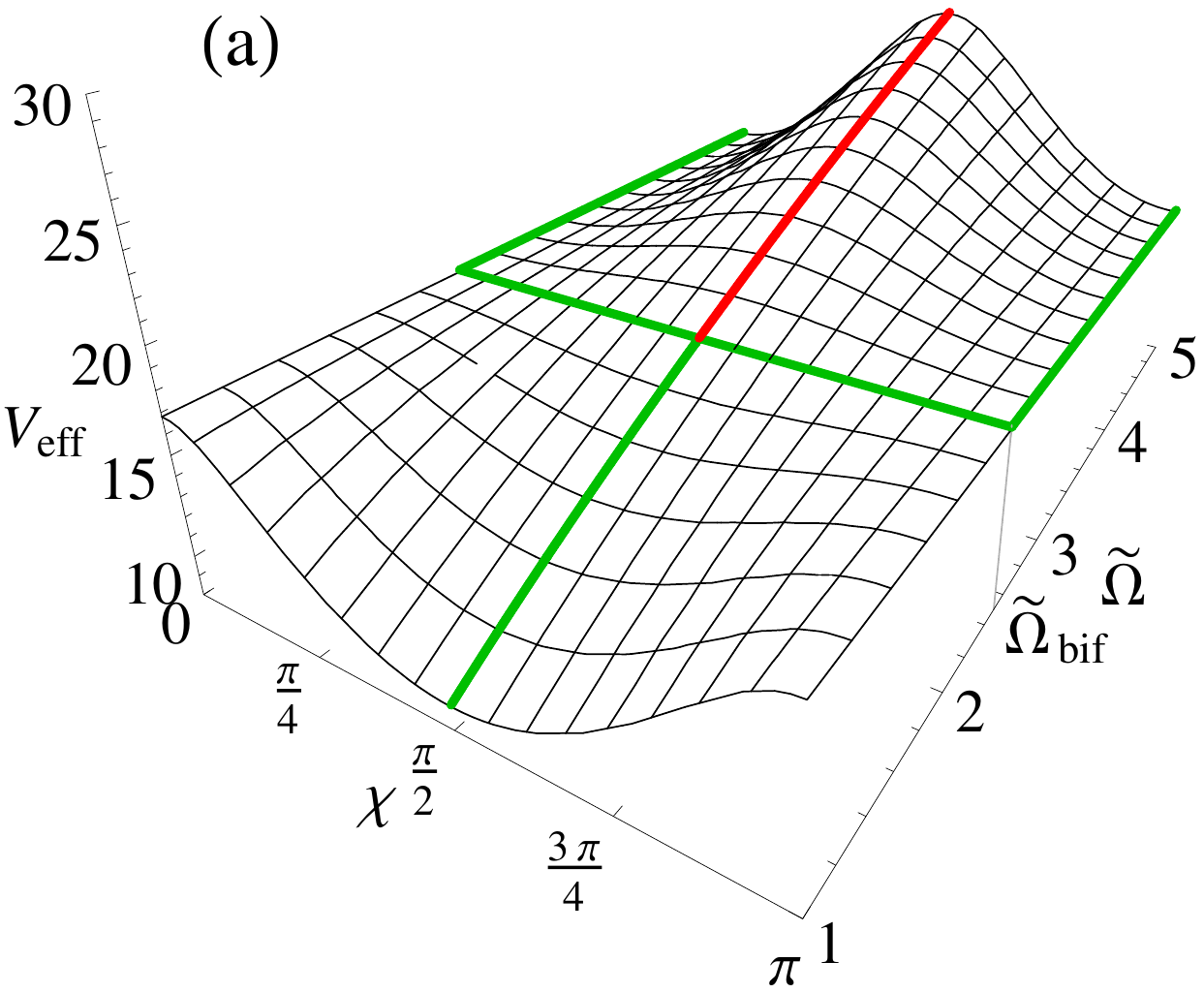}
\hspace{-2cm}
\includegraphics[scale=.42]{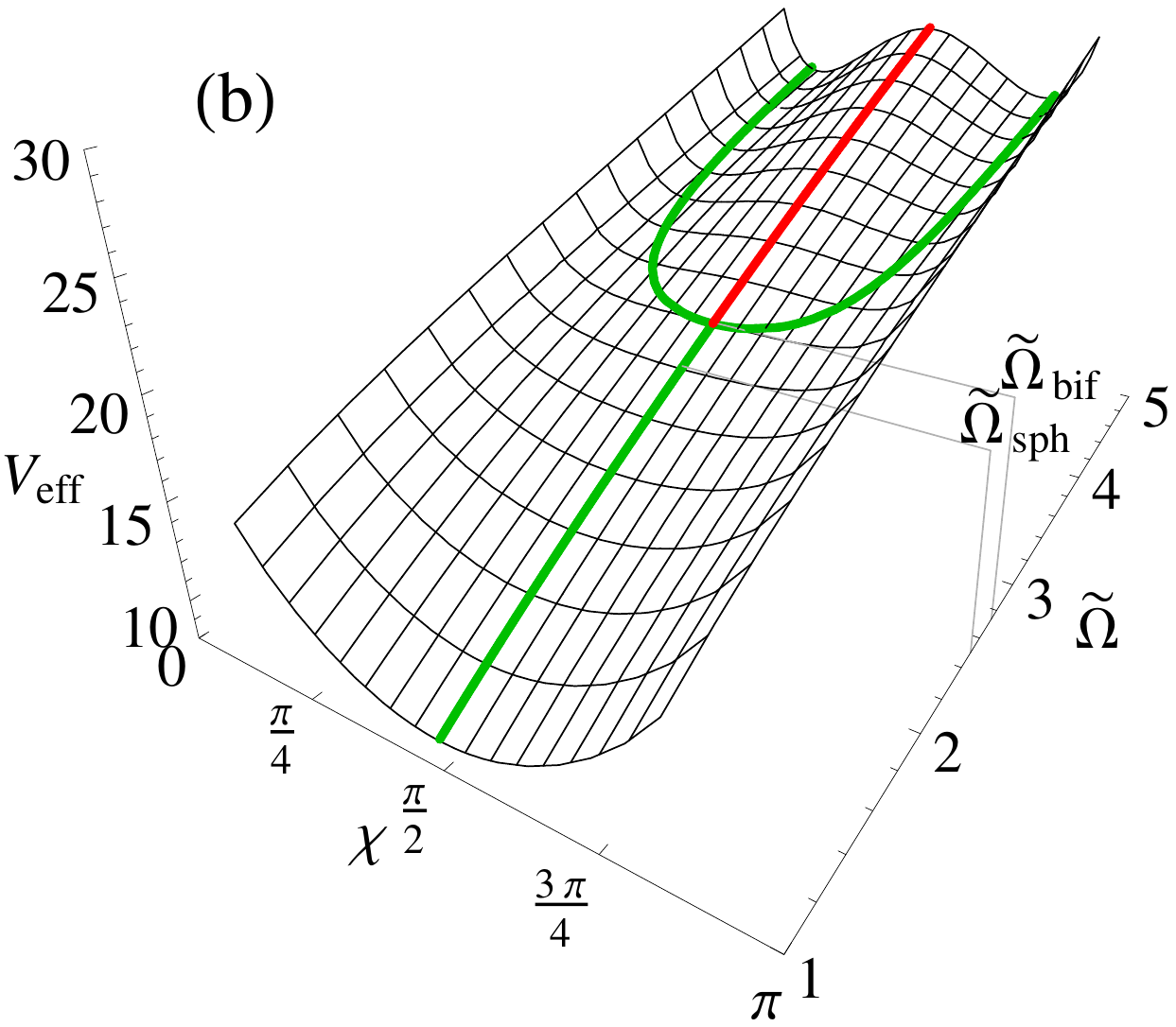}
\\
\vspace{-8cm}
\hspace{0.2cm}
\includegraphics[scale=.4]{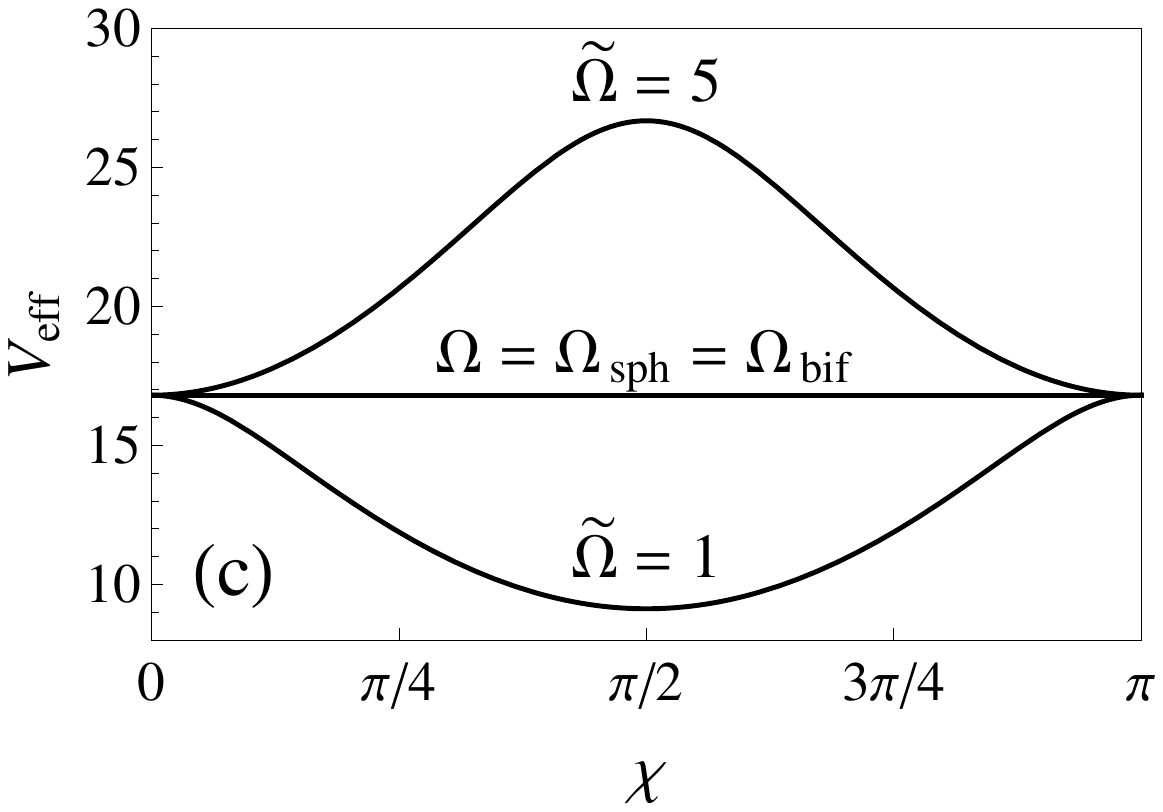}
\hspace{-2cm}
\includegraphics[scale=.4]{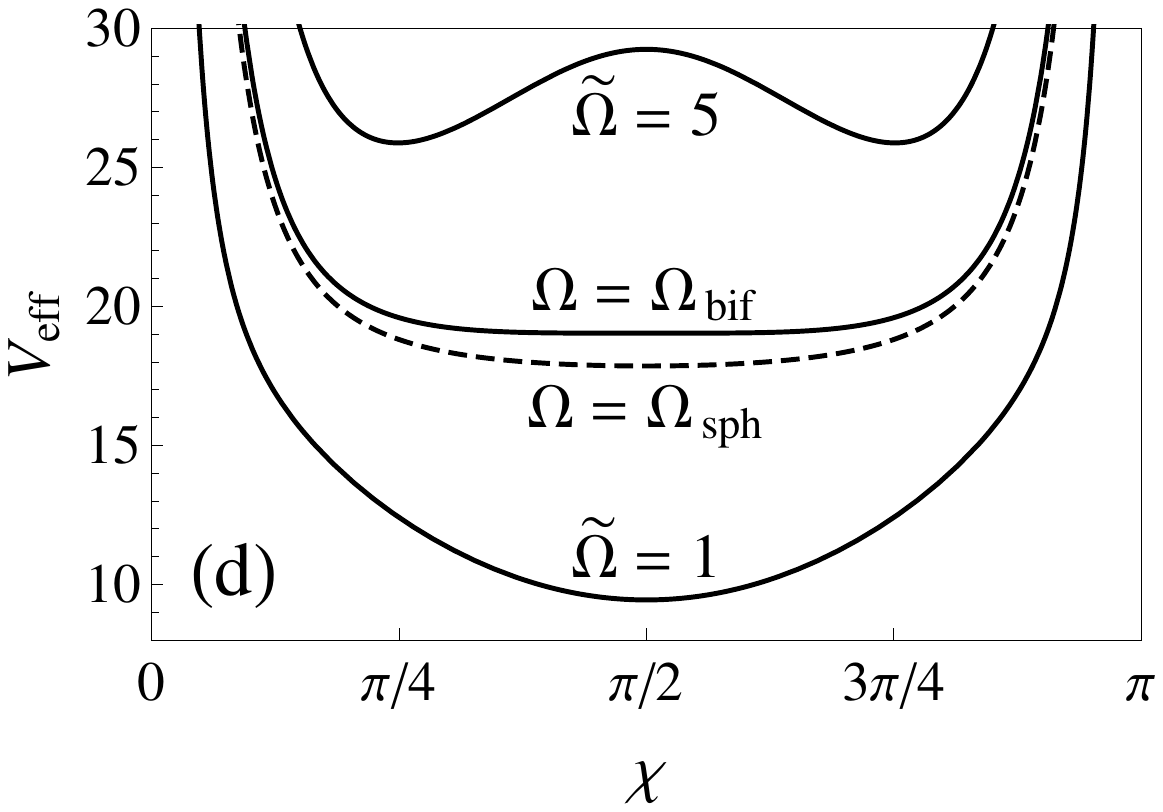}
\end{center}
\caption{(Color online) Top: Cuts of the potential
$V_\mathrm{eff}(\rho_{12}, z_{12})$ along the contour (\ref{ellipse})
at different values $\tilde\Omega$ for: (a) $m = 0$, and (b) $m = 2$. The
$\tilde\Omega$-dependence of the positions of stable and unstable
stationary points (green and red lines, respectively) represents
the bifurcation of the potential minimum. Bottom: The cuts of the potential
$V_\mathrm{eff}$ at several values of $\tilde\Omega$ ($1$,
$\tilde\Omega_\mathrm{sph}$, $\tilde\Omega_\mathrm{bif}$ and $5$),
for: (c) $m = 0$, and (d) $m = 2$.} \label{Veff-cut}
\end{figure*}

The bifurcation can be visualized by plotting the values of the 
potential (\ref{effpot})  along the MEP for different values of the
parameter $\tilde\Omega$. However, since the MEP varies by
changing $\tilde\Omega$ in a complicate manner, instead of the cut
along this path, we plot the cut of the potential (\ref{effpot}) along
a semi-elliptic contour, that connects the stationary points. This
contour can be expressed in the form
\begin{equation}
\rho_{12} = a\sin\chi,\quad z_{12} = b\cos\chi, \label{ellipse}
\end{equation}
where $\chi \in [0,\pi]$, and the semi-axes $a$ and $b$ are related
to the positions of stationary points
\begin{equation}
a = \rho_a, \quad b = \frac{\rho_a z_b}{\sqrt{\rho_a^2 -
\rho_b^2}}. \label{semi-axes}
\end{equation}
Note that the contour (\ref{ellipse}) and the MEP almost coincide when
$m = 0$ and $\Omega > \Omega_\mathrm{bif}$ (see
Fig.~\ref{fig:pot+ff-m=0}(b)). In other cases they are different,
but the cut of the potential (\ref{effpot}) along the contour
(\ref{ellipse}) still represents efficiently the behaviour of the
potential around stationary points. (It should be mentioned that
for ${\tilde\Omega} < {\tilde\Omega}_\mathrm{bif}$ and $m > 0$ the
solution (ii) is not real. In this case we replace $\rho_b \to
|\rho_b| = 0$ and $z_b \to r_b$, which give $b = r_b$ as in the
case $m = 0$.) The cuts of the potential (\ref{effpot}) with
$\tilde\omega_z = 2.5$ and $\kappa = 15$ for $m = 0$ and $m = 2$
as functions of $\tilde\Omega$ are shown in Fig.~\ref{Veff-cut}.
The dependence of stationary point positions ($\chi$
coordinate) on $\tilde\Omega$ yields a typical bifurcation diagram.

\begin{figure}
\vspace{-11.8cm}
\begin{center}
\includegraphics[scale=.6]{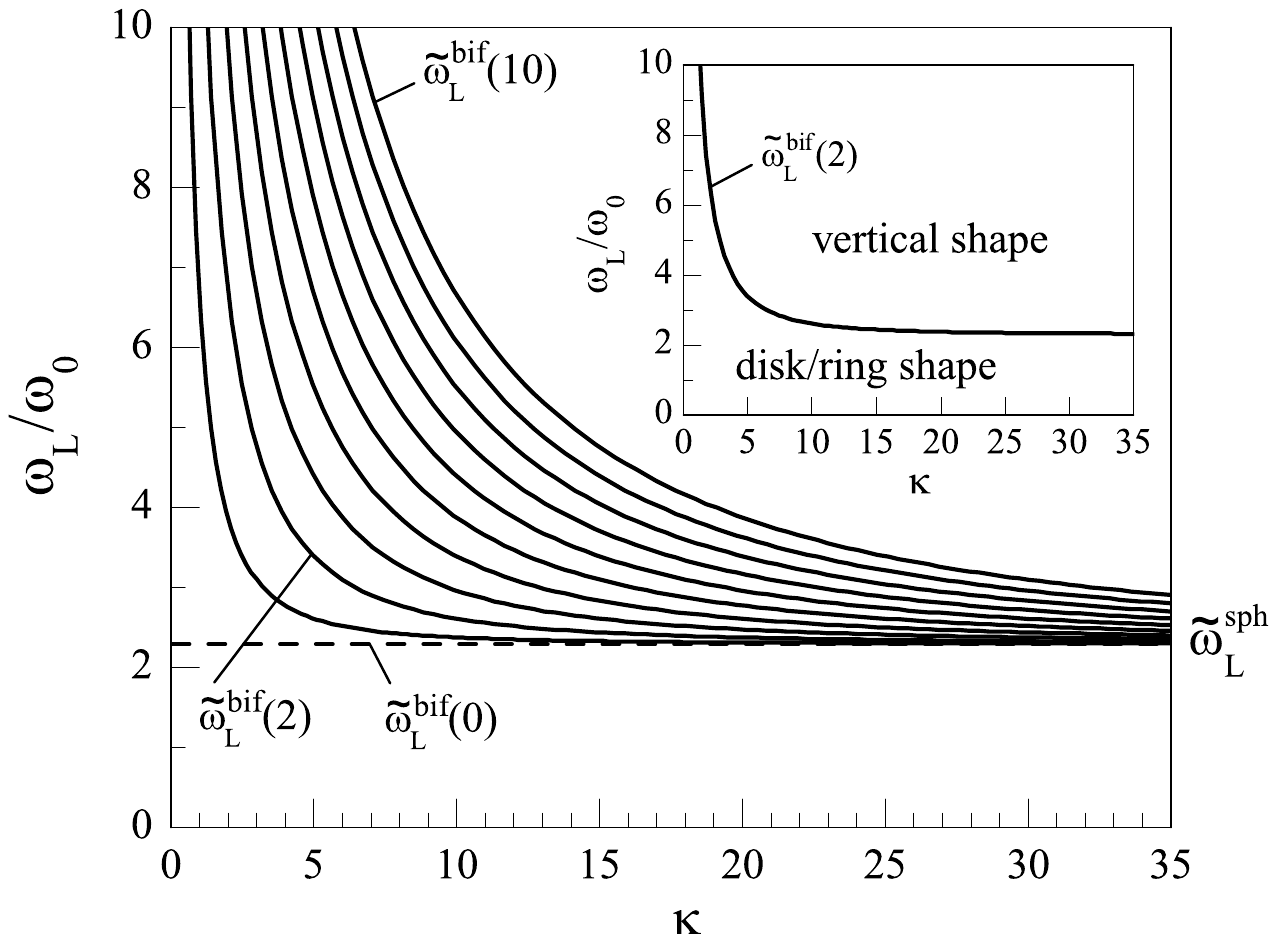}
\end{center}
\vspace{-.5cm} \caption{Dependence ${\tilde\omega}_L^{\rm bif} =
({\tilde\Omega}_\mathrm{bif}^2 - 1)^{1/2}$ (where
${\tilde\Omega}_\mathrm{bif}$ is given by Eq.~(\ref{bifur})) on
the parameter $\kappa$ for $\omega_z/\omega_0 = 2.5$ and different
values of $m$. The lines ${\tilde\omega}_L^{\rm bif}(m)$ separate
the areas in the $(\kappa,{\tilde\omega}_L)$-plane, where the
effective potential has the minimum (i) or the minima (ii). These
areas are related to two different shapes of the two-electron QD
(the disk/ring type and the vertical type, respectively). As an
example, the case $m = 2$ is shown separately in the inset.}
\label{fig-bif}
\end{figure}

Particularly, for $m = 0$ Eq.~(\ref{bifur}) gives
${\tilde\Omega}_\mathrm{bif} = {\tilde\omega}_z$ and
${\tilde\omega}_L^{\rm bif}(0) = ({\tilde\omega}_z^2 - 1)^{1/2}
\equiv {\tilde\omega}_L^{\rm sph}$. In our earlier studies
\cite{NSR,SN} it has been shown that at ${\tilde\omega}_L =
{\tilde\omega}_L^{\rm sph}$ the magnetic field gives rise to the
effective spherical symmetry ($\omega_z/\Omega=1$) in the axially symmetric
QD. Thus, for $m = 0$ the bifurcation of stationary point (i)
occurs when the potential (\ref{effpot}) becomes spherically
symmetric. The value ${\tilde\omega}_L^{\rm bif}(0) \equiv
{\tilde\omega}_L^{\rm sph}$ (dashed line in Fig.~\ref{fig-bif})
does not depend on the Coulomb interaction strength but only on
the ratio ${\tilde{\omega}}_z=\omega_z/\omega_0$. The corresponding value of the magnetic
field $B_{\rm sph}$, however, depends on $\omega_0$, too ($B \sim
\omega_L = \omega_0\, {\tilde\omega}_L$). For $m \neq 0$ the
bifurcation occurs at ${\tilde\omega}_L^{\rm bif} >
{\tilde\omega}_L^{\rm sph}$, which depends on $\kappa$ and on
$|m|$ (see Fig.~\ref{fig-bif}). The later dependence can be
explained by the influence of the centrifugal term
$m^2/2\rho_{12}^2$ in the effective potential (\ref{effpot}).

\subsection{Small oscillations around the potential minimum at
$\Omega < \Omega_\mathrm{bif}$.}\label{sec:smallosc_a}

As we have already seen, when $\Omega < \Omega_\mathrm{bif}$
the effective potential (\ref{effpot}) has a minimum at ($\rho_a,0$)
(see Figs.~\ref{fig:pot+ff-m=0}(a) and \ref{fig:pot+ff-m>0}(a)).
In the vicinity of this minimum the potential can be approximated by
the expansion in terms of $\Delta\rho = \rho_{12} - \rho_a$ and
$\Delta z = z_{12}$, in a similar way as it was done in
Refs.~\cite{Matulis, PSN} for the 2D case. Keeping all terms up to
the second order, we obtain the effective potential
\beq
V_{\rm eff} = V_a + \frac{1}{2}\,{\tilde\omega}_{1a}^2
\Delta\rho^2 + \frac{1}{2}\,{\tilde\omega}_{2a}^2 \Delta z^2.
\label{eq:Veff-expan-a}
\eeq
The first term in the potential $V_a $, i.e.,
\beq V_a = \frac{3}{2}\,{\tilde\Omega}^2 \rho_a^2
-\frac{m^2}{2\rho_a^2} \label{eq:Va} \eeq
is the classical minimum of the potential energy,
whereas the
second term yields the rotational series.
The frequencies 
\begin{eqnarray}
&&{\tilde\omega}_{1a} = \sqrt{3{\tilde\Omega}^2 +
\frac{m^2}{\rho_a^4}}, \label{eq:omega1a}
\\
&&{\tilde\omega}_{2a} = \sqrt{{\tilde\omega}_z^2 -
{\tilde\Omega}^2 + \frac{m^2}{\rho_a^4}} \label{eq:omega2a}
\end{eqnarray}
determine small oscillations  around this minimum in
$\rho_{12}$ and $z_{12}$ directions, respectively.

It can be shown that the positive root of Eq.~(\ref{sp1}), when
$\Omega = \Omega_\mathrm{bif}$, is $\rho_a^\mathrm{bif} = r_b$. As
a consequence, at the bifurcation point 
${\tilde\Omega} = {\tilde\Omega_\mathrm{bif}}=\sqrt{{\tilde\omega}_z^2+m^2/r_b^4}$
the frequencies (\ref{eq:omega1a}), (\ref{eq:omega2a}), become
\begin{equation}
{\tilde\omega}_{1a}^\mathrm{bif} = \sqrt{3{\tilde\omega}_z^2 +
4\frac{m^2}{r_b^4}},\quad {\tilde\omega}_{2a}^\mathrm{bif} = 0\,.
\end{equation}
Note that the vanishing frequency corresponds to the oscillatory
mode in $z_{12}$-direction, which is tangential to the MEP at the point
$(\rho_a,0)$ for $\Omega =\Omega_\mathrm{bif}$.

\subsection{Small oscillations around the potential minima when
$\Omega > \Omega_\mathrm{bif}$}\label{sec:smallosc_b}

When $\Omega > \Omega_\mathrm{bif}$, the effective potential
$V_\mathrm{eff}(\rho_{12},z_{12})$ has the minima at ($\rho_b,\pm
z_b$), see Fig.~\ref{fig:pot+ff-m>0}(b). In this case, it is
convenient to expand this potential into a series around the point
$(\rho_b,z_b)$ (or $(\rho_b,-z_b)$). If $\Delta\rho = \rho_{12} -
\rho_b$ and $\Delta z = z_{12} - z_b$, keeping all terms up to the
second order, one obtains
\begin{equation}
V_\mathrm{eff} \approx V_b +
\frac{1}{2}\,\tilde\Omega_\mathrm{eff}^2 \Delta\rho^2 +
\frac{1}{2}\,\tilde\omega_\mathrm{eff}^2 \Delta z^2 + \lambda
\Delta\rho\Delta z, \label{Vexpanded}
\end{equation}
where
\begin{eqnarray}
&&V_b = ({\tilde\Omega}^2\!-\!{\tilde\omega_z}^2)\rho_b^2 +
\hbox{$\frac{3}{2}$}\,{\tilde\omega_z}^2 r_b^2, \label{eq:Vb}
\\[1.5ex]
&&\tilde\Omega_\mathrm{eff}^2 =
4({\tilde\Omega}^2\!\!-\!{\tilde\omega_z}^2) +
3\,{\tilde\omega_z}^2 \rho_b^2/r_b^2,
\\[1.5ex]
&&\tilde\omega_\mathrm{eff}^2 = 3\,{\tilde\omega_z}^2 z_b^2/r_b^2,
\\[1.5ex]
&&\lambda = 3\,{\tilde\omega_z}^2 \rho_b z_b/r_b^2 \label{lambda}.
\end{eqnarray}

Note that in the special case, when $m = 0$ one has $\rho_b = 0$
and $z_b = r_b$ (see Eqs.~(\ref{rhob}), and (\ref{rb})), and the
coupling term $\lambda$ vanishes. Consequently, the potential
(\ref{Vexpanded}) separates immediately, i.e.,
\begin{equation}
V_\mathrm{eff} \approx V_b +
\frac{1}{2}\,\tilde\Omega_\mathrm{eff}^2 \Delta\rho^2 +
\frac{1}{2}\,\tilde\omega_\mathrm{eff}^2 \Delta z^2,
\label{Vexpand-m0}
\end{equation}
where $V_b = \frac{3}{2}\,{\tilde\omega_z}^2 r_b^2$,
$\tilde\Omega_\mathrm{eff} = 2\sqrt{{\tilde\Omega}^2 -
{\tilde\omega_z}^2}$ and $\tilde\omega_\mathrm{eff} =
\sqrt{3}\,{\tilde\omega}_z$.

In the general case (i.e., for arbitrary $m$) the motions in the
$\rho_{12}$ and $z_{12}$ directions are coupled. It is
possible, however, to find a new set of coordinates $(q_1,q_2)$, that
correspond to normal oscillatory modes of this system.
By means of the transformation of the variables $(\Delta\rho,\Delta z)$
to the local coordinate frame (see Fig.~\ref{fig:pot+ff-m>0}(b))
\beq
\bigg(
\begin{array}{cc}
\!\!\Delta\rho\!\! \\ \!\!\Delta z\!\!
\end{array}
\bigg) = {\cal R}(-\theta) \bigg(
\begin{array}{cc}
\!\!q_1\!\! \\ \!\!q_2\!\!
\end{array}
\bigg) = \bigg(
\begin{array}{cc}
\!\!\cos\theta & \sin\theta\!\!
\\
\!\!-\sin\theta & \cos\theta\!\!
\end{array}
\bigg)\bigg(
\begin{array}{cc}
\!\!q_1\!\! \\ \!\!q_2\!\!
\end{array}
\bigg).
\eeq
and introducing the notation $\cos\theta = c$, we have
\bea
&&\Delta\rho = c\,q_1 + \sqrt{1-c^2}\,q_2, \label{drho-q1q2}
\\
&&\Delta z = -\sqrt{1-c^2}\,q_1 + c\,q_2. \label{dz-q1q2}
\eea
Here it is important that the (reduced) kinetic energy $T =
\frac{1}{2}\,(p_{\rho}^2 + p_{z}^2)$ in the Hamiltonian
(\ref{relham}), after the rotation on an arbitrary angle
$\theta$, keeps the standard form, i.e., $T = \frac{1}{2}(p_1^2 +
p_2^2)$, where $p_1$ and $p_2$ are the momenta conjugated to the
coordinates $q_1$ and $q_2$, respectively.

Since for $m = 0$ the variables $\Delta\rho$ and $\Delta z$ are
decoupled, we can choose the angle $\theta = 0$ ($c = 1$) that
gives $q_1 = \Delta\rho$ and $q_2 = \Delta z$. Note, however, that
when $m = 0$ the rotations on $\theta = \pi/2$ and $\theta = \pi$
(then $c = 0$ and $c = -1$, respectively) also provide the
separability (see Appendix \ref{sec:rotpar}).

It is shown (see Appendix \ref{sec:rotpar}) that for $m \neq 0$ the
effective potential takes the normal form
\begin{equation}
V_\mathrm{eff} = V_b + \frac{1}{2}\,\tilde\omega_{1b}^2 q_1^2 +
\frac{1}{2}\,\tilde\omega_{2b}^2 q_2^2 \label{Veff-normf}
\end{equation}
at
\begin{equation}
c = \frac{1}{\sqrt{2}}\bigg( 1 -
\frac{\tilde\Omega_\mathrm{eff}^2-\tilde\omega_\mathrm{eff}^2}
{\sqrt{(\tilde\Omega_\mathrm{eff}^2-\tilde\omega_\mathrm{eff}^2)^2
+ 4\lambda^2}} \bigg)^{1/2}. \label{c-diff}
\end{equation}
The normal frequencies in Eq.~(\ref{Veff-normf}) are defined as
\begin{eqnarray}
&& \tilde\omega_{1b}^2 = c^2\, \tilde\Omega_\mathrm{eff}^2 +
(1-c^2) \tilde\omega_\mathrm{eff}^2 - 2c\sqrt{1-c^2}\,\lambda,
\label{om1}
\\
&& \tilde\omega_{2b}^2 = (1-c^2) \tilde\Omega_\mathrm{eff}^2 + c^2
\tilde\omega_\mathrm{eff}^2 + 2c\sqrt{1-c^2}\,\lambda. \label{om2}
\end{eqnarray}

Note that formula (\ref{c-diff}), when $m = 0$, does not give the
correct solution (one obtains $c = 0$ instead of $c = 1$). This is
a consequence of the fact that the cases with $m \neq 0$ cannot be
continuously connected to the case $m = 0$. Thus, $\lim_{m\to 0}
c(\tilde\Omega,m) \neq c(\tilde\Omega,0)$. Anyway, for $m = 0$ we
choose $c = 1$, that gives $\tilde\omega_{1b} =
\tilde\Omega_\mathrm{eff}$ and $\tilde\omega_{2b} =
\tilde\omega_\mathrm{eff}$, and formula (\ref{Veff-normf}) reduces
to Eq.~(\ref{Vexpand-m0}). Then, at the bifurcation point ($\Omega
= \omega_z$)
\begin{equation}
\tilde\omega_{1b}^\mathrm{bif} = 0,\quad
\tilde\omega_{2b}^\mathrm{bif} = \tilde\omega_\mathrm{eff} =
\sqrt{3}\,{\tilde\omega}_z.
\end{equation}

For $m \neq 0$, however, the bifurcation occurs when
$\tilde\Omega^2 = \tilde\Omega_\mathrm{bif}^2 \equiv
\tilde\omega_z^2 + m^2/r_b^4$. Then $\rho_b = r_b$, $z_b = 0$, and
we have $(\tilde\Omega_\mathrm{eff}^\mathrm{bif})^2 =
4{\tilde\Omega}_\mathrm{bif}^2 - \tilde{\omega}_z^2 =
\sqrt{3{\tilde\omega}_z^2 + 4m^2/r_b^4} $,
$\tilde\omega_\mathrm{eff}^\mathrm{bif} = 0$ and
$\lambda_\mathrm{bif} = 0$, that gives $c_\mathrm{bif} = 0$.
Consequently, we obtain
\begin{equation}
\tilde\omega_{1b}^\mathrm{bif} = 0,\quad
\tilde\omega_{2b}^\mathrm{bif} =
\tilde\Omega_\mathrm{eff}^\mathrm{bif} = \sqrt{3{\tilde\omega}_z^2
+ 4\frac{m^2}{r_b^4}}.
\end{equation}

Thus, for $m = 0$ and $m \neq 0$, the frequency
$\omega_{1b}$ tends to zero at $\Omega \to \Omega_\mathrm{bif}+0$. 
We recall that it corresponds to the oscillatory mode along the
MEP (blue line) in domain $\Omega > \Omega_\mathrm{bif}$. At the end of
Sec.~\ref{sec:smallosc_a} we found  that the frequency $\omega_{2a}$
(which corresponds to the same mode when $\Omega \to
\Omega_\mathrm{bif}-0$) also tends to zero.  
Therefore, we conclude that
 {\em when
the normal frequency along the MEP tends to zero, the bifurcation
of the potential minimum takes place: a symmetry breaking occurs.
In our case this is an
indication that the shape transition in the QD takes place.}
In fact,  similar phenomenon have been discussed in context of fast rotating nuclei
around a symmetry axis in \cite{m1,h1}. This analogy becomes evident once one compares
the Hamiltonian of QDs in a perpendicular magnetic field and a cranking Hamiltonian used
in nuclear structure physics (see, e.g., \cite{bla}). In an isolated small QD,
the external magnetic field acts like rotation of a nucleus (see discussion in \cite{bir,n1}).

We recall that in this section only the classical properties of the effective potential have been analysed.
On the other hand, the effective potential can be considered as the effective mean field 
of our quantum system. The next step is to establish the connection between the classical and quantum 
properties of the QD.

\section{Localization of electrons in the QD}
\label{sec:localization}

The analysis of the effective potential $V_\mathrm{eff}$ (
Sec.~\ref{sec:effpot}) is done under the assumption that
maxima of the probability density
$|\psi(\mathbf{r}_{12})|^2$ of low-lying eigenstates,
associated with the relative motion Hamiltonian
 $H_\mathrm{rel}$,  are located approximately at the
positions of minima of this potential. 
In order to clear up this point we calculate numerically
the electron density distribution in the QD. It depends 
on the relative motion and the CM motion parts of the total wave
function. The relative location of electrons in the QD
under the crystallization regime (Wigner molecule) is determined
by the quantity $|\psi(\mathbf{r}_{12})|^2$. In the laboratory
frame this arrangement is partially shaded by the CM motion. 
Therefore, we analyze the probability density for the
relative position of electrons and the total electron density. The
eigenstates of the Hamiltonian (\ref{relhamsc}) are determined by
diagonalizing this Hamiltonian in the oscillator basis. We recall that it
consists of the products of Fock-Darwin states,
corresponding to lateral degrees of freedom $(\rho_{12},
\varphi_{12})$, and oscillator functions in $z_{12}$-direction (see
Appendix \ref{sec:diag}). The details of calculation of the electron densities are 
given in Appendix \ref{sec:density}.

\begin{figure}[thb]
\vspace{-4cm}
\begin{center}
\begin{minipage}{1.2in}
\includegraphics[scale=.25]{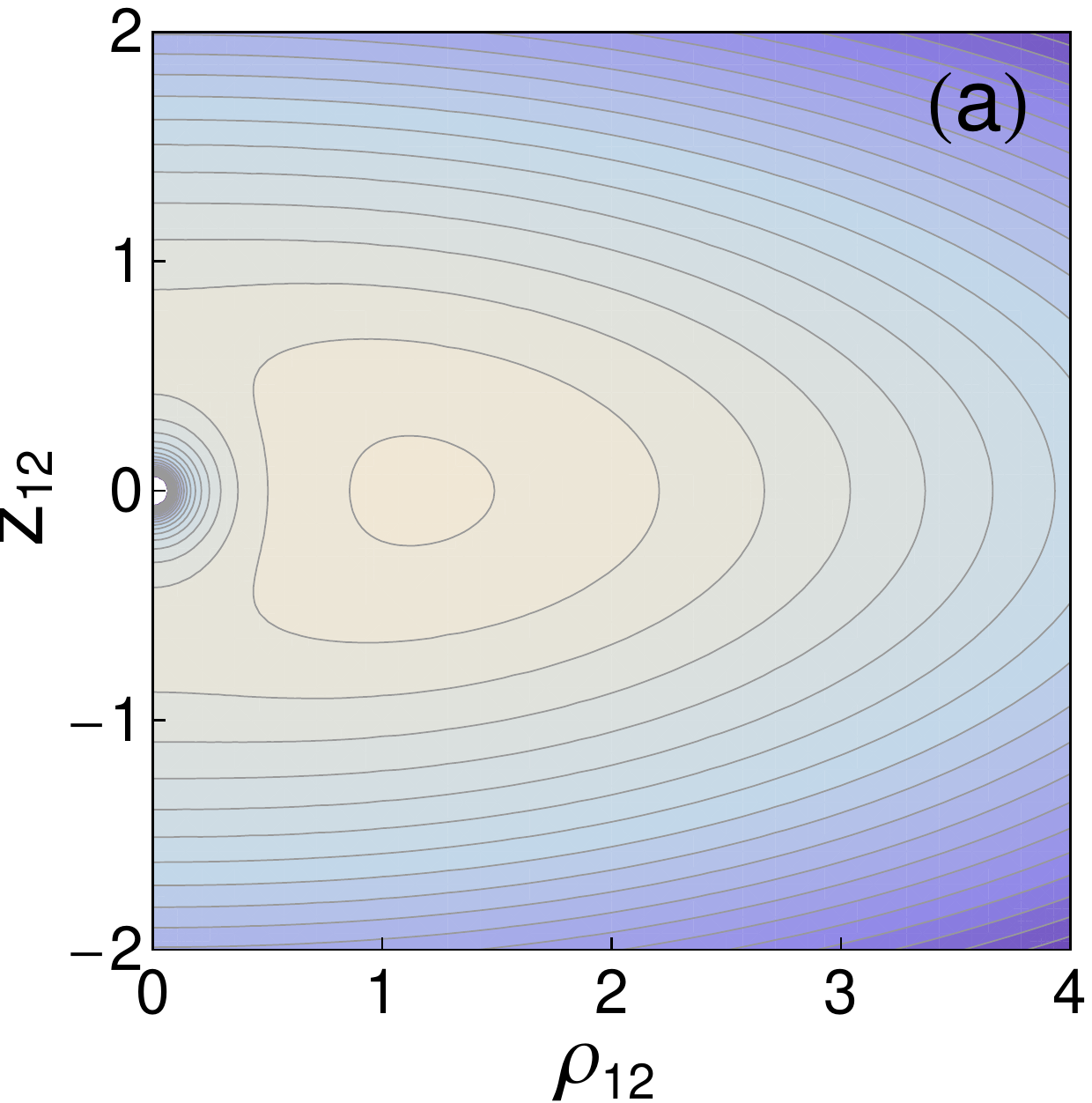}
\end{minipage}
\hspace{0.025in}
\begin{minipage}{.7in}
\includegraphics[scale=.25]{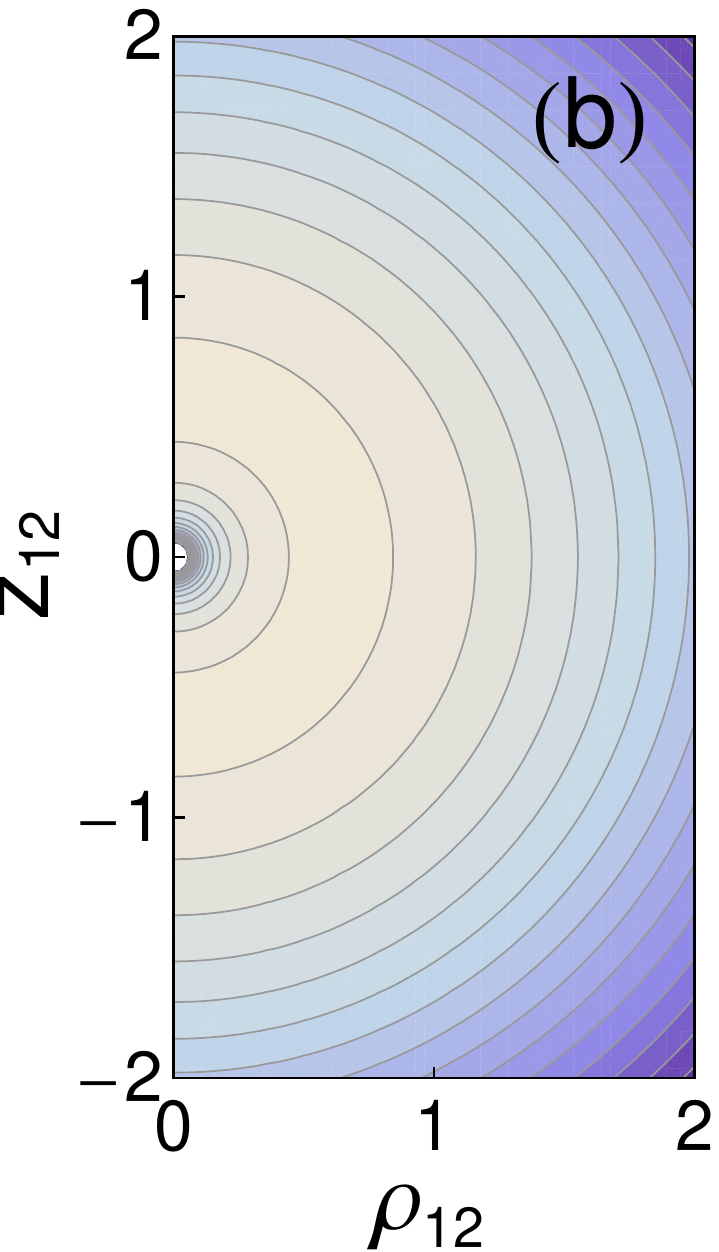}
\end{minipage}
\hspace{0.025in}
\begin{minipage}{.7in}
\includegraphics[scale=.25]{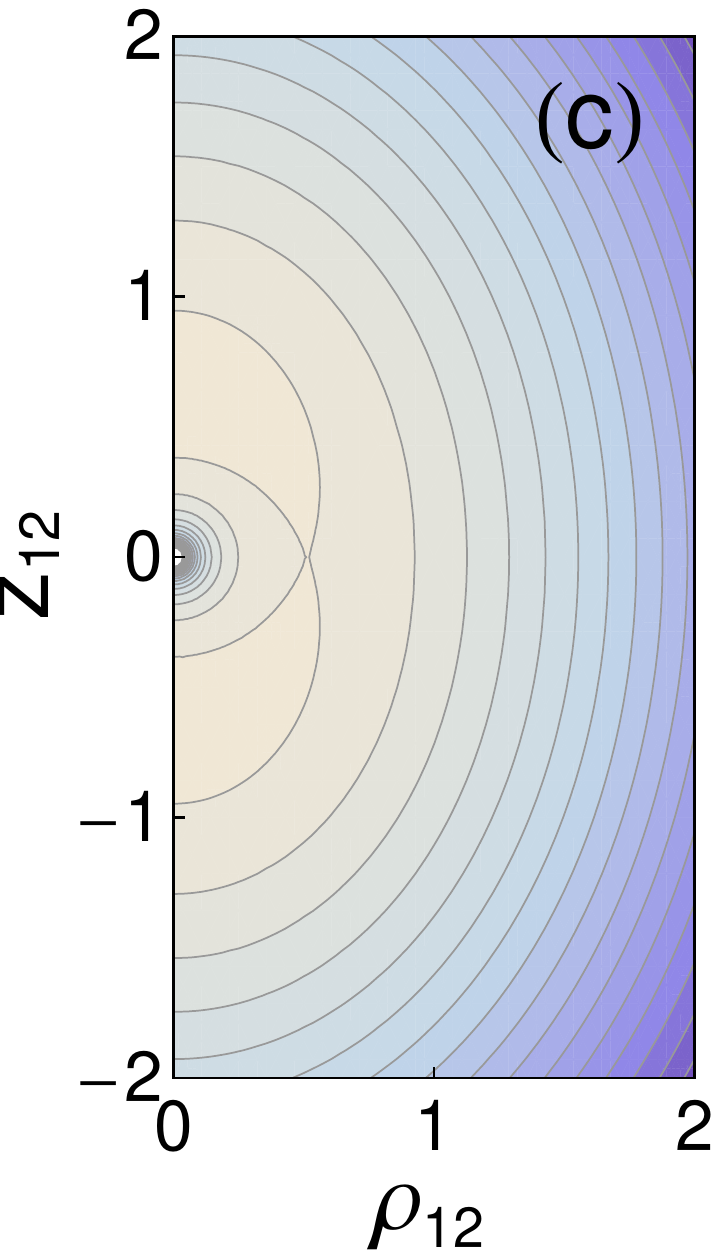}
\end{minipage}
\end{center}
\vspace{-5cm}
\begin{center}
\begin{minipage}{1.2in}
\includegraphics[scale=.25]{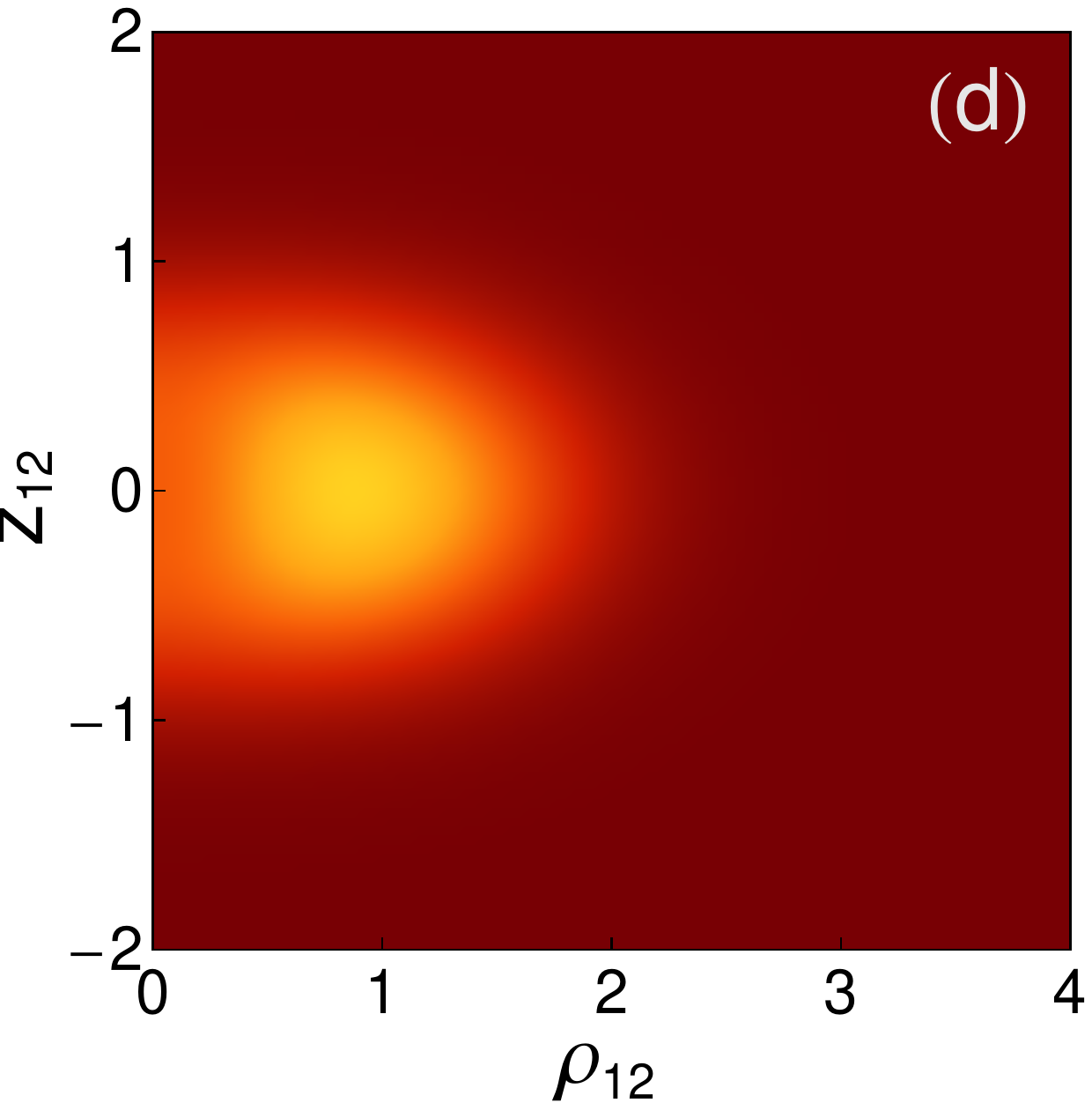}
\end{minipage}
\hspace{0.025in}
\begin{minipage}{.7in}
\includegraphics[scale=.25]{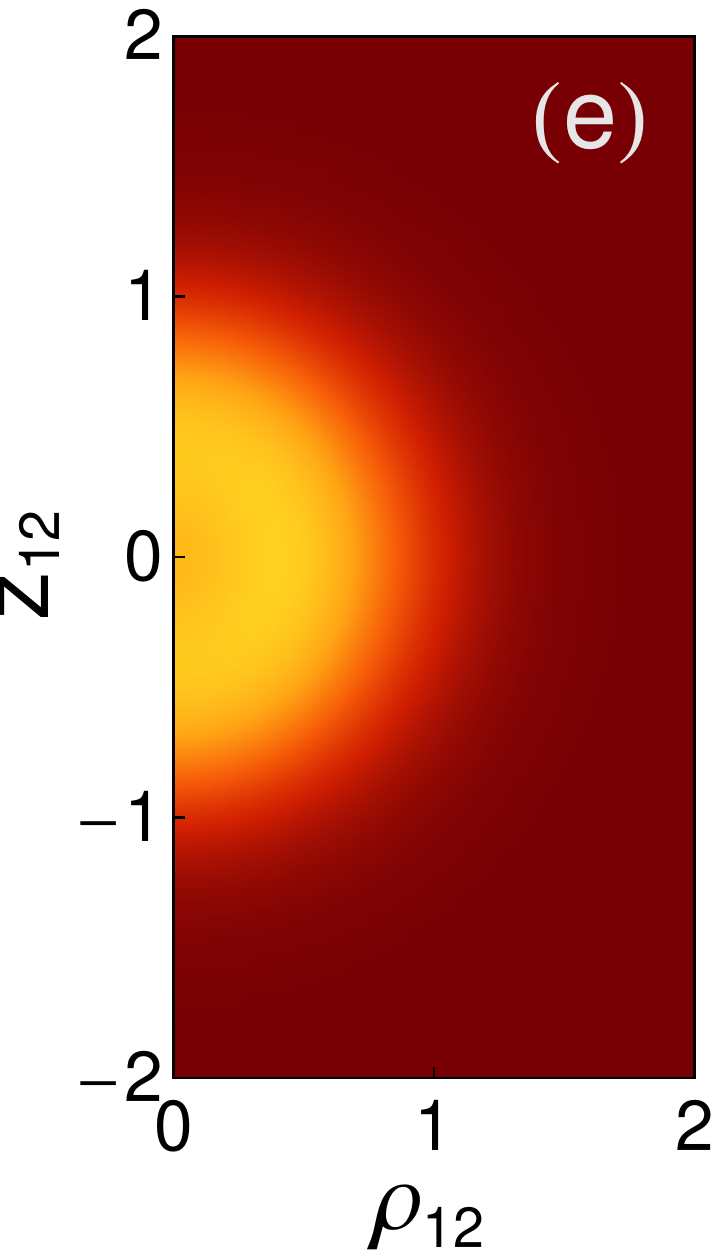}
\end{minipage}
\hspace{0.025in}
\begin{minipage}{.7in}
\includegraphics[scale=.25]{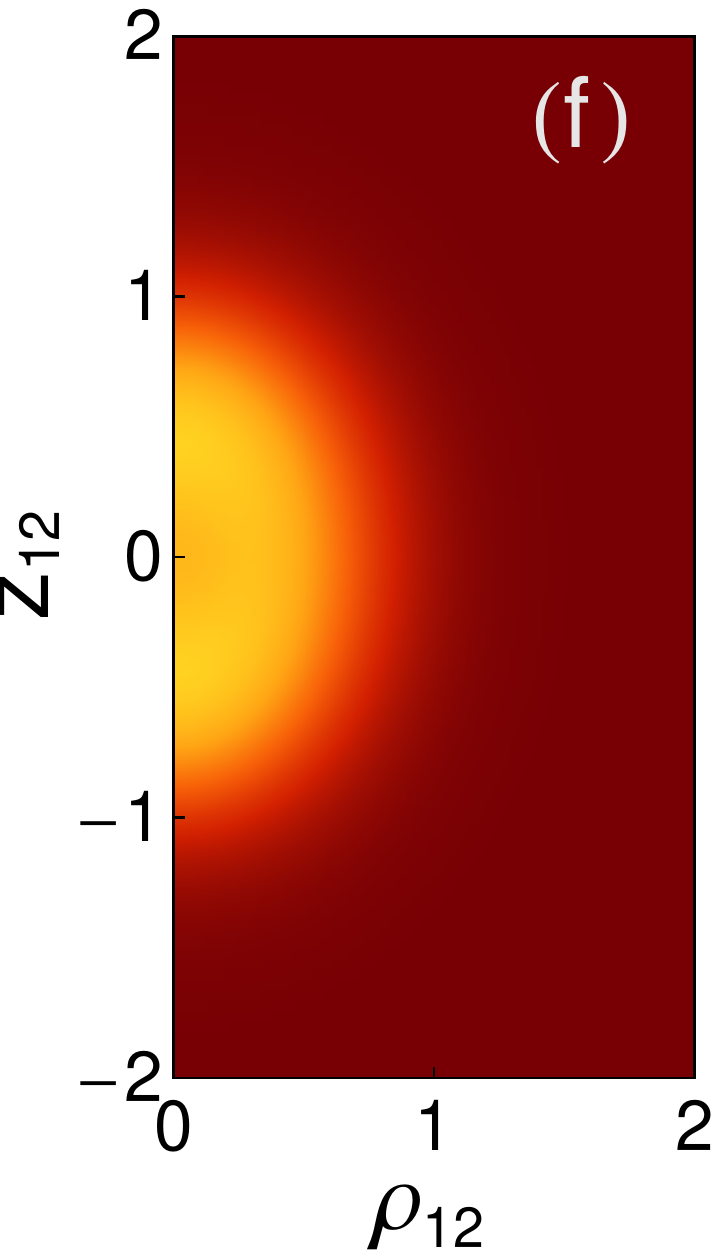}
\end{minipage}
\end{center}
\vspace{-5cm}
\begin{center}
\begin{minipage}{1.2in}
\includegraphics[scale=.25]{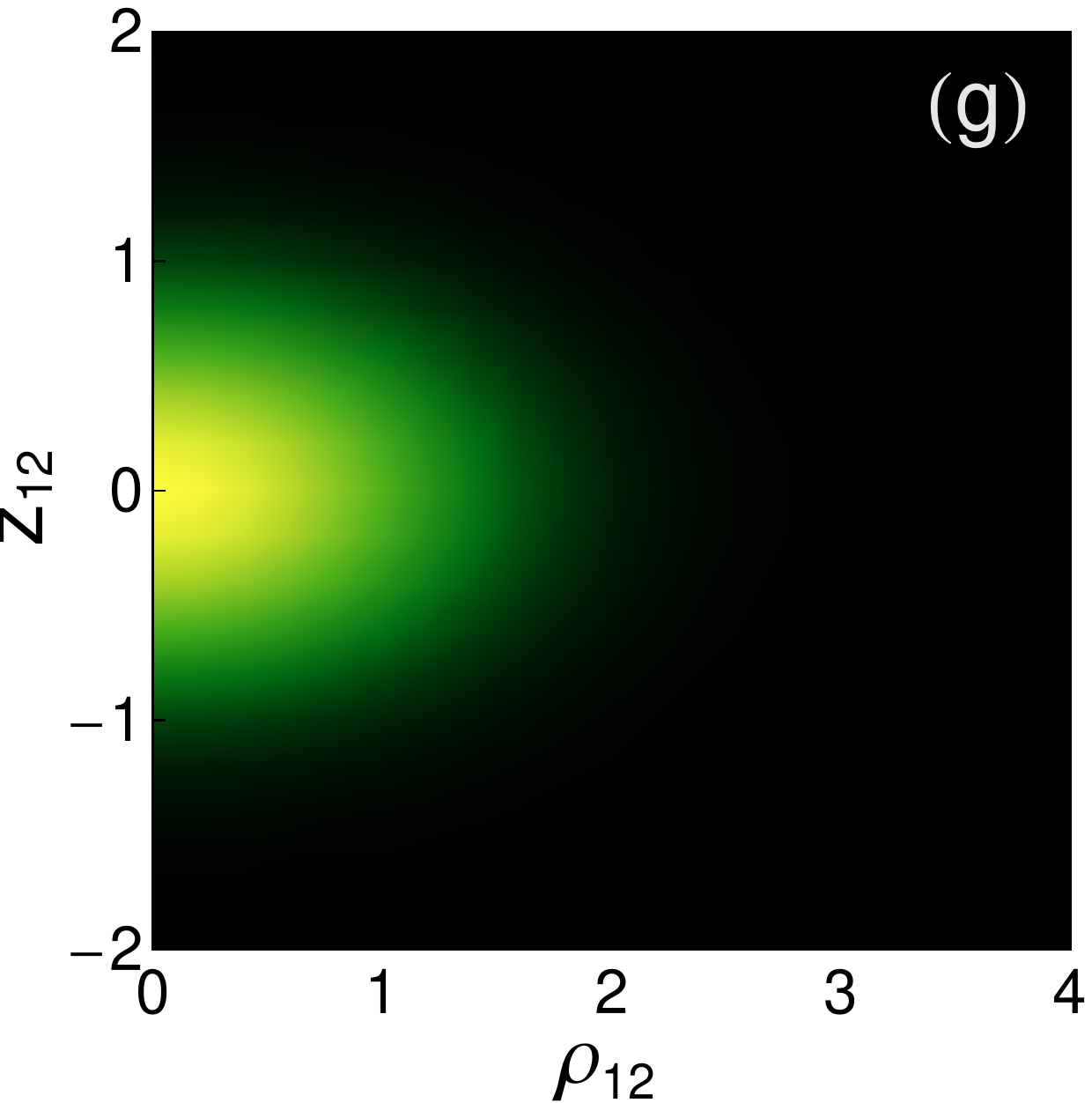}
\end{minipage}
\hspace{0.025in}
\begin{minipage}{.7in}
\includegraphics[scale=.25]{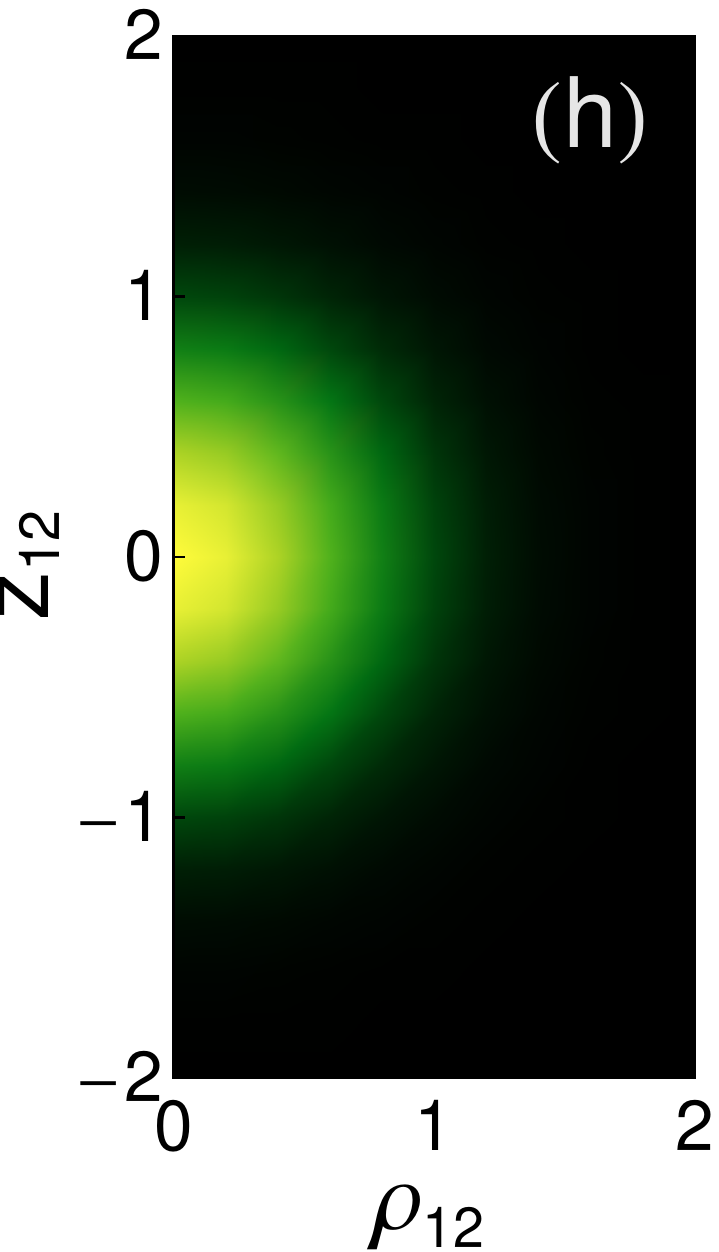}
\end{minipage}
\hspace{0.025in}
\begin{minipage}{.7in}
\includegraphics[scale=.25]{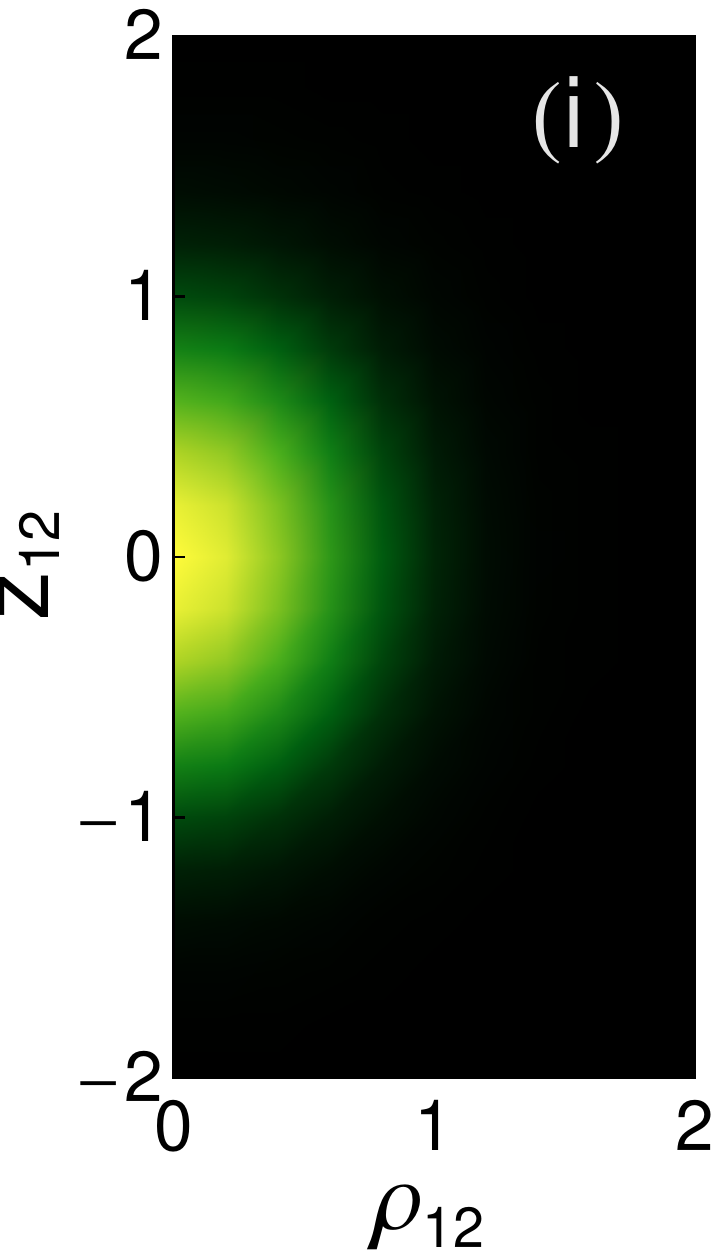}
\end{minipage}
\end{center}
\vspace{-.2in}
\caption{(Color online) The effective potential (\ref{effpot})
with $\omega_z/\omega_0 = 2.5$, $\kappa = 1.5$ and $m = 0$
(top), the probability density $\vert\psi(\mathbf{r}_{12})\vert^2$
(middle), and the electron density $n(\mathbf{r})$ (bottom) for the
lowest state with $m = M = 0$ at three values of the
magnetic field strength: (a,d,g) $B = 0$ (${\tilde\omega}_L = 0$,
$\omega_z/\Omega = 2.5$); (b,e,h) $B = B_\mathrm{sph}$
(${\tilde\omega}_L = {\tilde\omega}_L^{\rm sph} = 2.2913$,
$\omega_z/\Omega = 1$); (c,f,i) $B > B_\mathrm{sph}$
(${\tilde\omega}_L = 3.1798$, $\omega_z/\Omega
= 0.75$).} \label{fig-wfk1.5-m0}
\end{figure}
\begin{figure}[thb]
\vspace{-4cm}
\begin{center}
\begin{minipage}{1.2in}
\includegraphics[scale=.25]{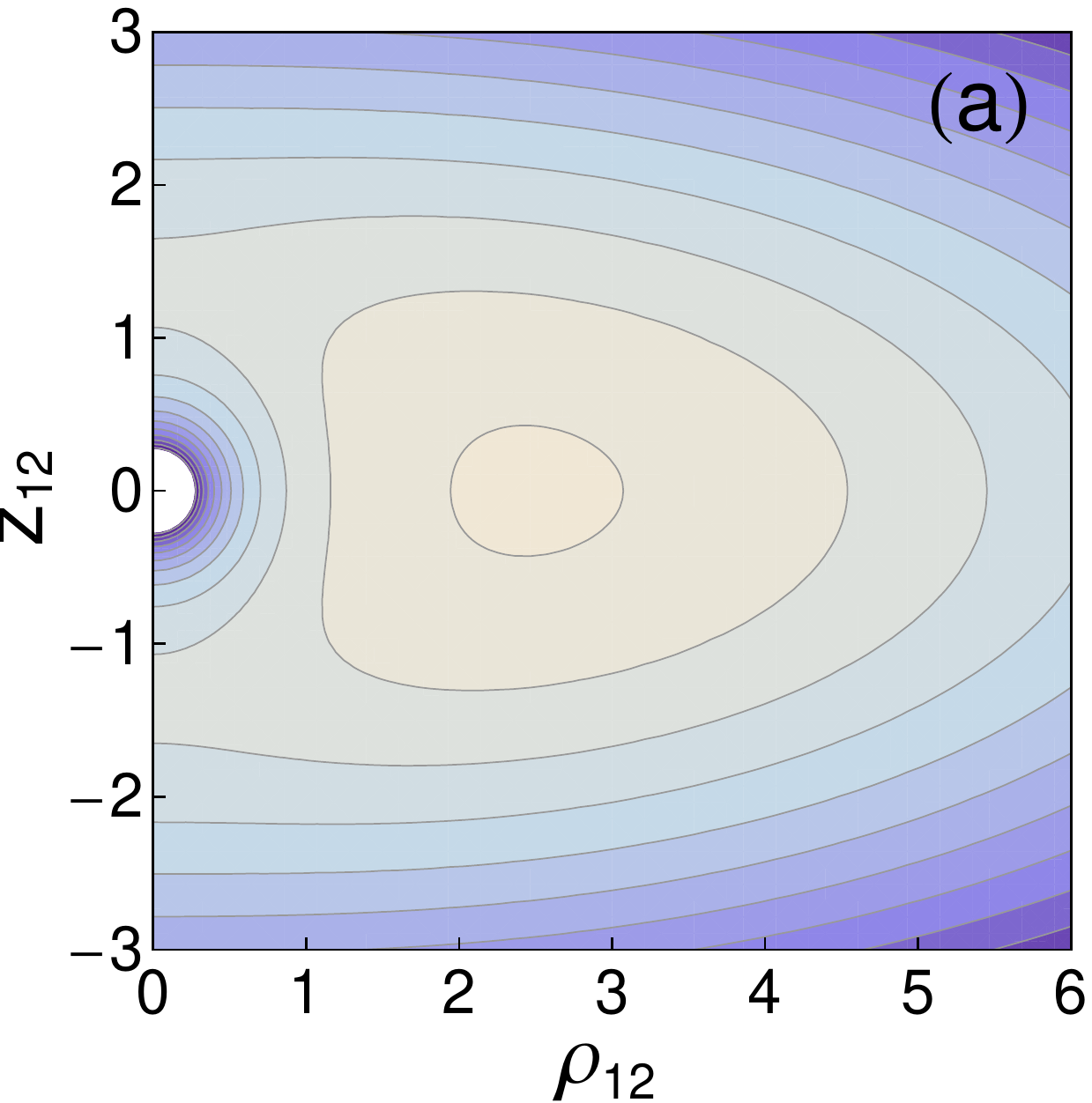}
\end{minipage}
\hspace{0.025in}
\begin{minipage}{.7in}
\includegraphics[scale=.25]{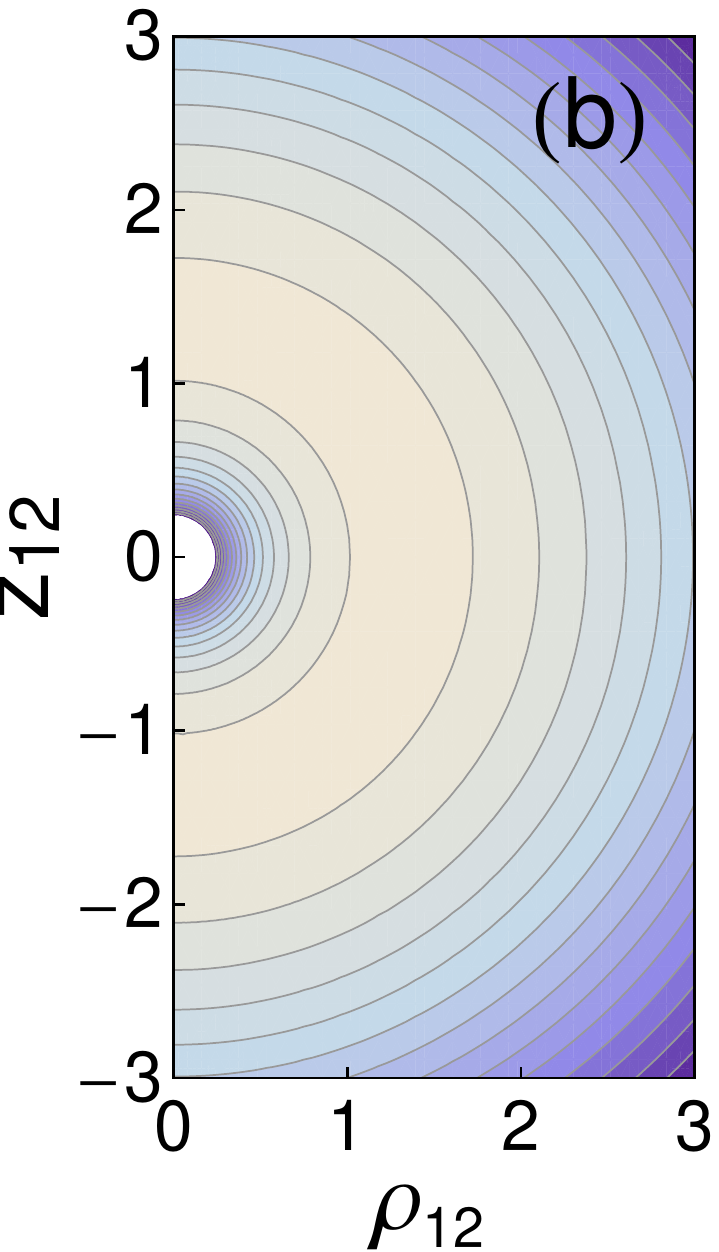}
\end{minipage}
\hspace{0.025in}
\begin{minipage}{.7in}
\includegraphics[scale=.25]{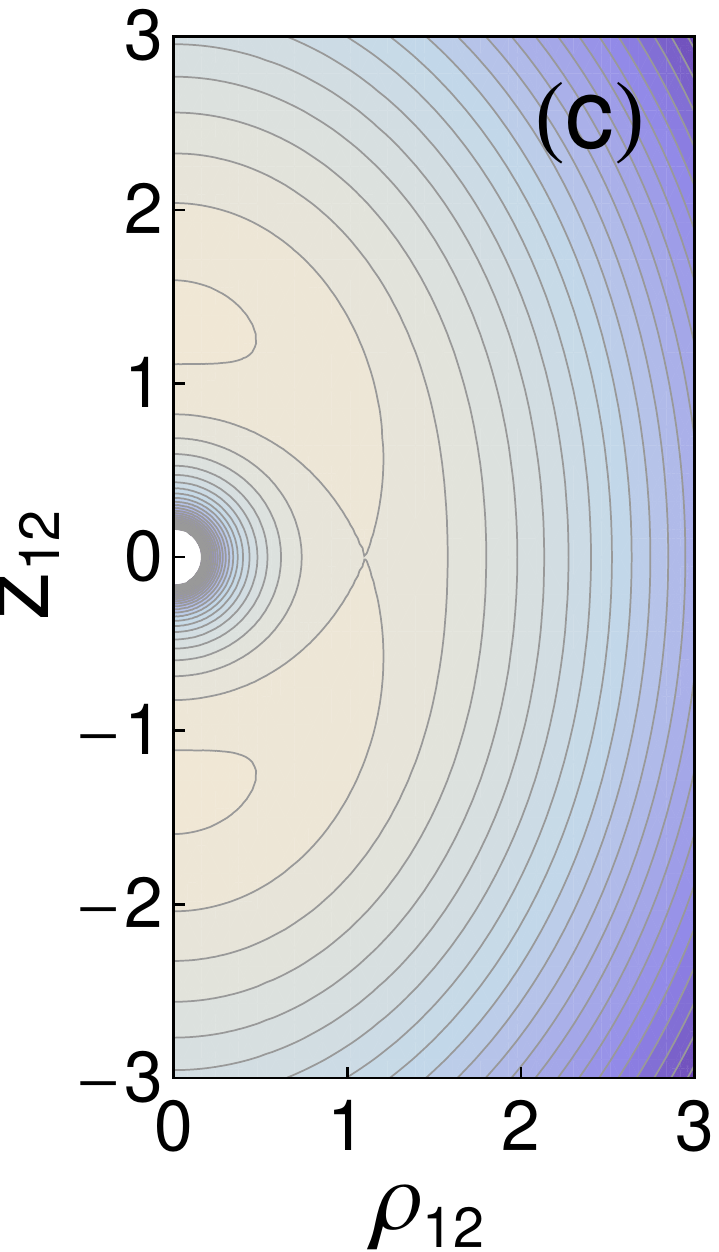}
\end{minipage}
\end{center}
\vspace{-5cm}
\begin{center}
\begin{minipage}{1.2in}
\includegraphics[scale=.25]{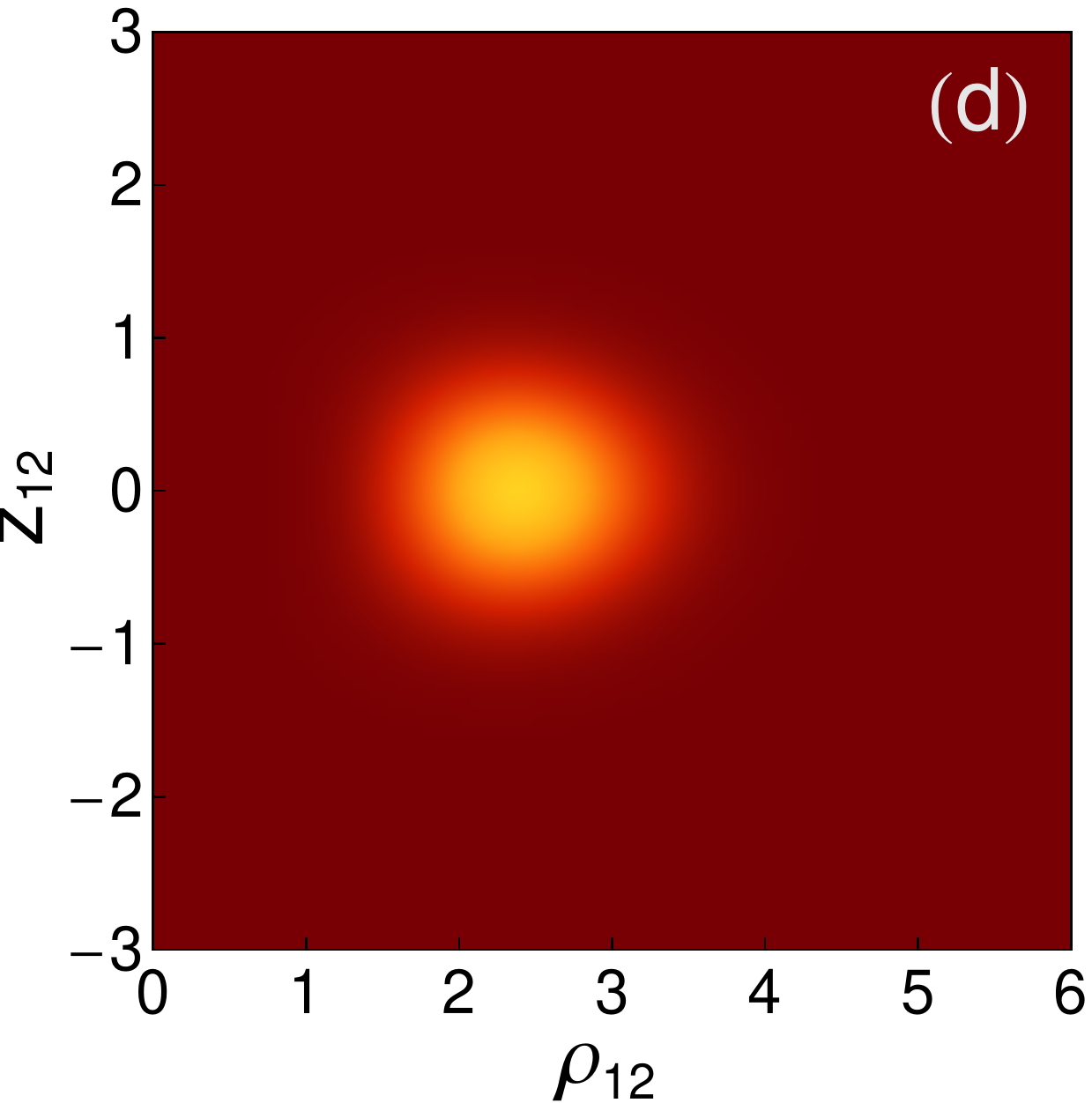}
\end{minipage}
\hspace{0.025in}
\begin{minipage}{.7in}
\includegraphics[scale=.25]{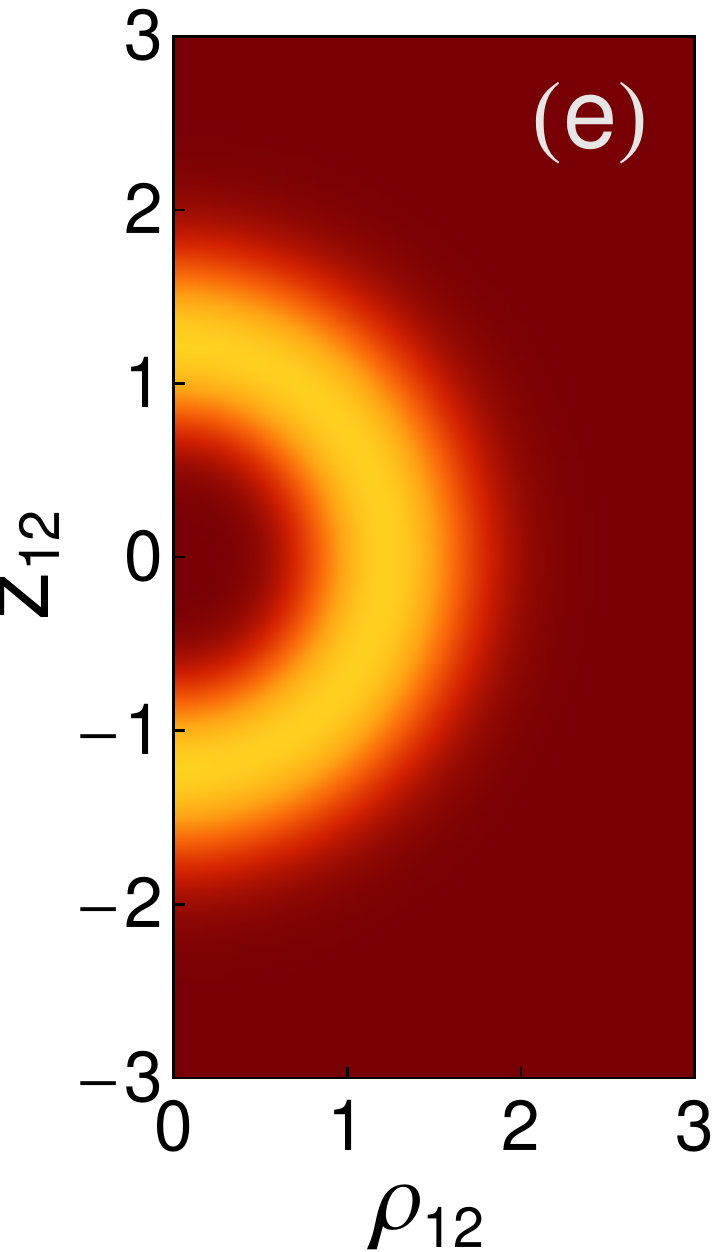}
\end{minipage}
\hspace{0.025in}
\begin{minipage}{.7in}
\includegraphics[scale=.25]{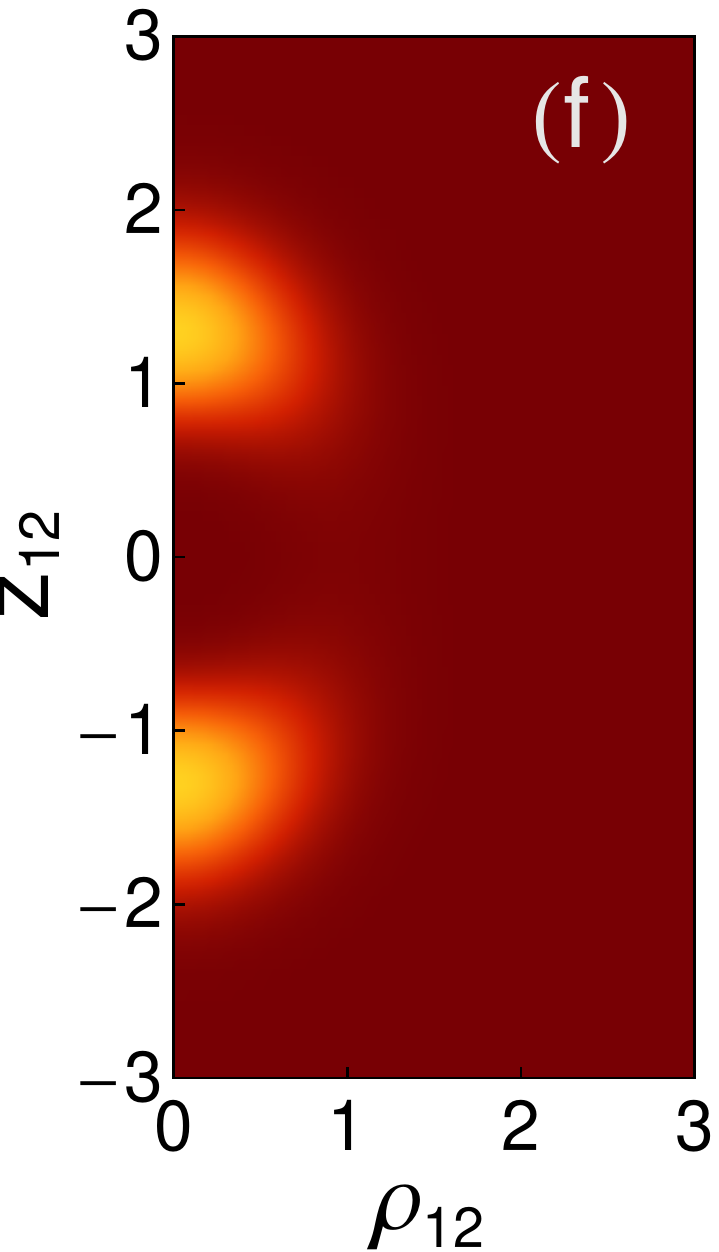}
\end{minipage}
\end{center}
\vspace{-5cm}
\begin{center}
\begin{minipage}{1.2in}
\includegraphics[scale=.25]{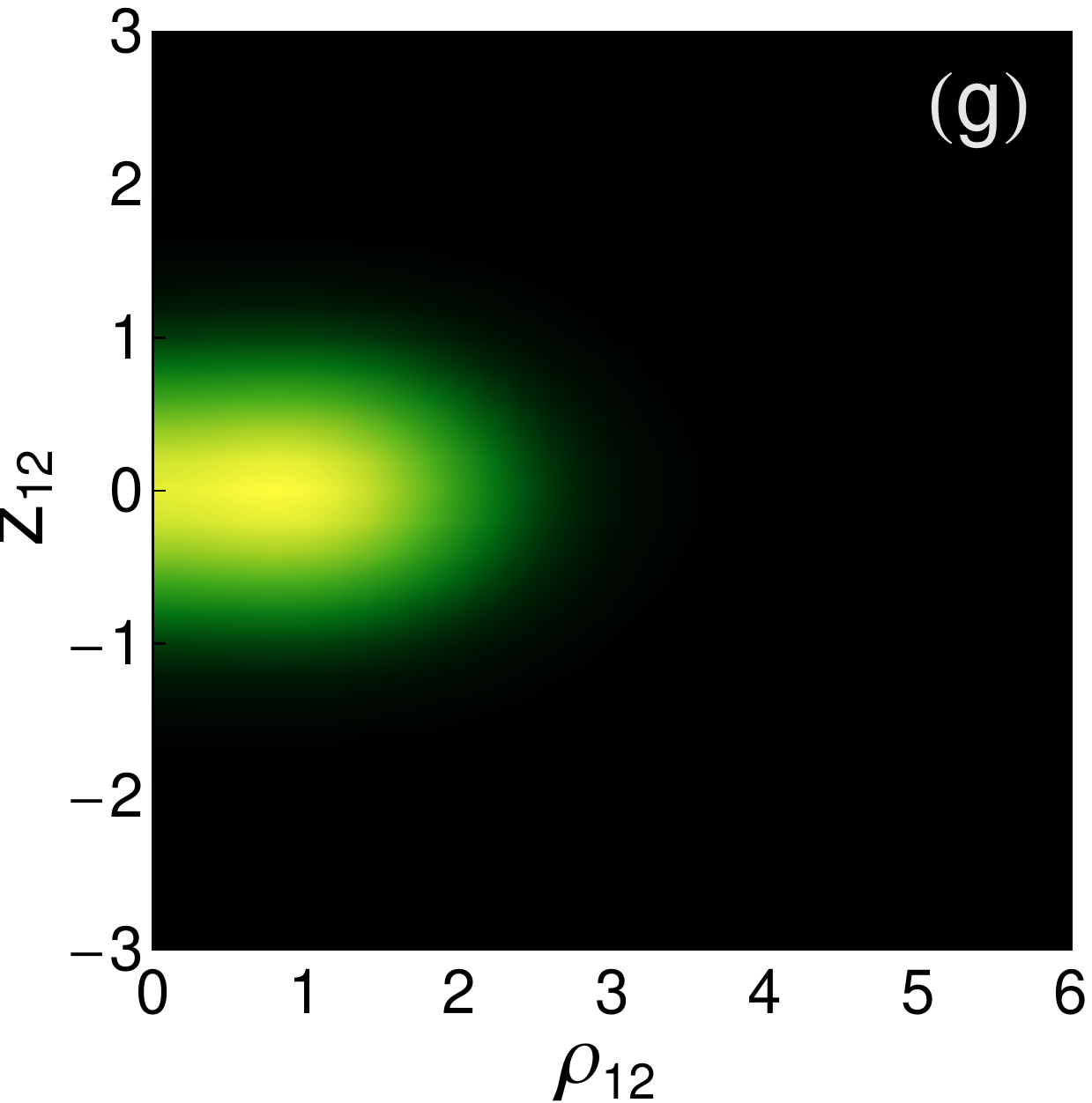}
\end{minipage}
\hspace{0.025in}
\begin{minipage}{.7in}
\includegraphics[scale=.25]{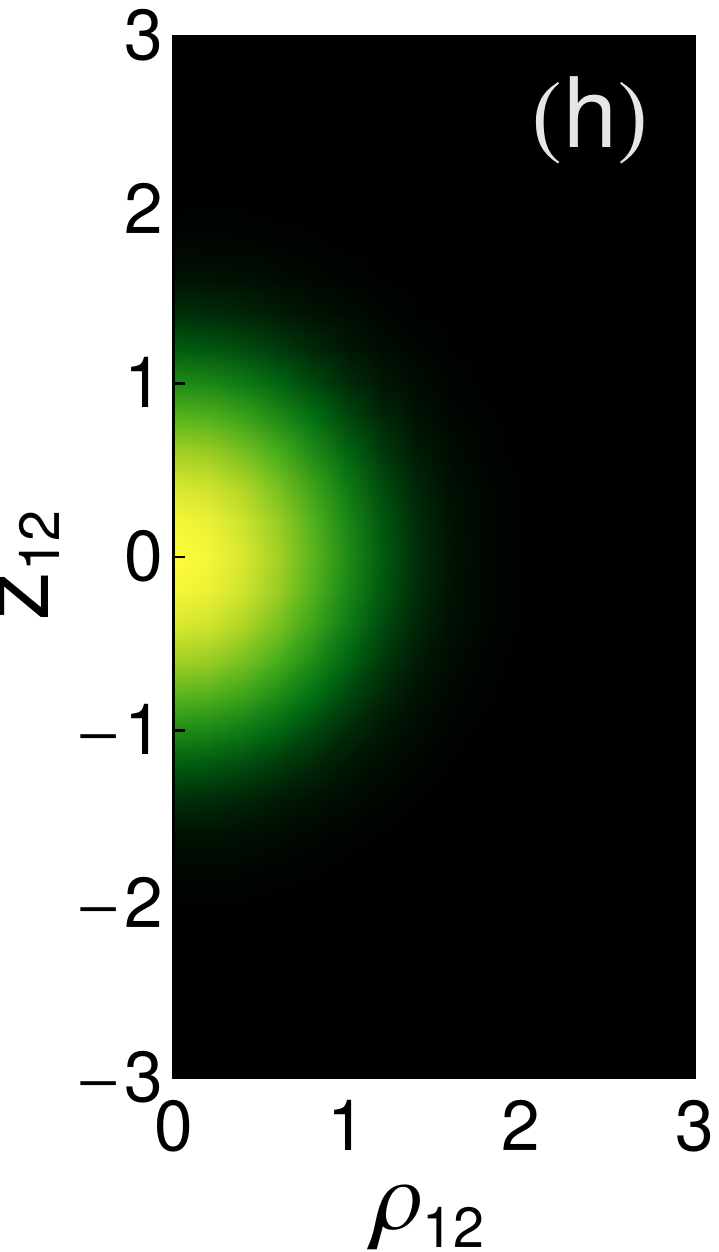}
\end{minipage}
\hspace{0.025in}
\begin{minipage}{.7in}
\includegraphics[scale=.25]{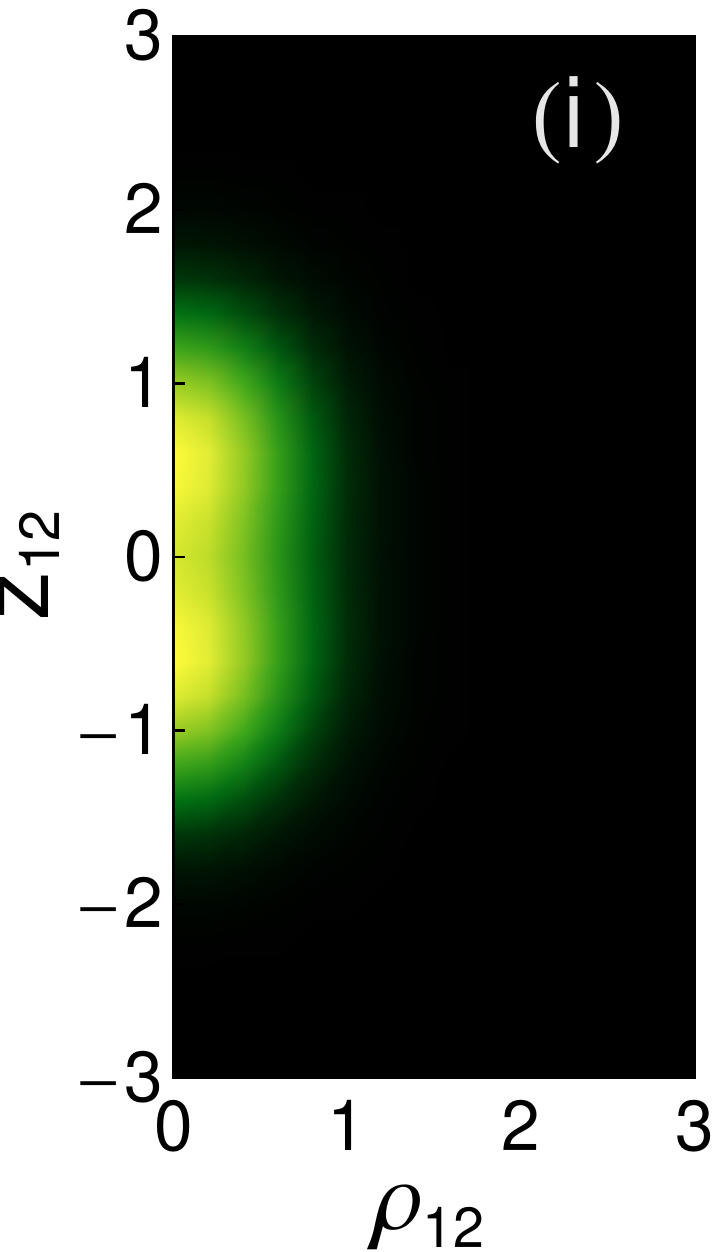}
\end{minipage}
\end{center}
\vspace{-.2in}
\caption{(Color online)
Similar to Fig.\ref{fig-wfk1.5-m0} for $\kappa = 15$.
} \label{fig-potwf-m0}
\end{figure}

For the sake of comparison, on the top of Figs.~\ref{fig-wfk1.5-m0},
\ref{fig-potwf-m0}, we display the equipotential surfaces of 
the effective potential $V_\mathrm{eff}$.
We start the quantum analysis of the electron localization
in the ground state at zero magnetic field ($m = M = 0$).
Figs.~\ref{fig-wfk1.5-m0}(d),(g), display
 the distributions $|\psi(\mathbf{r}_{12})|^2$ and
$n(\mathbf{r})$, respectively, at $\omega_z/\omega_0 =
2.5$, $\kappa = 1.5$ ($R_W = 2.12132$) in the ground state at
$B = 0$. This value of $\kappa$ (and $R_W$) corresponds to the
GaAs QD with $\hbar\omega_0 \approx 2.8$\,meV, which is a typical
lateral confinement strength for QDs created in semiconductor
layers. One observes that, although $|\psi(\mathbf{r}_{12})|^2$ has
a minimum at $\rho_{12} = z_{12} = 0$, the electron density
$n(\mathbf{r})$ has the maximum at $\rho = z = 0$. In other words, 
with given parameters of the QD, the criterion for the onset 
of the Wigner crystallization (see Introduction) is not fulfilled. By
increasing the magnetic field strength $B$ from zero to values
$B > B_\mathrm{sph}$, the ratio $\omega_z/\Omega$ decreases from
$\omega_z/\omega_0$ to values $\omega_z/\Omega < 1$. As a result, the
distributions $|\psi(\mathbf{r}_{12})|^2$ and $n(\mathbf{r})$
change the form. In the example shown in Fig.~\ref{fig-wfk1.5-m0},
however, the electron density $n(\mathbf{r})$ has the maximum at
$\rho = z = 0$ at all field strengths, i.e., there is no a
crystallization. The reason for this absence is a too small value of
the Wigner parameter $R_W$.

We found that 
in a circular two-electron QD with the parabolic confinement in the 2D
approximation (i.e., when $\omega_z/\omega_0 \to \infty$)
the criterion for the crystallization
is fulfilled at $R_W \ge 12.8$. For QDs with a finite thickness the
critical value for $R_W$ is additionally shifted upward. For
example, at the ratio $\omega_z/\omega_0 = 2.5$ and
zero magnetic field, the critical value is approximately $14$. The
GaAs QD with this value of $R_W$ has the lateral confinement
strength $\hbar\omega_0 \approx 0.065$\,meV. Below we consider an
example where $R_W$ is sufficiently large for the formation of
Wigner molecule at $B = 0$.

Fig.~\ref{fig-potwf-m0} displays the results discussed above for $\kappa =1.5$,
at $\kappa =15$ ($R_W = 21.2132$). For any value of $B$ the maxima of
$\vert\psi(\mathbf{r}_{12})\vert^2$ are located at the positions
of minima of the effective potential as in a typical QD.
For large $R_W$ the electrons are more separated, and
these maxima are more prominent. As a consequence, the maxima of
electron density $n(\mathbf{r})$ are now dislocated from the
center of the QD. If $B < B_\mathrm{sph}$, we have a typical Wigner
molecule of a ring type. For $B > B_\mathrm{sph}$ the electron
density has two maxima in the vertical direction, located
symmetrically from the opposite sides of the center (vertical
type). (The full 3D shapes can be obtained by rotating plots
(g)-(i) in Fig.~\ref{fig-potwf-m0} around $z$-axis.) The shape
transition of the Wigner molecule occurs at $B \approx B_\mathrm{sph}
\equiv  B_\mathrm{bif}(0)$.

Thus, our assumption on the correspondence between the classical potential minima and
the probability density
$|\psi(\mathbf{r}_{12})|^2$ maxima of the low-lying eigenstates  are confirmed, indeed, by 
 direct comparisons of the corresponding plots (see
the top and middle rows of Figs.~\ref{fig-wfk1.5-m0},
\ref{fig-potwf-m0}).
For $B <B_\mathrm{bif}$ the stationary point (i) of the potential, which
is located at $(\rho_a,0)$, is the minimum (see
Sec.~\ref{sec:effpot}), while the probability density
$|\psi(\mathbf{r}_{12})|^2$ has there a maximum (see Figs.6(a),(d); 7(a),(d)).
For $B > B_\mathrm{bif}$, however, there are
two minima of the potential at $(\rho_b,\pm z_b)$, divided by the
potential barrier. Consequently, the probability density is distributed
symmetrically in the $z_{12} < 0$ and $z_{12} > 0$ half-planes at
the positions of these minima (see Figs.6(c),(f); 7(c),(f)).
Therefore, for low-lying states a splitting of the probability density maximum
occurs simultaneously with the bifurcation of the stationary point (i)
of the effective potential.

We recall that the so-called
singlet-triplet transitions take place (see below
Sec.~\ref{sec:STvsSHT}) with the increase of the magnetic field strength.
The ground state, which is initially
characterized by $m = M = 0$, at higher field strengths is
characterized by different non-zero values of $m$ (keeping $M =
0$). To trace the localization of electrons in the ground states
at $B > 0$, we have to consider also the lowest states with $m >
0$. We found that, similar to the $m = 0$ case, the maxima of
$|\psi(\mathbf{r}_{12})|^2$ for $m \neq 0$ are located at
the positions of the effective potential minima. On the other
hand, the electron density $n(\mathbf{r})$ maxima are allocated on
the ring around the dot center (associated with the formation of
the Wigner molecule) at $B < B_\mathrm{bif}(m)$;
 and it is of the vertical type distribution at $B > B_\mathrm{bif}(m)$.

In general, in an axially symmetric QD,  
 for the lowest state with $m>0$
at the field strength  $B$ the corresponding electron density distribution 
changes the shape when the magnetic field exceeds 
the value $B_\mathrm{bif}(m)$.
 We will show below that in such a QD {\em the shape transitions can not
take place in the ground states, in principle}.

\section{Energy spectra}
\label{sec:spectra}

As we pointed out in Sec.~\ref{sec:model}, the QD energy levels  can
be written as sums $E = E_\mathrm{CM} + E_\mathrm{rel}$. Here
$E_\mathrm{CM}$ are the eigenenergies of the Hamiltonian $H_\mathrm{CM}$, given by
Eq.~(\ref{eq:Ecm}), and $E_\mathrm{rel}$ are the eigenenergies of
the Hamiltonian $H_\mathrm{rel}$ that can be determined either numerically or
using an approximate method. Since the Coulomb interaction does
not affect the CM motion, manifestations of the strength of this interaction
are expected in the energy spectrum of the Hamiltonian $H_\mathrm{rel}$. Below we
analyze the magnetic field dependence of the energy levels
$E_\mathrm{rel}$ at typical (small)
and large values of the strength parameter $\kappa$.

Numerical values for the eigenenergies of the Hamiltonian $H_\mathrm{rel}$ are
calculated by diagonalizing this Hamiltonian in the oscillator
basis.  Details of the method are given in
Appendix \ref{sec:diag}. Alternatively, low-lying energy levels for the
relative motion can be calculated by means of analytical expressions
discussed below.

\begin{figure*}
\vspace{-7cm}
\begin{center}
\hspace{1.5cm}
\includegraphics[scale=.45]{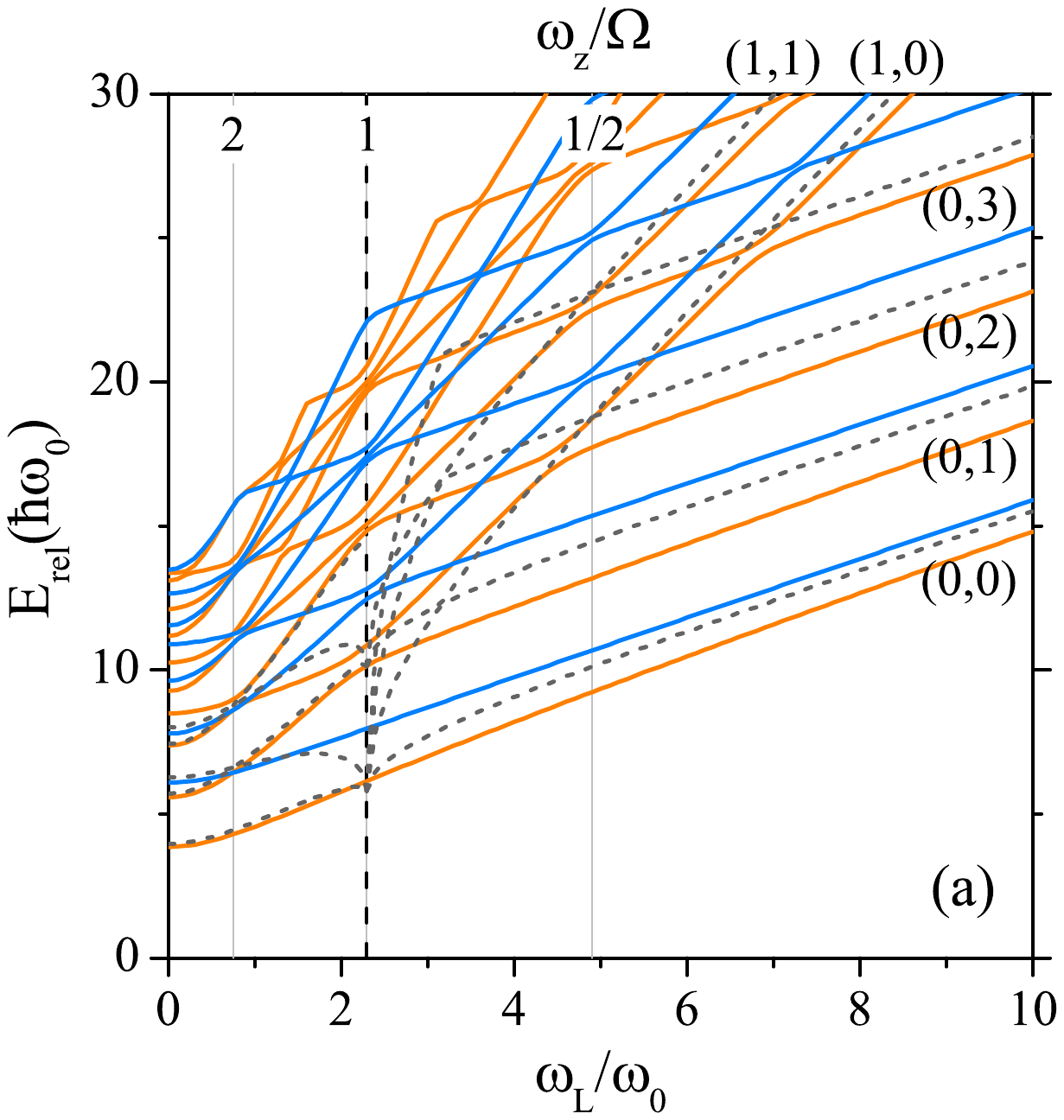}
\hspace{-3.5cm}
\includegraphics[scale=.45]{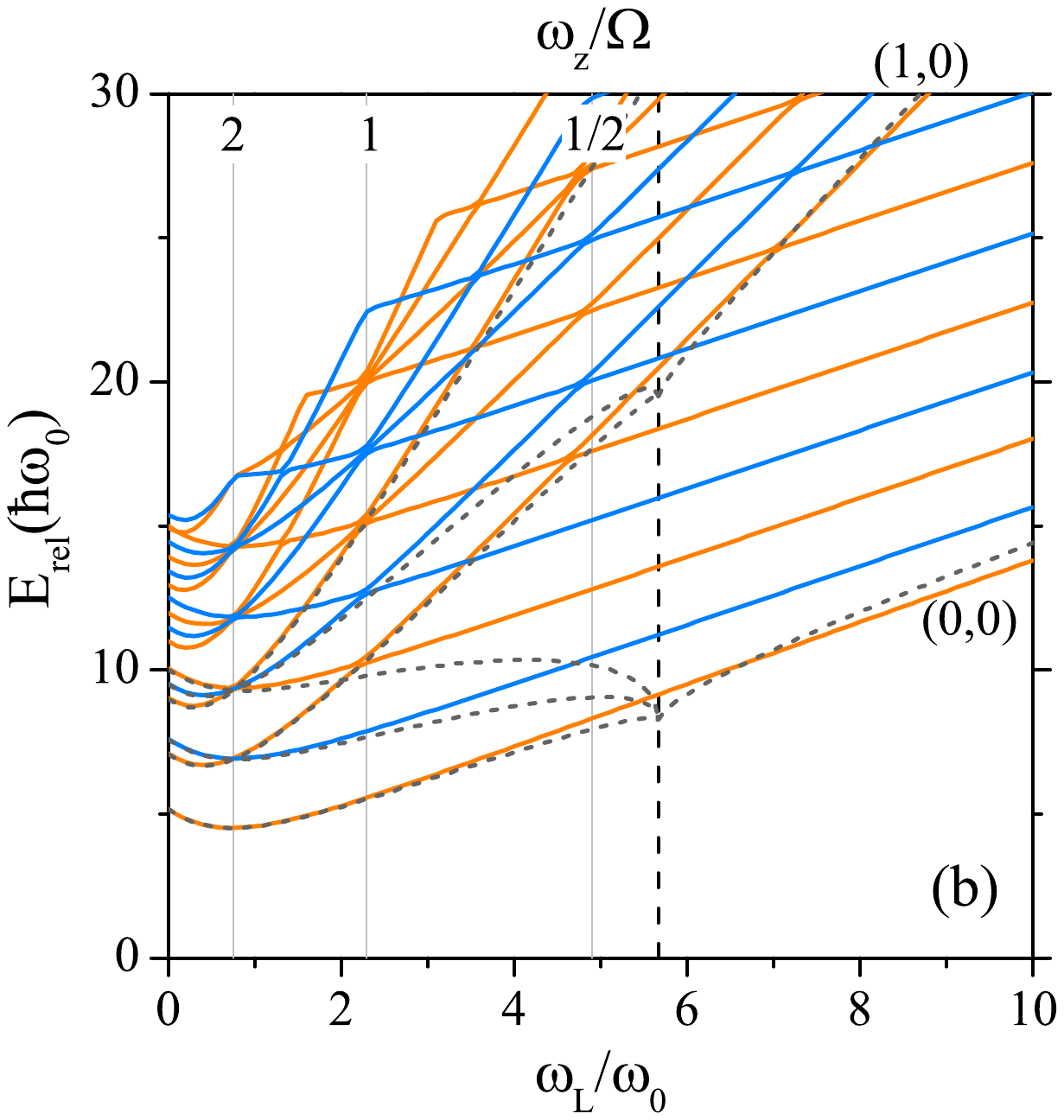}
\end{center}
\vspace{-.2in} \caption{(Color online) Low-lying energy levels corresponding to
the relative motion of two electrons in the axially symmetric QD
with $\omega_z/\omega_0 = 2.5$ and $\kappa = 1.5$ at different
magnetic field strengths  ($\omega_L/\omega_0$) for: (a) $m = 0$
and (b) $m = 2$. Orange and blue full lines represent numerically
calculated levels that correspond to even and odd states,
respectively; the dotted lines represent the approximate
levels, obtained with the aid of Eqs.~(\ref{Erel-a}) and (\ref{Erel-b}). The
vertical dashed lines mark ${\tilde\omega}_L^{\rm bif}(m)$-values
(here ${\tilde\omega}_L^{\rm bif}(0) \equiv {\tilde\omega}_L^{\rm
sph} = 2.2913$ and ${\tilde\omega}_L^{\rm bif}(2) = 5.66296$). The
integrable cases: $\omega_z/\Omega = 2$, $1$, and $1/2$ are
indicated by vertical gray lines.} \label{fig-enlevs-k1p5}
\end{figure*}

\subsection{Approximate calculations of low-lying energy levels:
Classification of states.} \label{sec:approx}

As we have seen in Sec.~\ref{sec:effpot}, when $\omega_L <
\omega_L^{\rm bif}(m)$, the minimum of the effective potential
(\ref{effpot}) is located at $(\rho_{12},z_{12}) = (\rho_a,0$)
(see Figs.~\ref{fig:pot+ff-m=0}(a) and \ref{fig:pot+ff-m>0}(a)),
and the expansion around this point is given by
Eq.~(\ref{eq:Veff-expan-a}). The low-lying eigenenergies of the Hamiltonian
$H_\mathrm{rel}$ (in $\hbar\omega_0$ units) are determined
approximately by formula
\beq
E_{\rm rel} = V_a + {\tilde\omega}_{1a}(n_\rho +
\hbox{$\frac{1}{2}$}) + {\tilde\omega}_{2a}(n_z +
\hbox{$\frac{1}{2}$}) - {\tilde\omega}_L m, \label{Erel-a}
\eeq
where $V_a$, ${\tilde\omega}_{1a}$ and ${\tilde\omega}_{2a}$ are
given by Eqs.~(\ref{eq:Va}), (\ref{eq:omega1a}) and
(\ref{eq:omega2a}), respectively.

When $\omega_L > \omega_L^{\rm bif}$, the effective potential
$V_\mathrm{eff}(\rho,z)$ has the minima at ($\rho_b,\pm z_b$) (see
Figs.~\ref{fig:pot+ff-m=0}(b) and \ref{fig:pot+ff-m>0}(b)), and we
expand the potential around one of them.
With the aid of the results obtained in Sec.\ref{sec:smallosc_b}, we determine
the low-lying eigenenergies of $H_\mathrm{rel}$ (in
$\hbar\omega_0$ units) by the approximative formula
\begin{equation}
E_{\rm rel} =  V_b + \tilde\omega_{1b} (n_1 +
\hbox{$\frac{1}{2}$}) + \tilde\omega_{2b} (n_2 +
\hbox{$\frac{1}{2}$}) - \tilde\omega_L m, \label{Erel-b}
\end{equation}
where $V_b$, $\tilde\omega_{1b}$ and $\tilde\omega_{2b}$ are given
by Eqs.~(\ref{eq:Vb}), (\ref{om1}) and (\ref{om2}), respectively.

Therefore, the low-lying energy levels can be classified
by the set of quantum numbers $\{N, N_z, M, n, n_z, m\}$, when
$\omega_L < \omega_L^{\rm bif}$; and by the set $\{N, N_z, M, n_1,
n_2, m\}$, when $\omega_L > \omega_L^{\rm bif}$. Here $N$, $N_z$,
$M$ and $m$ are good quantum numbers, and $n$, $n_z$, $n_1$ and
$n_2$ are approximate quantum numbers.

\subsection{Manifestations of shape transitions in the energy spectrum}
\label{sec:manifest}

Fig.~\ref{fig-enlevs-k1p5} demonstrates a good agreement between
the numerical and analytical results
obtained by means of Eq.~(\ref{Erel-a})  for $B \ll B_\mathrm{bif}(m)$
at the small value of $\kappa$. The spectra
exhibit typical features, as (anti)crossings of energy levels at
the field strengths when the system becomes integrable \cite{SN}. 
For small values of $\kappa$ ($\sim 1$),
however, there is no any indication in the spectra on
shape transitions.

\begin{figure*}
\vspace{-7cm}
\begin{center}
\hspace{1.6cm}
\includegraphics[scale=.45]{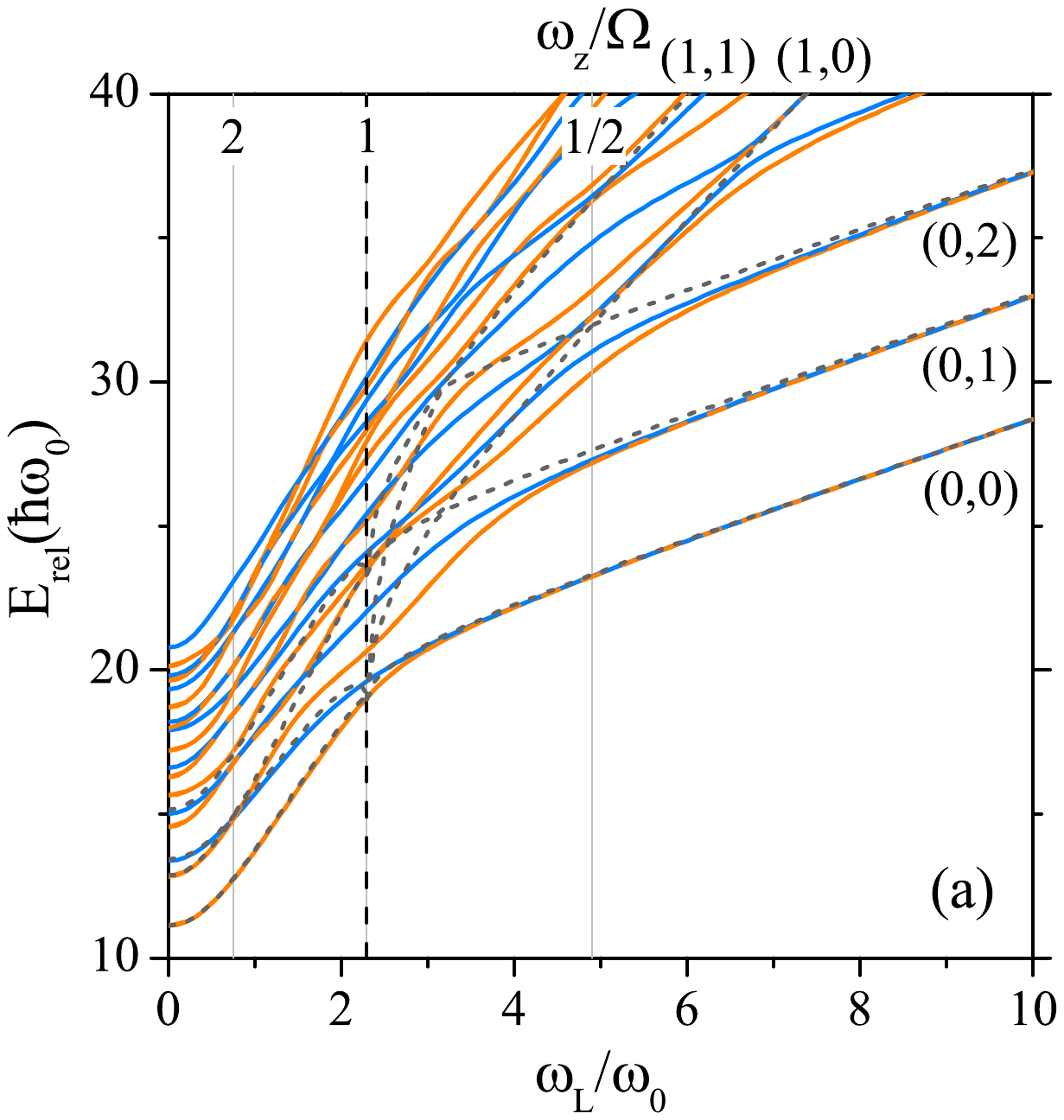}
\hspace{-3cm}
\includegraphics[scale=.45]{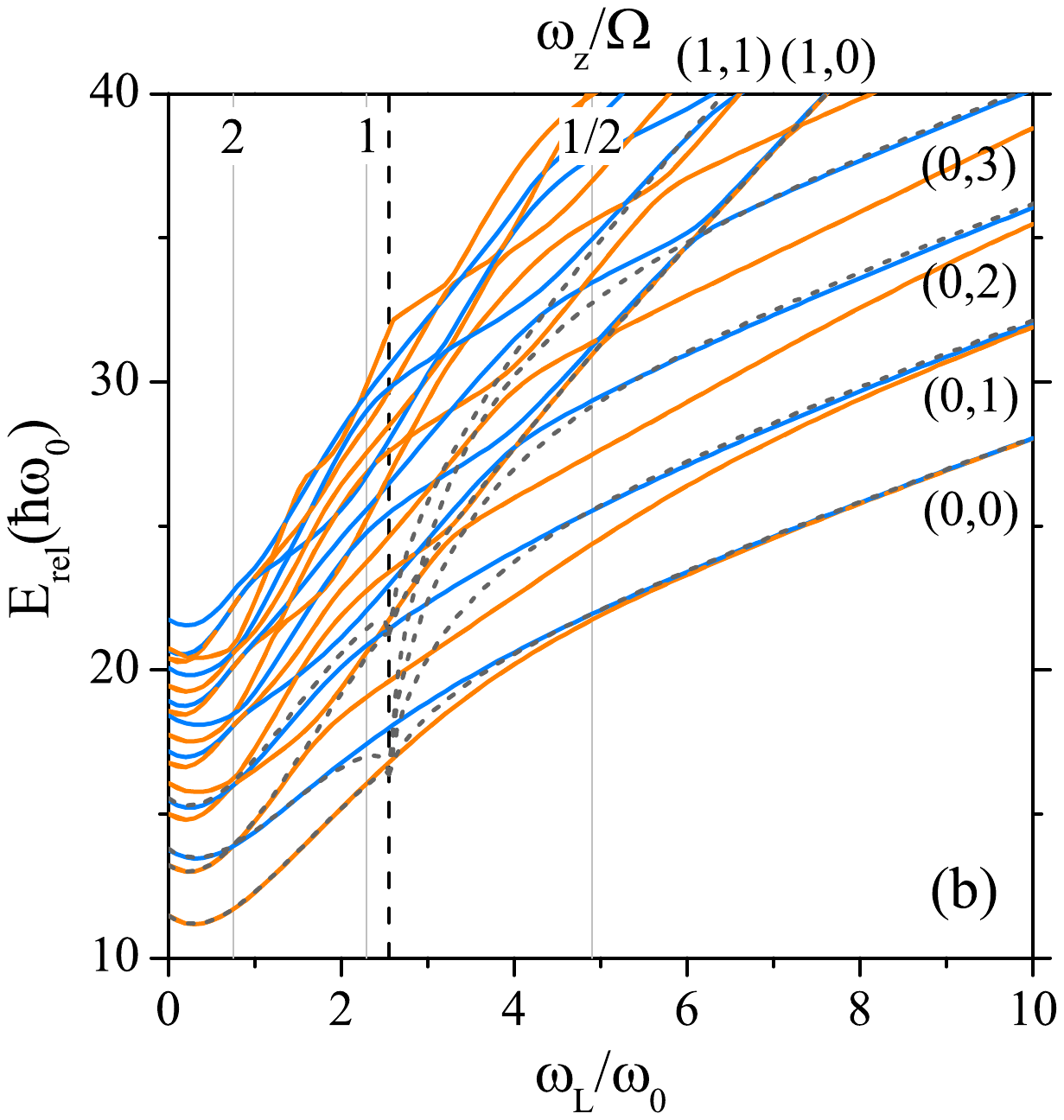}
\end{center}
\vspace{-.2in}
\caption{(Color online) Similar to Fig. \ref{fig-enlevs-k1p5} at $\kappa = 15$. The
vertical dashed lines mark ${\tilde\omega}_L^{\rm bif}(m)$-values
(here ${\tilde\omega}_L^{\rm bif}(0) \equiv {\tilde\omega}_L^{\rm
sph} = 2.2913$ and ${\tilde\omega}_L^{\rm bif}(2) = 2.5485$).}
\label{fig-enlevs-k15}
\end{figure*}

\begin{figure}
\vspace{-3cm}
\begin{center}
\includegraphics[scale=.3]{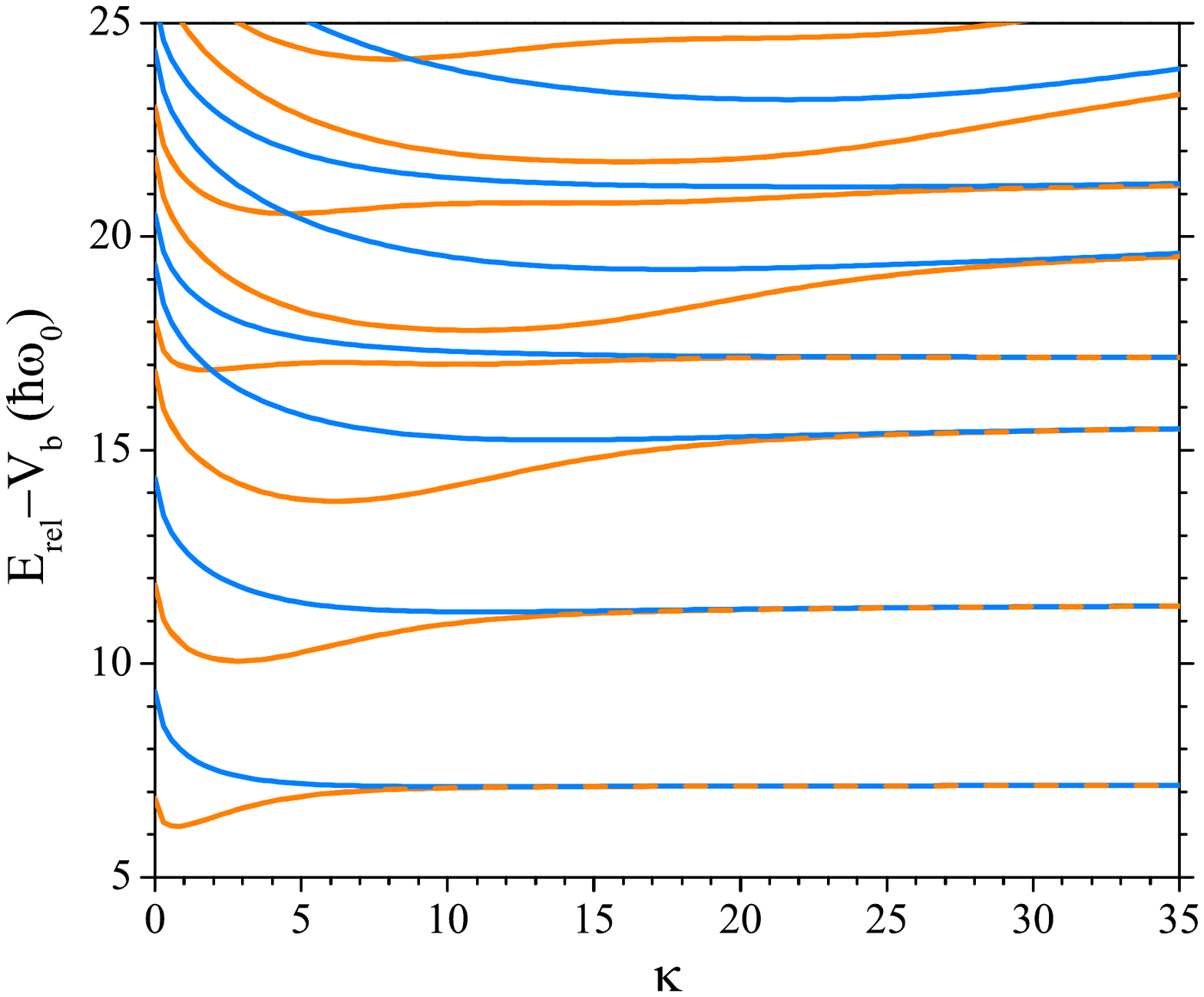}
\end{center}
\vspace{-.2in}
\caption{(Color online) Dependence of the low-lying energy levels (with $m = 0$) on
the parameter $\kappa$ at $\omega_z/\omega_0
= 2.5$ and $\omega_L/\omega_0 = 5.5$. The value $V_b$ is the minimum of
the effective potential (\ref{effpot}) at $(\rho_b,z_b)$, which is for
$m = 0$ equal $\frac{3}{2} {\tilde\omega_z}^2 r_b^2$. The levels
corresponding to even and odd states are denoted by orange and
blue lines, respectively.} \label{fig-denk}
\end{figure}

The situation is different for large values of the strength $\kappa$.
Fig.~\ref{fig-enlevs-k15} displays the same levels as shown in
Fig.~\ref{fig-enlevs-k1p5}, with ${\tilde\omega}_z
= 2.5$ and much stronger strength $\kappa = 15$. For the magnetic field strengths, 
i.e., $\omega_L < \omega_L^\mathrm{bif}(m)$
the spectra have similar structure at small and at large values of
$\kappa$. However, by increasing $\omega_L$ over
$\omega_L^\mathrm{bif}(m)$ and more, if $\kappa$ is sufficiently
large, the pairs of nearby levels with different parities
(even/odd) start converging each to other, forming doubly
degenerate levels. A similar effect occurs if, at the fixed
magnetic field strength such that $\omega_L > \omega_L^\mathrm{bif}(m)$,
the parameter $\kappa$ increases (see Fig.~\ref{fig-denk}).

\begin{figure*}[thb]
\hspace{.2in}
\vspace{-6cm}
\begin{center}
\begin{minipage}{2in}
\includegraphics[scale=.4]{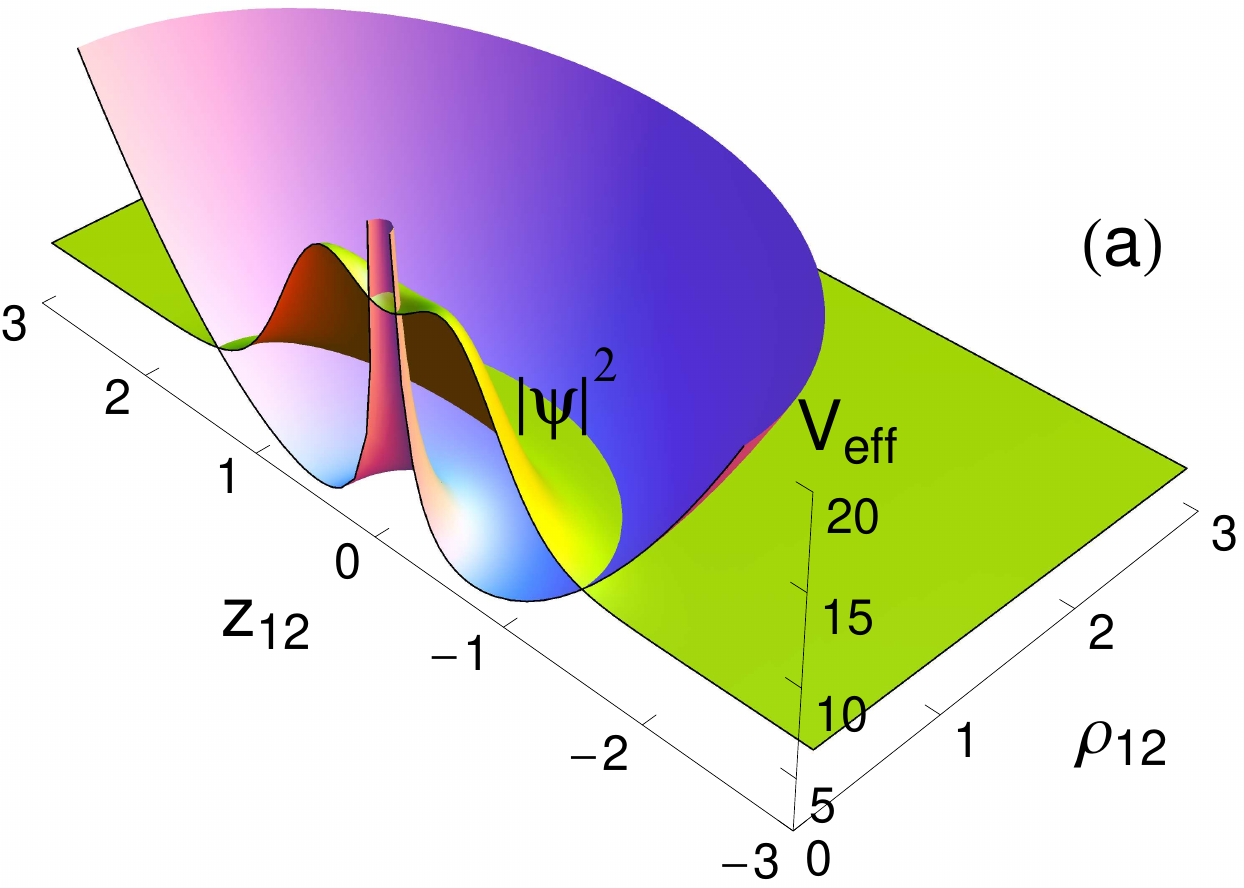}
\end{minipage}
\hspace{1.5cm}
\begin{minipage}{2in}
\includegraphics[scale=.4]{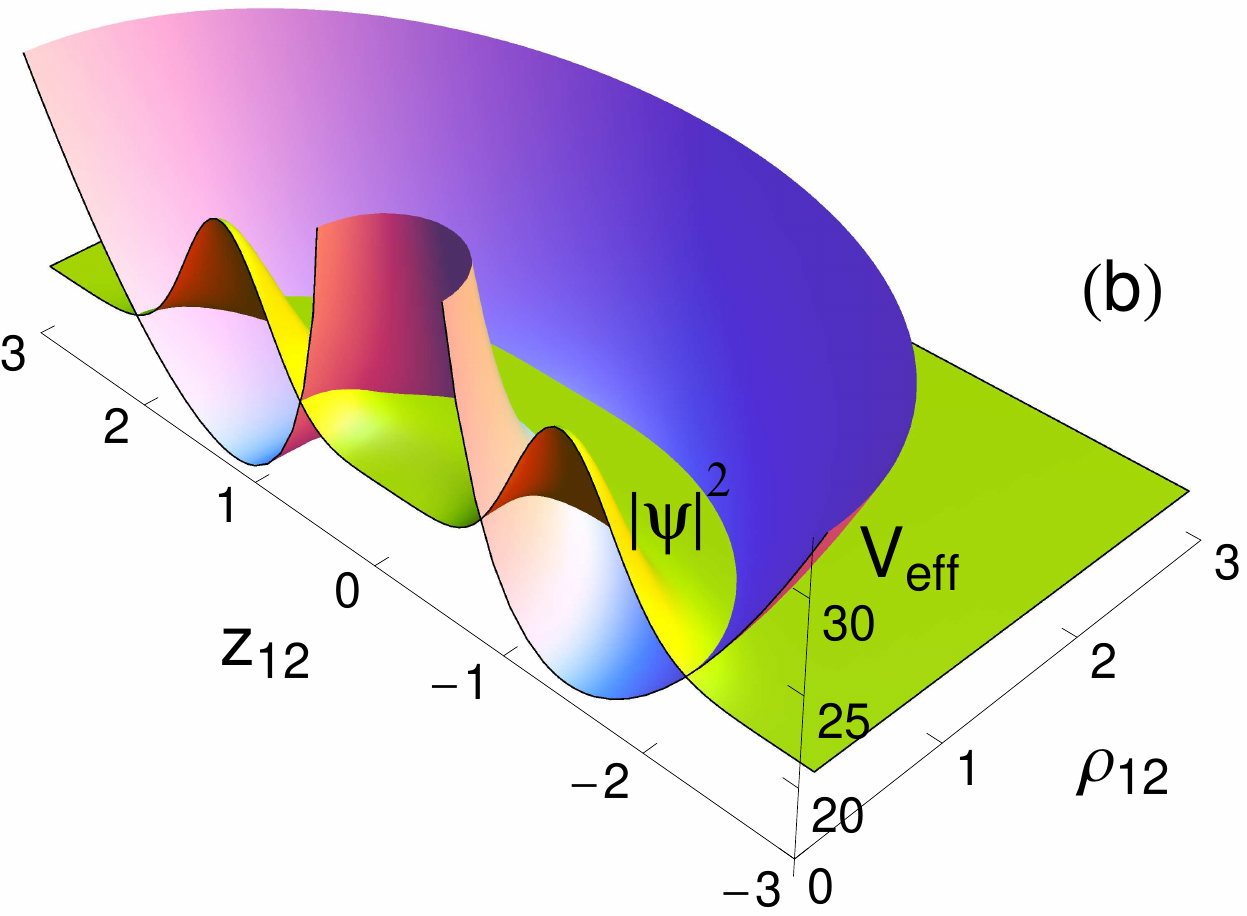}
\end{minipage}
\end{center}
\caption{(Color online) The surfaces (blue) of the effective potential~(\ref{effpot})
with $m = 0$, $\omega_z/\omega_0 = 2.5$ and $\omega_L/\omega_0 =
3.1798$ ($\omega_z/\Omega = 0.75$) and the probability densities
$|\psi(\mathbf{r}_{12})|^2$ (green) for the corresponding lowest
states for: (a) $\kappa = 1.5$, and (b)
$\kappa = 15$. In the case (b) the potential barrier with the ridge
along line $z_{12} = 0$ is sufficiently wide and high, and it
separates the motions in vicinities of the potential minima. It results in 
the double degeneracy of the energy levels.}
\label{potsurf}
\end{figure*}

These effects can be explained by the reflection symmetry of the
effective potential with respect to line $z_{12} = 0$
($\rho_{12}$-axis) and by the existence of the potential barrier
(ridge) along this line, when $\omega_L >
\omega_L^\mathrm{bif}(m)$. In QDs with small $\kappa$ ($\sim 1$)
the barrier width, as well as its height at the saddle point ($V_a
- V_b$), are small. As a result, the motions around two minima at
$(\rho_b,\pm z_b)$ are coupled, either over the barrier or by
tunnelling through the barrier (see Fig.~\ref{potsurf}(a)). A
consequence of this coupling is the splitting of the energy levels,
corresponding to even and odd states with the same values of quantum
numbers ($n_1, n_2, m$). Contrary, for large $\kappa$ ($\gg 1$)
and a sufficiently large $\omega_L$ ($> \omega_L^\mathrm{bif}(m)$)
the barrier is large enough to separate almost exactly the motions
in vicinities of the minima (see Fig.~\ref{potsurf}(b)). It results in
the double degeneracy of lowest levels.

Turning back to Fig.~\ref{fig-enlevs-k15}, one observes that at
large values of $\kappa$ the energies, obtained with the aid of the approximate
expressions (\ref{Erel-a}) and (\ref{Erel-b}), are in a good agreement with the
numerical results if our system is far from the shape transition, i.e.,
when $B \ll B_\mathrm{bif}(m)$ or $B \gg B_\mathrm{bif}(m)$. In
the later case this agreement is a consequence of the fact that
the motions around minima $(\rho_b,\pm z_b)$ are efficiently
decoupled, that was not the case at small values of $\kappa$.
Since Eq.~(\ref{Erel-b}) is derived by considering small
oscillations around one of the minima, i.e., by neglecting any
coupling, it is clear that this formula agree well with numerical
results only in the domain of double degeneracy.

The formation of doubly degenerate levels
occurs for any value of the quantum number $m$. For $m \neq
0$, however, this effect takes place at larger values of the Larmor frequency $\omega_L$
in comparison to the one for $m = 0$ (see Fig.~\ref{fig-enlevs-k15}(b)).
Evidently, the increase of the  Larmor frequency yields
the increase of values $|m|$. It results in the increase of
$\omega_L^\mathrm{bif}(m)$ for larger $|m|$.

Another manifestation of the shape transitions in the QDs with
large values of $\kappa$ is the change of slope of the lowest
energy levels $E_\mathrm{rel}(B)$ with small values of $m$,
particularly for $m = 0$, around the values $B_\mathrm{bif}(m)$
(see Figs.~\ref{fig-enlevs-k15} and \ref{low-lev}(b)). The
explanation for this effect lies in the change of the form of the wave
function $\psi(\mathbf{r}_{12})$ around the bifurcation points.

 We conclude, {\em the double degeneracy and the change of slope
of the lowest energy levels are indications of the redistributon of the electron density
in the vertical direction of the QD with well separated electrons.} From the
correlation diagram given in Fig.~\ref{fig-denk}, showing the
lowest levels with $m = 0$  at the magnetic
field larger than $B_\mathrm{bif}(0)$, it follows that doubly
degenerate levels appear if $\kappa\geq 10$.

\subsection{Singlet-triplet transitions versus shape transitions}
\label{sec:STvsSHT}

\begin{figure*}[thb]
\vspace{-7cm}
\begin{center}
\hspace{1cm}
\includegraphics[scale=.45]{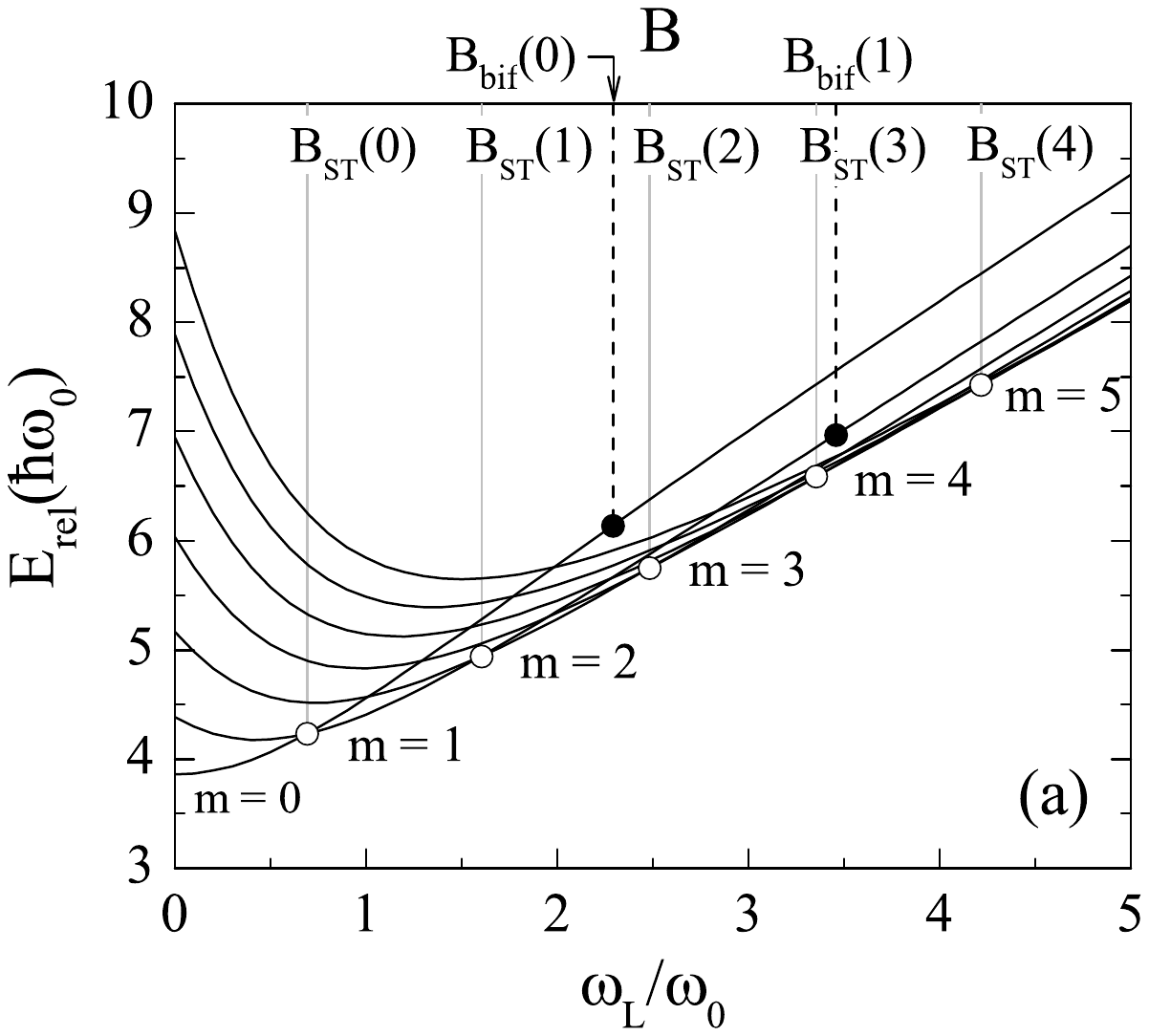}
\hspace{-3cmcm}
\includegraphics[scale=.45]{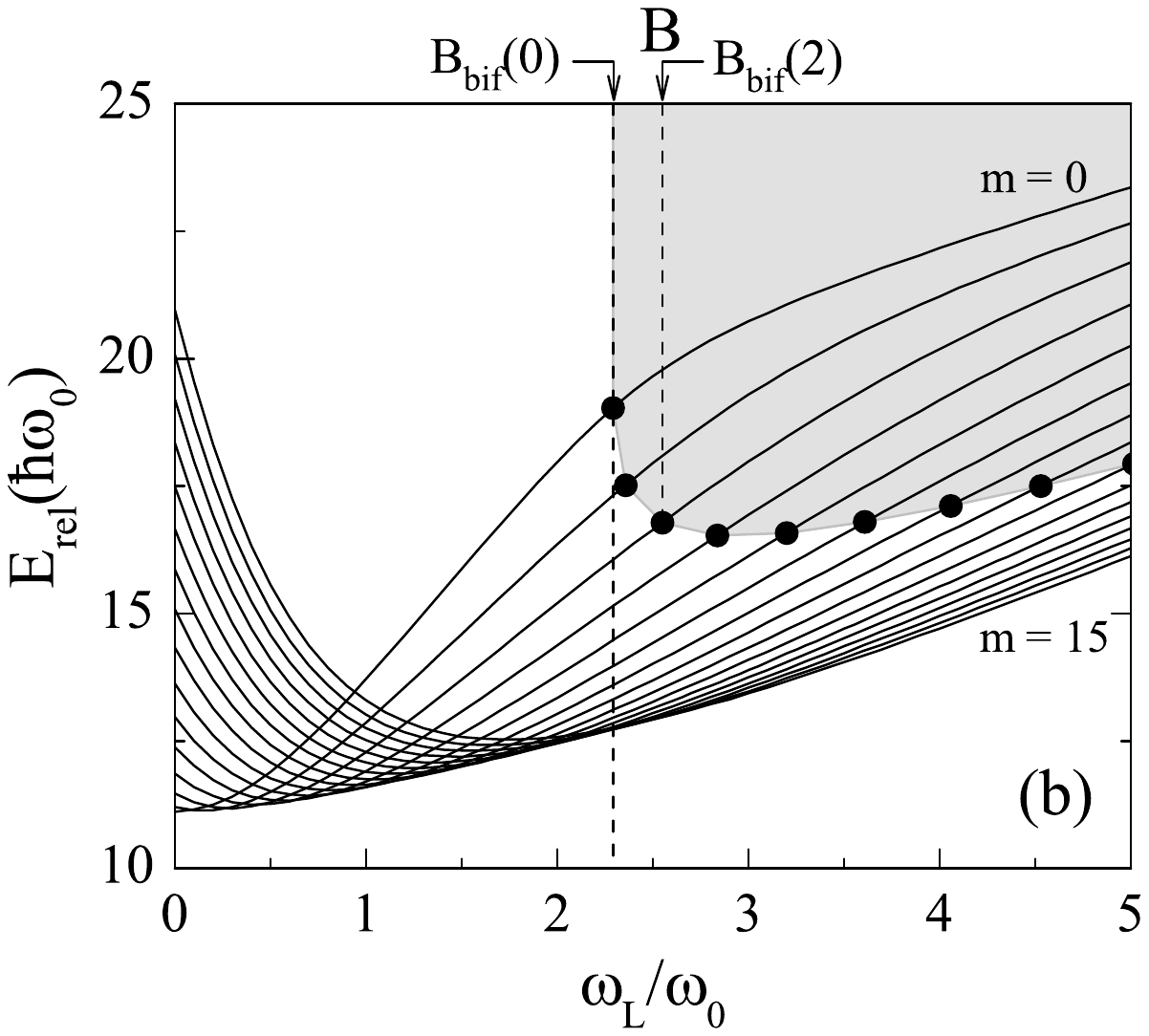}
\end{center}
\caption{The lowest energy levels with different values of quantum
number $m$ of the QD with $\omega_z/\omega_0 = 2.5$ at:
(a) $\kappa = 1.5$ ($R_W = 2.12132$) and (b) $\kappa = 15$
($R_W = 21.2132$), as functions of the scaled Larmor frequency
$\omega_L/\omega_0$ ($\sim B$).  In the panel (a) open dots mark the
positions of crossings of the lowest levels that correspond to the
singlet-triplet transitions in the ground state. The solid dots
show the positions of bifurcation points for different values of
$m$ (particularly $B_\mathrm{bif}(0) \equiv B_\mathrm{sph}$). The
shaded area covers the part of spectrum, related to the
vertical shape of QD.} \label{low-lev}
\end{figure*}

 Fig.~\ref{low-lev} displays the lowest energy levels $E_\mathrm{rel}$ with different values of
$m$ at: (a) ${\tilde\omega}_z = 2.5$ and $\kappa =
1.5$; (b) ${\tilde\omega}_z = 2.5$ and $\kappa = 15$, as
functions of the magnetic field.
From the point of view of the roto-vibrational model
(Secs.~\ref{sec:smallosc_a}, \ref{sec:smallosc_b} and
\ref{sec:approx}) this part of spectrum shows the rotational
levels around the lowest vibrational level ($n_\rho = n_z = 0$
when $B < B_\mathrm{bif}$, and $n_1 = n_2 = 0$ when $B >
B_\mathrm{bif}$). At zero magnetic field the lowest vibrational
state with $m = 0$ is the ground state (we assume here that the CM
quantum numbers are $N = N_z = M = 0$). However, with the increase of the
magnetic field strength the states with $m \neq 0$ become
the ground states at different intervals of the field
strength (see e.g. Fig.~\ref{low-lev}(a)).
This effect leads to the well-known spin oscillations or
singlet-triplet (ST) transitions in the ground state
\cite{chak,Wagner,din,thick}.

Let us denote by $B_\mathrm{ST}(m)$ the magnetic field strength, when 
the lowest level with  the magnetic quantum
number $m$  crosses the lowest level with another value of this quantum
number. We assume that these levels represent the ground state
energies at $B < B_\mathrm{ST}(m)$ and $B > B_\mathrm{ST}(m)$,
respectively. Fig.~\ref{low-lev} demonstrates that for small
and large values of the strength $\kappa$
\begin{equation}
B_\mathrm{bif}(m) > B_\mathrm{ST}(m) \label{eq:bif-vs-st}
\end{equation}
for all $m$. For example, when $m = 0$ the bifurcation point is at
$B = B_\mathrm{sph}$, but in the interval $(0,B_\mathrm{sph})$
there are several ST transitions, and the ground state of the QD
with ${\tilde\omega}_z = 2.5$ at $B = B_\mathrm{sph}$ is the
lowest vibrational state with $m = 2$ if $\kappa = 1.5$, or with
$m = 13$ if $\kappa = 15$. We have verified the generality of the rule
(\ref{eq:bif-vs-st}) and found that for the QDs with $\omega_z >
\omega _0$ it is always fulfilled, with the exception of near
spherical cases \cite{NS2013}. Thus, we can conclude that: (i)
{\em the shape transitions of axially symmetric two-electron QDs
do not occur at the ground state and takes place only in the excited states};
(ii) {\em if the Wigner
molecule is formed in the ground state it is always of the ring
type.} Of course, the vertical type of the Wigner molecule can be
formed in the lowest vibrational states with different values of
quantum number $m$ (see Sec.~\ref{sec:localization}), but at field
strengths $B > B_\mathrm{bif}(m)$ when they are not the ground
state.

\section{Conclusions}

To illuminate shape transitions in axially symmetric two-electron QDs found in
\cite{NSPC, NS2013}, we perform a thorough analysis of the
effective potential, created by the external parabolic potential, the Coulomb
interaction and the external magnetic field. The classical analysis enables
to us to identify all stationary points of the considered
potential and their properties (at fixed different values of the magnetic field). 
The electron localization is
explained by the existence of the minimum in the effective potential
at weak magnetic fields, and the shape transition is a consequence
of the bifurcation. At a specific value $B_\mathrm{bif}$ of the
magnetic field, the bifurcation  splits the potential minimum in two minima, 
separated vertically by the potential barrier.

To elucidate the quantum features of the shape transformation we
calculate the electron density with the aid of the wave function.
It is shown that with the increase of the  magnetic field
strength, the  density distribution  changes  from
the ring type in the plane to the two vertical maxima, located symmetrically around
$z_{12} = 0$. We have determined the values of the magnetic field
that trigger this shape transition at corresponding magnetic
quantum numbers $m$ for excited states, and found that these
values are related to $B_\mathrm{bif}(m)$ values.

As a matter of fact, for  $m = 0$, the shape transition takes place at
the excited state, when the effective lateral and the vertical confinement strengths
are equal, i.e., at the onset of the effective spherical symmetry.
Although the third component of the angular momentum is conserved
in the both shapes, the electron density distribution is
different. This fact is also reflected in the variation of the
cuts of the effective potential (see Fig.\ref{Veff-cut}), which
resemble in appearance  a quantum phase transition of the second
order in many-body systems (see textbook \cite{sad}).

It is noteworthy that the entanglement of two-electron QD states
(a genuine quantum property) is strongly affected by the shape transition. In our analysis
\cite{NSPC,NS2013} we have found that the entanglement decreases
by increasing the magnetic field till the bifurcation point,
reaching its minimum, and then rising again. In the considered
case of the shape ("phase") transition we speculate that the order
parameter might be associated with a moment of inertia, or with the
magnetic susceptibility of the QD. The optical response might be another
source on the shape transitions.
However, this subject is beyond
the scope of the present  paper and deserves itself a separate
study.

In the considered system the shape transition yields a family of
doubly degenerate levels with different parities and for different
values of the magnetic quantum numbers $m$ in the energy spectrum.
This transition manifests itself also as the change of the slope of
the lowest energy levels with small values $m$ with the increase
of the interaction strength. Consequently, it seems that the
appearance of the doubly  degenerate levels at some magnetic field
strength could indicate the shape transition from the lateral
distribution to the well separated electrons in the vertical
direction (depending also on the the thickness of QDs). If one
turns to the ground state properties and shape transitions, we
found that {\em there is no any shape transition in the ground
states of the axially symmetric two-electron QDs}. The Wigner molecule (if it will occur
eventually) will form a distribution of the ring type in the
lateral interface for any strength of the magnetic field. In
contrast, the shape transition occurs to the vertical density
distribution  only in {\em the excited states of the QD}
(see Fig. \ref{low-lev}).

Since for $B < B_\mathrm{bif}(m)$ and for $B >B_\mathrm{bif}(m)$ 
the effective potential has minima (which are different
in these two domains), we have used the small
oscillation approximation in the both cases. With the aid of the transformation to
the rotating frame, we determine normal coordinates and normal
frequencies. By means of these frequencies  we derive  the
analytical results for quantum eigenenergies that remarkably well
reproduce the results of quantum-mechanical numerical calculations
at strong magnetic field strengths, independently on the strength of the
Coulomb interaction $\kappa$. We have demonstrated that when one
of the normal frequencies  tends to zero the
bifurcation of the potential minimum occurs, which yields the
shape transition. Since this phenomenon takes place in different
quantum systems (e.g., \cite{m1,h1}), we suggest to use this
criterium to identify shape transitions in finite quantum systems.

\appendix

\section{Exact solution for the position of stationary point ($\mathbf{i}$)}
\label{sec:rhoa}

The positive solution of the four-order algebraic
equation~(\ref{sp1}) is
\beq
\rho_a = \frac{A}{2\sqrt{3}} +
\frac{1}{2}\sqrt{\frac{8m^2{\tilde\Omega}^2}{2^{1/3}B} -
\frac{2^{1/3}B}{6{\tilde\Omega}^4} +
\frac{\sqrt{6}\,R_W}{{\tilde\Omega}^2A}}\,, \label{rhoa}
\eeq
where $R_W = \sqrt{2}\,\kappa$ and
\beq
A = \sqrt{\frac{B^2 - 24\,2^{1/3} m^2
{\tilde\Omega}^6}{2^{2/3}{\tilde\Omega}^4 B}},
\eeq
\beq
B = {\tilde\Omega}^{8/3}\bigg[27 R_W^2 +
3\sqrt{3}\sqrt{1024\,m^6 {\tilde\Omega}^2 + 27
R_W^4}\,\bigg]^{1/3}.
\eeq

\section{The rotation parameter $c$ for normal coordinates}
\label{sec:rotpar}

With the aid of the relations (\ref{drho-q1q2}) and (\ref{dz-q1q2}), the
expansion (\ref{Vexpanded}) transforms to the form
\begin{eqnarray}
V_\mathrm{eff} \!\!\!\!&=&\!\!\!\! V_0 + \frac{1}{2}
\Big[\,c^2\Omega_\mathrm{eff}^2 + (1-c^2)\,\omega_\mathrm{eff}^2 -
2c\sqrt{1-c^2}\,\lambda\, \Big] q_1^2 \nonumber
\\
\!\!\!\!&+&\!\!\!\! \frac{1}{2}
\Big[\,(1-c^2)\,\Omega_\mathrm{eff}^2 + c^2\omega_\mathrm{eff}^2 +
2c\sqrt{1-c^2}\,\lambda\, \Big] q_2^2 \nonumber
\\
\!\!\!\!&+&\!\!\!\! \Big[\, c\sqrt{1-c^2}\,(\Omega_\mathrm{eff}^2
- \omega_\mathrm{eff}^2) + (2c^2-1)\,\lambda \,\Big] q_1 q_2.
\end{eqnarray}
In order to eliminate the mixing term ($\sim q_1 q_2$), we 
put an equated constraint
\beq
c\sqrt{1-c^2}\,(\Omega_\mathrm{eff}^2 -
\omega_\mathrm{eff}^2) + (2c^2-1)\,\lambda = 0. \label{eq4c}
\eeq
The solution of this equation yields
\begin{equation}
c = \pm \frac{1}{\sqrt{2}}\bigg( 1 \pm
\frac{|\Omega_\mathrm{eff}^2-\omega_\mathrm{eff}^2|}
{\sqrt{(\Omega_\mathrm{eff}^2-\omega_\mathrm{eff}^2)^2 +
4\lambda^2}} \bigg)^{1/2}. \label{sol4c}
\end{equation}
Using one of these values for $c$, the effective potential takes
the normal form (\ref{Veff-normf}), where the frequencies
$\omega_1$ and $\omega_2$ are given by Eqs.~(\ref{om1}),
(\ref{om2}).

For $m = 0$ one has $\rho_b = 0$, $z_b = r_b$ and $\lambda = 0$
(see Eqs.~(\ref{rhob}), (\ref{rb}) and (\ref{lambda})). In this
case the solutions of Eq.~(\ref{eq4c}) are $c = 0, \pm 1$. If we
choose $c = 1$, one obtains $q_1 = \Delta\rho$, $q_2 = \Delta z$ and
$\omega_1 = \Omega_\mathrm{eff}$, $\omega_2 =
\omega_\mathrm{eff}$.

For $m \neq 0$ the solutions (\ref{sol4c}) are not differentiable
functions of $\tilde\Omega$ at the point 
$\Omega_\mathrm{eff} = \omega_\mathrm{eff}$. A differentiable
function can be constructed if in front of the ratio
$|\Omega_\mathrm{eff}^2-\omega_\mathrm{eff}^2|/
\sqrt{(\Omega_\mathrm{eff}^2-\omega_\mathrm{eff}^2)^2 +
4\lambda^2}$ we 
put the sign '$+$' in the domain $\Omega_\mathrm{eff}
< \omega_\mathrm{eff}$, and the sign '$-$' in the domain
$\Omega_\mathrm{eff} > \omega_\mathrm{eff}$. As a result, the positive
differentiable solution can be written in the form (\ref{c-diff}).

\section{Diagonalization in the 3D oscillator basis}
\label{sec:diag}

The eigenenergies and eigenstates  of the
Hamiltonian~(\ref{relham}) are calculated by the exact diagonalization
in the oscillator basis. The basis functions are the products
of the Fock-Darwin states and oscillator functions in $z$-direction
(i.e., the eigenstates of the Hamiltonian~(\ref{relham}) for $R_W =
0$)
\bea
\label{oscfun} \psi^{(0)}_\alpha({\mathbf r}_{12}) &=&
\frac{e^{{\mathrm i}m\varphi_{12}}}{\sqrt{2\pi}}\,
f_{n_{\rho},m}(\rho_{12})\,g_{n_z}(z_{12}),
\\
\label{ffun} f_{n_\rho,m}(\rho_{12}) &=&
\sqrt{\frac{2\,\tilde\Omega\, n_\rho!} {(n_\rho\!+\!|m|)!}}\,
(\sqrt{\tilde\Omega}\,\rho_{12})^{|m|} \times \nonumber
\\
&&e^{-\frac{1}{2}{\tilde\Omega}\,\rho_{12}^2}\,
L_{n_\rho}^{|m|}({\tilde\Omega}\,\rho_{12}^2),
\\
\label{gfun} g_{n_z}(z_{12}) &=&
\frac{(\tilde\omega_z/\pi)^{1/4}}{\sqrt{2^{n_z}n_z!}}\,
e^{-\frac{1}{2}{\tilde\omega_z}z_{12}^2} \times \nonumber
\\
&&H_{n_z}(\sqrt{\tilde\omega_z}\,z_{12}),
\eea
where $\alpha = \{n_\rho,m,n_z\}$, and $L_{n_\rho}^{|m|}$ and
$H_{n_z}$ are Laguerre and Hermite polynomials, respectively. The
corresponding eigenenergies are
\beq
E_\alpha^{(0)} = \tilde\Omega\,(2 n_\rho + \vert m\vert + 1)
+ \tilde\omega_z (n_z + \hbox{$\frac{1}{2}$}) - \tilde\omega_L m.
\label{fock}
\eeq
The Hamiltonian matrix elements  are
\beq
{\cal H}_{\alpha\beta} = E^{(0)}_\alpha \delta_{\alpha\beta}
+ \frac{R_W}{\sqrt{2}}\,\langle\psi^{(0)}_\alpha\vert
r_{12}^{-1}\vert\psi^{(0)}_\beta\rangle,
\eeq
where $\beta = \{n^\prime_\rho,m^\prime,n^\prime_z\}$. 
The functions~(\ref{oscfun}) enable to us to express 
the interaction matrix elements in the analytical
form (see below).

Due to the axial symmetry of the system the $z$-component of
angular momentum of the relative motion $l_z$ is the integral of
motion (beside the energy). Consequently, the diagonalization can be
performed in the subspaces with a given value of the
magnetic quantum number $m$. In this case, it is convenient to present 
the eigenfunctions of the Hamiltonian (\ref{relham}) in the form
\beq
\psi({\bf r}_{12}) = \frac{e^{{\rm i}m\varphi_{12}}}
{\sqrt{2\pi}}\,F(\rho_{12},z_{12}), \label{psirel}
\eeq
where the functions $F$ are linear combinations of the products
$f_{n_{\rho},m}(\rho_{12})\,g_{n_z}(z_{12})$ with weights,
determined by the diagonalization procedure.

By expressing the functions (\ref{ffun}) and (\ref{gfun}) in
spherical coordinates, the interaction matrix elements take the form
\bea
\langle\psi^{(0)}_\alpha\vert
r^{-1}\vert\psi^{(0)}_\beta\rangle &=& \delta_{m m^\prime}
\int_0^\infty \!{\mathrm d}r \int_0^\pi \!{\mathrm d}\vartheta\,\,
r\sin\vartheta \times
\nonumber \\
&&f_{n_\rho,m}(r\sin\vartheta)\,
f_{n_\rho^\prime,m}(r\sin\vartheta) \times
\nonumber \\
&& g_{n_z}(r\cos\vartheta)\, g_{n_z^\prime}(r\cos\vartheta).
\eea
Using the expansions for Laguerre and Hermite polynomials
\bea
&&L_{n}^m(x) = \sum_{k=0}^n (-1)^k
\frac{(n+m)!}{k!(n-k)!\,(m+k)!}\,x^k,
\\
&&H_{n}(x) = \sum_{l=0}^{[\frac{n}{2}]} (-1)^l
\frac{n!\,(2l-1)!!}{(2l)!\,(n-2l)!}\, 2^{n-l} x^{n-2l},
\eea
where $[n/2]$ is the integer part of $n/2$, we obtain
\bea
&&\langle\psi^{(0)}_\alpha\vert
r^{-1}\vert\psi^{(0)}_\beta\rangle = \delta_{m m^\prime}\,
A_{\alpha, \beta} (\tilde\Omega,\tilde\omega_z) \times
\nonumber \\
&&\sum_{k=0}^{n_\rho} \sum_{k^\prime=0}^{n_\rho^\prime}
\sum_{l=0}^{[\frac{n_z}{2}]}
\sum_{l^\prime=0}^{[\frac{n_z^\prime}{2}]}
B_{k,k^\prime,l,l^\prime}^{\alpha, \beta}
\frac{{\tilde\Omega}^{k+k^\prime}}{{\tilde\omega}_z^{l+l^\prime}}\,
I_{k,k^\prime,l,l^\prime}(\tilde\Omega,\tilde\omega_z),\qquad
\eea
where
\bea
&&A_{\alpha, \beta} (\tilde\Omega,\tilde\omega_z) =
\tilde\Omega^{|m|+1}
\sqrt{\frac{\tilde\omega_z^{n_z+n_z^\prime+1}} {\pi\,
2^{n_z+n_z^\prime}}} \times
\nonumber \\
&&\qquad \sqrt{\frac{n_\rho!\, n_\rho^\prime!} {(n_\rho+|m|)!\,
(n_\rho^\prime+|m|)!\, n_z!\, n_z^\prime!}}, \eea
\bea &&B_{k,k^\prime,l,l^\prime}^{\alpha, \beta} =
(-1)^{k+k^\prime+l+l^\prime} \times
\nonumber \\
&&\frac{(n_\rho+|m|)!\, (n_\rho^\prime+|m|)!} {k!\, k^\prime!\,
(n_\rho-k)!\, (n_\rho^\prime-k^\prime)!\, (k+|m|)!\,
(k^\prime+|m|)!} \times
\nonumber \\
&&\frac{n_z!\, n_z^\prime!\, (2l-1)!!\, (2l^\prime-1)!!} {(2l)!\,
(2l^\prime)!\, (n_z-2l)!\, (n_z^\prime-2l^\prime)!}\,
2^{n_z+n_z^\prime-l-l^\prime},
\eea
and
\bea
&&I_{k,k^\prime,l,l^\prime}^{\alpha, \beta}
(\tilde\Omega,\tilde\omega_z) =
\Gamma(\hbox{$\frac{1}{2}$}(N_1+N_2)+1) \times \nonumber \\
&&\qquad \int_{-1}^1 \frac{(1-t^2)^{N_1/2}\, t^{N_2}}
{[\tilde\Omega(1-t^2)+\tilde\omega_z
t^2]^{\frac{1}{2}(N_1+N_2)+1}}\, {\rm d}t, \label{I-expression}
\eea
where
\beq
N_1 = 2(k+k^\prime+|m|),\quad N_2 =
n_z+n_z^\prime-2(l+l^\prime).
\eeq
The integral in Eq.~(\ref{I-expression}) is equal to zero if $N_2$
is odd. Otherwise, it can be expressed in terms of the
hypergeometric $_2F_1$-function
\bea
&&I_{k,k^\prime,l,l^\prime}^{\alpha, \beta}
(\tilde\Omega,\tilde\omega_z) = \nonumber
\\
&&\quad {\tilde\Omega}^{-\frac{N_1+N_2}{2}-1}\,
\frac{\big(\frac{N_1+N_2}{2}\big)!\,\big(\frac{N_1}{2}\big)!\,
\Gamma\big(\frac{N_2+1}{2}\big)}
{\Gamma(\hbox{$\frac{1}{2}$}(N_1+N_2+3))} \times \nonumber
\\
&&\quad _2F_1\big(\hbox{$\frac{N_2+1}{2}, \frac{N_1+N_2}{2}+1,
\frac{N_1+N_2+3}{2},
1-\frac{{\tilde\omega}_z}{\tilde\Omega}$}\big).
\eea

\section{The electron density in a two-electron QD}
\label{sec:density}

The electron density of a $n$-electron system in a state $\Psi$ is
defined as
\begin{equation}
n(\mathbf{r}) = \langle\Psi|\sum_{i=1}^n
\delta(\mathbf{r}-\mathbf{r}_i)|\Psi\rangle.
\label{eld1}
\end{equation}
For a two-electron system, characterised by
a wave function $\Psi(\mathbf{r}_1,\mathbf{r}_2)$, Eq.(\ref{eld1}) takes the form
\beq
n(\mathbf{r}) = \int
\left[\,|\Psi(\mathbf{r},\mathbf{r}^\prime)|^2+|\Psi(\mathbf{r}^\prime,\mathbf{r})|^2
\right] \mathrm{d}\mathbf{r}^\prime. \label{el-density}
\eeq

In  a two-electron QD with the parabolic confinement,
the CM and the relative motions are separated (see
Sec.~\ref{sec:model}). Consequently, the wave function reads
$\Psi(\mathbf{r}_1,\mathbf{r}_2) = \psi_\mathrm{CM}(\mathbf{R})\,
\psi(\mathbf{r}_{12})$. Using this separation and replacement
$\mathbf{r}^\prime \to \mathbf{r}^\prime + \mathbf{r}$ (under
which the integral in (\ref{el-density}) is invariant),
Eq.~(\ref{el-density}) transforms to the form
\beq
n({\bf r}) = \int |\psi_{\rm CM}({\bf r} + {\bf
r}^\prime/2)|^2 \left[\,|\psi(-{\bf r}^\prime)|^2 + |\psi({\bf
r}^\prime)|^2 \right] {\rm d}{\bf r}^\prime.
\label{el-density-sep}
\eeq

If the QD is in a state with the lowest values of quantum numbers
for the CM motion ($N = N_z = M = 0$), i.e., if
\beq
\psi_{\rm CM}({\bf R}) = \bigg(\frac{{\tilde\Omega}^2
{\tilde\omega_z}}{\pi^3}\bigg)^{1/4}
e^{-\frac{1}{2}{\tilde\Omega}(X^2\!+\!Y^2) -
\frac{1}{2}{\tilde\omega}_z Z^2},
\eeq
the electron density (\ref{el-density-sep}) takes the form
\bea
&&n(x,y,z) = \frac{\tilde\Omega}{2\pi^2}
\sqrt{\frac{{\tilde\omega}_z}{\pi}} \int\!\!\int\!\!\int {\rm
d}x^\prime {\rm d}y^\prime {\rm d}z^\prime \times \nonumber
\\
&&\qquad e^{-{\tilde\Omega}[(x+x^\prime/2)^2 +
(y+y^\prime/2)^2]}\, e^{-{\tilde\omega}_z (z+z^\prime/2)^2} \times
\\
&&\qquad \left[F^2(\sqrt{x^\prime+y^\prime},-z^\prime) +
F^2(\sqrt{x^\prime+y^\prime},z^\prime)\right], \nonumber
\label{el-density-xyz}
\eea
where $F$ is the radial part of the wave function (\ref{psirel}).
The transformation to cylindrical coordinates yields 
\bea
&&n(\rho,z) = \frac{\tilde\Omega}{\pi}
\sqrt{\frac{{\tilde\omega}_z}{\pi}}\,e^{-{\tilde\Omega}\rho^2}
\int_0^\infty\!\!\!\int_{-\infty}^{\infty} \rho^\prime {\rm
d}\rho^\prime {\rm d}z^\prime \times \nonumber
\\
&&\qquad
I_0(-\tilde\Omega\rho\rho^\prime)\,e^{-{\tilde\Omega}{\rho^\prime}^2/4}\,
e^{-{\tilde\omega}_z (z+z^\prime/2)^2} \times
\label{el-density-cyl}
\\
&&\qquad \left[F^2(\rho^\prime,-z^\prime) +
F^2(\rho^\prime,z^\prime)\right], \nonumber
\eea
where $I_0(x)$ is the modified Bessel function of the first kind
(of order zero). Using the expansion $I_0(x) = \sum_{k=0}^\infty
(x/2)^{2k}/(k!)^2$, the evaluation of 
(\ref{el-density-cyl}) is reduced to the summation over $k$ and
numerical integration over $\rho^\prime$ and $z^\prime$
coordinates.

\vfil

\end{document}